\documentclass[aps,superscriptaddress,eqsecnum,nofootinbib,preprintnumbers]{revtex4}
\usepackage{graphicx,epsfig}
\usepackage{amsmath}
\usepackage {amssymb}
\usepackage{subfigure}

\usepackage{relsize}
\newcommand{\be}{\begin{equation}}
\newcommand{\ee}{\end{equation}}
\newcommand{\bea}{\begin{eqnarray}}
\newcommand{\eea}{\end{eqnarray}}
\newcommand{\nn}{\nonumber}

\begin{document}

\title{Self-gravitating branes again}

\author{Georgios Kofinas}\email{gkofinas@aegean.gr}
\affiliation{Research Group of Geometry, Dynamical Systems and
Cosmology\\
Department of Information and Communication Systems Engineering\\
University of the Aegean, Karlovassi 83200, Samos, Greece}

\author{Maria Irakleidou}\email{maria.irakleidou@tuwien.ac.at}
\affiliation{Institute for Theoretical Physics, Vienna University of Technology,\\
Wiedner Hauptstrasse 8–10/136, A-1040 Vienna, Austria}

\date{\today}

\begin{abstract}
In view of the absence of direct observational evidence of regular gravitating defects, we raise on theoretical grounds
the question of the physical relevance of Israel matching conditions and their generalizations to higher
dimensions and codimensions, the standard cornerstone of the braneworld paradigm and other membrane scenarios.
Our reasoning is based on two points: First, the incapability of the conventional matching conditions to
accept the Nambu-Goto probe limit (even the geodesic limit of the Israel matching conditions is not
acceptable since being the geodesic equation a kinematical fact it should be preserved independent
of the gravitational theory or the codimension of the defect, which is not the case for these matching
conditions). Second, in the $D$-dimensional spacetime we live (maybe $D=4$), classical defects of any possible
codimension could in principle be constructed (even in the lab), and therefore, they should be compatible. The
standard matching conditions fail to accept codimension-2 and 3 defects for $D=4$ (which represents effectively
the spacetime at certain length and energy scales) and most probably fail to accept high enough codimensional
defects for any $D$ since there is no corresponding high enough Lovelock density to support them.

Here, we indicate that the problem is not the distributional character of the defects, neither the
gravitational theory, but mainly the equations of motion of the defects. We propose alternative matching conditions
which seem to satisfy all the previous criteria. Instead of varying the brane-bulk action with respect to the bulk metric at
the brane position, we vary with respect to the brane embedding fields in a way that takes into account the
gravitational back-reaction of the brane to the bulk (``gravitating Nambu-Goto matching conditions'').
In the present paper we consider in detail the case of a
codimension-2 brane in six-dimensional Einstein-Gauss-Bonnet gravity, prove its consistency for an axially
symmetric cosmological configuration and show that the theory possesses richer structure compared to the
standard theory. In all the cosmologies found there is the standard LFRW behaviour and extra correction terms.
In particular, one of these solutions for a radiation brane and for a range of the integration constants avoids
a cosmological singularity (both in density and curvature) and undergoes accelerated expansion near the minimum
scale factor. In the presence of an induced gravity term, there naturally appears in the theory
the effective cosmological constant scale $\lambda/(M_{6}^{4}r_{c}^{2})$, which for a value of the brane
tension $\lambda\sim M_{6}^{4}$ (e.g. $\text{TeV}^{4}$) and $r_{c}\sim H_{0}^{-1}$ gives the observed value
of the cosmological constant (a similar scale fails for other dimensions $D$).
Even if the constraint from the four-dimensional Newton's constant $G_{\!N}$
is not easily satisfied, there is still hope, since $G_{\!N}$ in general depends not only on the parameters of
the action but also on the integration constants of the considered solution.

\end{abstract}

\maketitle

\section{Introduction} \label{Introduction}

Many physical systems are described by a dynamics which models them
as relativistic membranes in an appropriate dimensional spacetime.
One example studied extensively during the last decade is the
universe itself within the context of various braneworld and other
scenarios. Distributional (thin) as well as finite thickness defects
have been studied. However, thin sources are expected to describe
the important main features of brane dynamics for an arbitrary
family of finite width regularizations in the limit of infinitesimal
thickness, and therefore, distributional description is independent
from the regularization scheme used.

It is admitted that the phenomenological action describing at lowest
order the dynamics of a classical infinitely thin test brane (probe)
with tension, moving in a given background spacetime, is
proportional to the intrinsic volume of the worldsheet, the
Nambu-Goto action \cite{CGC}. Variation of this action with respect to the
embedding fields of the brane position gives the Nambu-Goto
equations of motion which are geometrically described by the
vanishing of the trace of the extrinsic curvature, and therefore,
the worldsheet swept by the brane is extremal (minimal). Of course,
the induced metric on a Nambu-Goto brane is finite (with the
possible exception of a set of points with measure zero) which means
that the bulk metric is regular at the brane position, since only
then, the Nambu-Goto equations of motion are defined. When the
gravitational field of the defect is taken into account the
situation becomes very different and difficult since now both the
bulk metric and the brane position become dynamical. {\it{It is the
aim of this communication to examine anew the dynamics of a
classical self-gravitating brane with a bulk metric regular on the
brane}}. The situation with a bulk metric singular on the brane is
an interesting (not in the context of a braneworld) but very
different story since now the equation of motion of the defect
cannot contain the induced metric which is singular. In this case,
actually, the full non-perturbative equation of motion is unknown,
however, there is a consensus about the equation of motion for the
special case of a 0-brane (point particle) in four dimensions and
only at the linearized level around an arbitrary background
\cite{MiSaTa}, \cite{QuWa}.

Let us remind the standard/conventional treatment for obtaining the
equations of motion for the brane-bulk system. One starts from the
gravitational equations containing the Einstein term or some
modification on the left-hand side and all the matter content of the
bulk on the right-hand side with the localized brane energy-momentum
tensor included. Therefore, a source proportional to a delta
function of suitable codimension fuels the bulk gravitational
equations. Whenever a delta function appears on the right-hand side
of an equation it always tries to find an analogous distribution on
the left-hand side to balance. This balance, when possible, extracts
the discontinuity of an appropriate quantity, e.g. in
electrodynamics the discontinuities of electric/magnetic field.
However, in gravity this discontinuity refers to the (finite)
extrinsic curvature of the defect, and since the extrinsic curvature
is related to the embedding fields of the defect, it extracts as
boundary conditions (matching conditions) the equations of motion of
the defect. This is what is said that in gravity everything, bulk
dynamics as well as brane equations of motion, is contained in
gravity theory itself. By contrast, in electrodynamics one
postulates the Lorentz force equation of motion for the point
charge. Unfortunately, it is known that Einstein gravity being
non-linear does not succeed the previous balance of distributional
terms for all kinds of defects. Thin shells of matter (codimension-1
objects) do make mathematical sense within Einstein gravity and the
equations of motion for the domain wall are the well-known Israel
matching conditions \cite{Israel 1966} which are always consistent
with bulk gravity equations. Note that thin shells have always
regular metrics (analogously to the non-singular electric potential
of a charged plane). However, when a generic distributional
stress-energy tensor is supported on a codimension-2 defect the
analysis shows that it does not make mathematical sense to consider
solutions of Einstein's equations \cite{Israel 1977}, \cite{Geroch},
\cite{Garfinkle} (a pure brane tension is a special situation which
is consistent \cite{Vilenkin}, \cite{Frolov}). Early work on codimension-2 distributional
braneworlds also showed that they were pure tension objects \cite{cline}. As referred above,
possible singular solutions consistent with the underlying symmetry
(analogous to the logarithmic electric potential of a charged line)
will not be of our interest here \cite{Moss}. Not surprisingly, the problematic behaviour
still continues for other higher-codimension branes.

A nice idea to resolve the puzzle of inconsistency was to understand
that it was not the defect construction which was problematic,
rather the gravity theory itself did not have the relevant
differential complexity in order to describe complicated
distributional solutions. The clear-cut hint of this point was the
notice \cite{Ruth 2004} that upon considering the general second
derivative gravity theory in six dimensions (Einstein-Gauss-Bonnet)
one could have, at least in principle, a non trivial energy-momentum
tensor fueling geometric junction conditions for a codimension-2
conical defect. In \cite{CKP} the consistency of the whole set of
junction plus bulk field equations was explicitly shown for an
axially symmetric codimension-2 cosmological brane (not necessarily
totally geodesic) in six-dimensional EGB gravity. Presumably this
consistency will persist if axial symmetry is abandoned.
Analogously, e.g. a 5-brane in eight dimensions is again of
codimension-2 and an EGB theory would suffice, but for a 4-brane in
eight dimensions (codimension-3) the third Lovelock density \cite{Lovelock} would
need for consistency. However, it is not clear what is the situation
when codimensionality is even bigger, e.g. a 1-brane in six
dimensions (it is probably inconsistent since the spirit of the
proposal is to include higher and higher Lovelock densities to
accommodate higher codimension defects and there is no higher than
the second Lovelock density in six dimensions). In brief, the
generalization of the proposal is that in a $D$-dimensional
spacetime the maximal $[(D\!-\!1)/2]$ Lovelock density should be
included (possibly along with lower Lovelock densities) and the
branes with codimensions $\delta=1,2,...,[(D\!-\!1)/2]$ should be
consistent according to the standard treatment; for yet higher
codimensions the situation is not clear and probably inconsistent.

It is evident that the system of gravitational equations with a
delta source on the right-hand side, whenever consistent, is
equivalent to this system without the localized energy-momentum
tensor, plus the corresponding matching conditions. An equivalent
way to get all this set of equations is to take the variation of the
bulk action with respect to the bulk metric as far as the bulk
equations of motion are concerned and to take the variation of the
total brane-bulk action with respect to the bulk metric at the brane
position as far as the matching conditions are concerned. This
method of deriving the matching conditions, although formally
equivalent, presents in some cases a few technical differences
compared to the initial method where the appropriate distributional
terms are isolated. For example, for codimension-1 branes in
Einstein gravity the extra Gibbons-Hawking term has to be included
in the brane action in order for the variation to be well-defined, a
term that does not appear in the first method. Note also that
contrary to the situation in classical mechanics or field theory
where the variation of the coordinates or fields usually vanish on
the boundary, here, the variation of the bulk metric remains
arbitrary on the brane since much useful information stems from there.
Although the two methods for deriving the
matching conditions are formally equivalent, however, there is a
subtle but important point concerning branes with codimension larger
than one, which makes the variational method superior. To explain,
we note that for higher codimension branes there appear more than
one kinds of distributional terms in the full gravitational
equations. For example, for a codimension-2 brane there are
($r-$independent) terms multiplied with $\delta(r)/r$ and others
proportional to $\delta(r)$, where $r$ is the radial coordinate from
the brane. One could assume that both kinds of distributional terms
should be balanced among the two sides of the distributional
equation, arising therefore two different matching conditions. For
the six-dimensional EGB gravity the consistency mentioned above
includes only the matching conditions related to the wilder
distributional terms $\delta(r)/r$, while if additionally the extra
matching conditions \cite{Soda} arising from the distributional
terms $\delta(r)$ are included in the whole system of equations, the
problem in general becomes inconsistent \cite{CKP}. An explanation why the mild
distributional terms should be ignored is to multiply the initial
distributional equation by $r$, therefore $\delta(r)$ becomes
$r\delta(r)$ and vanishes. However, this is not a safe and
unambiguous manner to handle distributional equations and therefore
this is not the correct reasoning. Because a distribution is
defined through integration, the action and the variation of the
action naturally provides this integration. In our example of a
codimension-2 brane the volume element of integration close to the
brane is $rdrd\theta$, and therefore the quantity $r\delta(r)$
naturally and unambiguously appears sending the mild distributional
terms to zero.

To make a criticism on the standard treatment for obtaining the
equation of motion of a defect that was described in the previous
paragraphs, we mention the following points:
\par
(a) A test brane moving
in a curved background spacetime traces a minimal surface in the
lowest order approximation. When the self-gravitational field of the
brane starts to be taken into account it is natural to expect that a
small correction should result on top of the background minimal
surface motion of the test approximation. However, the equations of
motion for a gravitating defect derived following the standard
approach do not obey this condition of continuous deformation from
the Nambu-Goto probe limit. Indeed, since the matching conditions
described so far are non-perturbative, in order to realize the probe
limit one should split the brane quantities appearing in these
matching conditions (total induced metric, extrinsic
curvature and possible deficit angles) to their corresponding background plus perturbed parts.
It is not quite obvious what one should do in the probe limit
with the brane energy-momentum tensor. Either the total brane energy-momentum including the brane
tension (and the possible induced gravity term being an extra brane source) should go to zero, or the brane
tension being always non-vanishing should be kept fixed and small (as a sort of regulator) and only the
additional brane energy-momentum tensor should go to zero. More precisely, since the brane
tension has dimensions of energy per spatial volume, the product of the
brane tension with the brane volume should be much smaller than some definition of global energy of the
background space. Although there are extreme cases where either the brane volume could be infinite or the
gravitational energy attributed to a background space according to some definition of energy could be
infinite, however, a criterion of a sort of smallness of the brane tension should be necessary in the probe limit.
Under this condition, to suppress the gravitational character of the brane and go back to a probe brane
moving in the fixed background, one should formally take to zero the bulk gravity couplings which control
the gravitational back-reaction of the brane, as well as some extra matter sources which also contribute to the
back-reaction (then, all the perturbed quantities vanish) and the matching condition should reduce to the
Nambu-Goto equation.
In a similar context, in the singular case which is not the subject of our discussion here, the linearized
equation of motion of a point particle in four dimensions \cite{MiSaTa}, \cite{QuWa} is indeed a correction
of the geodesic equation of motion on a given background (of course, for a 0-brane the geodesic
equation coincides with the Nambu-Goto) and for a two-body system the probe limit is realized when one
mass is much smaller than the other.
When the brane tension is not small or the brane contains additional energy-momentum, the brane
back-reacts with the gravitational field and the full equations are needed.
Obviously, the Israel matching conditions do
not satisfy the requirement of the Nambu-Goto limit since they contain the extrinsic curvature instead of its
trace (as expected, the same also happens for
the matching condition of a codimension-1 brane in EGB gravity
theory \cite{Germani}). For the codimension-2 matching
condition in EGB theory discussed above \cite{Ruth 2004}, \cite{CKP}, \cite{Charmousis}, the situation is similar.
To go one step further, one could oppose by claiming that the correct probe limit of a defect is not
the Nambu-Goto equation of motion but the geodesic one, which means
that all the extrinsic curvatures have to become zero. This is
indeed the case for the Israel matching conditions under the
assumption that the total brane matter content goes to zero. However, this reasoning is not correct,
because if the geodesic was the correct probe limit, it would be so, independently
of the gravitational theory considered (for example, a
probe point mass moves on the geodesic of a background-solution of
any gravitational theory \cite{GerochJang}, \cite{Ehlers}), or also independently of the codimension
of the defect. But this is not the case, since other than
Einstein gravitational theories for codimension-1 or other codimension defects,
in the limit of vanishing brane energy-momentum, do not give
the geodesic equation \cite{Germani}, \cite{Ruth 2004}, \cite{Charmousis}.
\par
(b) There is an additional reason why the idea of adding higher Lovelock densities in order to get
consistency according to the standard approach cannot be the final word: generic codimension-2 or 3
branes are not allowed to reside in four dimensions since Einstein gravity is insufficient and there are
no any higher Lovelock densities in four dimensions to add. Even if four dimensions are not the actual
spacetime dimensionality, at certain length and energy scales it has been tested that four-dimensional
Einstein gravity represents effectively the spacetime to high accuracy.
\par
(c) Furthermore, it seems most probable that branes of high enough codimension (higher than $[(D\!-\!1)/2]$)
cannot reside in a $D$-dimensional spacetime according to the standard treatment since there is no corresponding
sufficiently high Lovelock density. In the $D$-dimensional spacetime we live (maybe $D=4$) classical defects
of any possible codimension should be compatible. Even if particular branes are not physically interesting
or do not appear in nature, they could in principle be constructed in the lab, and therefore the correct
mathematical framework should allow all kinds of branes (in analogy, all kinds of charged distributions exist).
It does not seem reasonable some hidden symmetry to prohibit the existence of any possible classical defect.
If one disregards this point and sticks to particular branes, the criteria of consistency of a given formulation
are reduced. Additionally, since both regular and singular solutions are in general allowed mathematically,
it would be peculiar if the regular solutions we investigate here are not permitted. Moreover, it seems impossible
the equations of motion of the various codimensional defects to be derived through different methods, but a
unified principle should give the equations of motion of all kinds of defects. Possible direct observational
evidence of codimension-1 defects in 4-dimensions could shed light on the issue of matching conditions.
\par
{\it The present paper studies, instead of the standard, alternative matching conditions which aim to satisfy
the previous three points}. In particular, concerning the first point above, these alternative matching
conditions always have the Nambu-Goto probe limit, independently of the gravitational theory considered, the
dimensionality of spacetime or the codimensionality of the defect. As far as the second point is concerned,
considering the alternative matching conditions, a codimension-2 brane is consistent in $D$-dimensional Einstein
gravity \cite{KofTom}, or in particular in four-dimensional Einstein gravity. Finally, for the third point, the
simple case of a codimension-1 brane is still, as in the standard approach, always consistent, independently
of the gravitational theory. The codimension-2 brane in EGB theory, examined in the present paper, is also proved
to be consistent. Accordingly, it is expected, although we do not have a proof, that any higher codimension
brane will also be consistent in either Einstein gravity or any Lovelock extension.

Let us describe now the method for deriving these alternative matching conditions. In the standard
method the variation of the bulk action with respect to the bulk metric gives the bulk
equations of motion and the variation of the total brane-bulk action with respect to the bulk metric
at the brane position gives the matching conditions. Although the bulk equations of motion are not debatable,
the sort of variation at the brane position is not unquestionable. The brane defines a sort of boundary
(not necessarily of codimension-1) and a boundary is an exceptional place whose position
is primarily determined by the embedding fields and secondarily/implicitly by its induced metric or the bulk
metric at the brane position. The Nambu-Goto equations of motion arise by varying the
Nambu-Goto action with respect to the embedding fields. Typically, the Nambu-Goto action being proportional
to the worldsheet volume is a function of the induced brane metric which depends explicitly and
implicitly (through the bulk metric at the brane position) on the embedding fields. If, furthermore, the brane
back-reacts with the bulk gravity, gravity will be also present at the brane position and the motion of the
brane will be influenced by the gravitational bulk action.
A variation of the embedding fields implies a variation of the bulk metric at the brane position, and
therefore, additional contributions to the brane equation of motion beyond the Nambu-Goto term arise from
the variation of the bulk action with respect to the embedding fields. The distributional terms are responsible
for this contribution. Again, as in the standard treatment, everything, bulk dynamics as well as
brane equations of motion is contained in gravity theory itself, but in a different manner than before. Here, the
brane equation of motion is the result of the variation of the brane-bulk action at the brane position with
respect to the position variables, what can be called ``gravitating Nambu-Goto matching conditions''.
Although the brane energy-momentum tensor is still defined by the variation
of the brane action with respect to the bulk metric at the brane position, however, this tensor enters the new
matching conditions in a different way than before. The present proposal is not based
primarily on the modification of the gravitational theory, but on the modification of the matching
conditions. Since the consistency here is not based crucially on the inclusion of the maximal Lovelock density,
it is plausible that the consistency will occur even for all higher codimension defects. Here, the distributional
terms are still present, not inside a distributional differential equation leading directly to inconsistencies
at certain cases, but rather smoothed out inside an integration. Of course, in a higher $D$-dimensional spacetime
higher Lovelock densities should in principle contribute to the bulk action, and actually, the consistency of
such a theory, along with the modified matching conditions described above, is the main subject of study in
the present paper. Moreover, the inclusion of such extra terms in the action leads probably to physically more
interesting and realistic solutions.

Our approach is reminiscent of the `Dirac style' variation performed in \cite{Davidson} in the study of
codimension-1 defects. It also resembles the Regge-Teitelboim brane gravitational theory \cite{Teitelboim}
which is an extension of the Nambu-Goto style of variation, with the crucial difference however that in
\cite{Teitelboim} there are no higher-dimensional gravity terms in the action and the bulk space is prefixed
(Minkowski) instead of dynamical which is here. In \cite{Davidson}, since the probe limit and the
consistency of the various codimensions were not considered, the standard approach was not set into doubt, but
rather the derivation was basically suggested as a formal treatment to unify different gravitational theories.
In \cite{KofTom}, by varying with respect to the embedding fields, the alternative matching conditions
of a 3-brane in six-dimensional Einstein gravity were derived and their consistency was shown for an
axially symmetric configuration (the same however is true for a codimension-2 brane in any dimension).
In the present paper we go one step further and derive the matching conditions of a codimension-2 brane
in EGB theory. Their consistency is checked for an axially symmetric cosmological configuration. The derived
cosmologies have various differences compared to the corresponding cosmologies derived using the standard
matching conditions \cite{CKP}. The main point stressed in the present paper is that in view of the three
points mentioned above, the alternative matching conditions derived by varying with respect to the embedding
fields instead of the bulk metric at the brane position, could well be closer to the correct direction for
deriving realistic matching conditions, compared to the Israel matching conditions and generalizations.

The set-up of the paper is as follows: In section \ref{General arguments and introduction of the method}
the method is introduced as an extension of the Nambu-Goto variation so that the contribution from the
gravitational back-reaction is included. In section \ref{General setup, matching conditions and effective equations}
we consider a 3-brane in six-dimensional Einstein-Gauss-Bonnet gravity and derive the generic alternative
matching conditions and the remaining effective equations on the brane. Similar or identical equations hold
for other codimension-2 branes in other spacetime dimensions, but we choose the 3-brane as it can
represent our world in the braneworld scenario. In section \ref{Cosmological equations and consistency} we
specialize to the cosmological configuration and demonstrate the consistency of the system. In section
\ref{cosmo evolution} we investigate the cosmological equations and derive solutions for the cosmic evolution.
In section \ref{specom} we study a few special characteristic cases, we discuss the Einstein limit of the theory
and compare with the standard treatment where the conventional matching conditions are used. Finally, in
section \ref{Conclusions} we conclude.

\section{General arguments and introduction of the method}
\label{General arguments and introduction of the method}

In order to understand how the proposed variation with respect to the embedding fields of the brane position
is performed, we start with a general four-dimensional action
of the form $s_{4}\!=\!\int_{\Sigma}d^{4}\chi\sqrt{|h|}\,L(h_{ij})$, where $L$ is any
scalar on $\Sigma$ built up from the induced metric $h_{ij}$. The brane coordinates are $\chi^{i}$
($i,j...$ are coordinate indices on the brane) and the bulk coordinates are $x^{\mu}$ ($\mu,\nu,...$ are
$D$-dimensional indices). In this section $D$ is arbitrary, while in the next sections of the paper we
will specialize to $D=6$. In the present paragraph the bulk metric $\textsl{g}_{\mu\nu}$ is fixed and
non-dynamical, while the treatment of a back-reacted metric will be given in the next paragraphs of this
section. The embedding fields are the external (bulk) coordinates of the brane, so they are some functions
$x^{\mu}(\chi^{i})$. We could use a capital case letter for the embedding fields to distinguish from the bulk
coordinates, but it is better not to do so because first, the embedding fields are the bulk coordinates at the
brane position, second, the functions $x^{\mu}(\chi^{i})$ do not define the brane position unless the bulk
coordinates $x^{\mu}$ in the full space are given, and third, since our concern is the brane equation of
motion, the bulk coordinates away from the brane only incidentally will be considered. Since
$h_{ij}=\textsl{g}_{\mu\nu}\,x^{\mu}_{\,\,,i}\,x^{\nu}_{\,\,,j}$ and on the brane it is
$\textsl{g}_{\mu\nu}(x^{\lambda})$, thus $h_{ij}$ has explicit and implicit dependence on the embedding fields
$x^{\mu}$. Let their variation is described by the displacement vector $\bar{\delta}x^{\mu}(x^{\nu})$
and the corresponding variation of the various quantities is denoted by $\bar{\delta}_{x}$. The quantities
$x^{\mu}_{\,\,,i}$ are tangent vectors on the brane and their variation is $\bar{\delta}_{x}(x^{\mu}_{\,\,,i})
=(\bar{\delta}x^{\mu})_{,i}\equiv \bar{\delta} x^{\mu}_{\,\,,i}$.
We must observe that the variation of $\textsl{g}_{\mu\nu}$ is
$\bar{\delta}_{x}\textsl{g}_{\mu\nu}\!=\!\textsl{g}_{\mu\nu,\lambda}\bar{\delta} x^{\lambda}$, so
$\textsl{g}_{\mu\nu}$ is considered as a simple scalar function of $x^{\lambda}$ ignoring the possible tensorial
indices. The reason is that the bulk coordinates, which determine the tensorial behaviour of the quantities with
spacetime indices, do not change, only the brane position changes, i.e. the embedding fields are varied.
So, the spirit of the $\bar{\delta}_{x}$ variation is that when an explicit embedding field $x^{\mu}$
is met, it is varied, while when a function of this $x^{\mu}$ is met, the partial derivative is taken.
Then, $\bar{\delta}_{x}h_{ij}=\textsl{g}_{\mu\nu,\lambda}x^{\mu}_{\,\,,i}x^{\nu}_{\,\,,j}\bar{\delta} x^{\lambda}
+\textsl{g}_{\mu\nu}x^{\mu}_{\,\,,i}\bar{\delta} x^{\nu}_{\,\,,j}+\textsl{g}_{\mu\nu}x^{\nu}_{\,\,,j}
\bar{\delta} x^{\mu}_{\,\,,i}=x^{\mu}_{\,\,,i}x^{\nu}_{\,\,,j}(\textsl{g}_{\mu\nu,\lambda}\bar{\delta}x^{\lambda}
+\textsl{g}_{\mu\lambda}\bar{\delta}x^{\lambda}_{\,\,,\nu}+\textsl{g}_{\nu\lambda}\bar{\delta}x^{\lambda}_{\,\,,\mu})$.
If we had wrongly considered the tensorial character of $\textsl{g}_{\mu\nu}$, its variation would be
$\textsl{g}\,'_{\mu\nu}(x^{\rho}+\bar{\delta} x^{\rho})-\textsl{g}_{\mu\nu}(x^{\rho})
=-\textsl{g}_{\mu\lambda}(x^{\rho})\bar{\delta} x^{\lambda}_{\,\,,\nu}-\textsl{g}_{\nu\lambda}
(x^{\rho})\bar{\delta} x^{\lambda}_{\,\,,\mu}$, so the variation of $h_{ij}$ would vanish, which is a trivial result expressing
no dynamics and is simply due to that $\textsl{g}_{\mu\nu}\,x^{\mu}_{\,\,,i}\,x^{\nu}_{\,\,,j}$ is scalar in the
spacetime indices $\mu,\nu$. The variation of $s_{4}$ is
$\bar{\delta}_{x}s_{4}\!=\!\int_{\Sigma}d^{4}\chi\sqrt{|h|}\,\tau^{ij}\bar{\delta}_{x}h_{ij}$, where
$\tau^{ij}\!=\!\frac{\delta L}{\delta h_{ij}}+\frac{L}{2}h^{ij}$. Substituting $\bar{\delta}_{x}h_{ij}$,
integrating by parts and imposing $\bar{\delta} x^{\mu}|_{\partial \Sigma}=0$, we get $\bar{\delta}_{x}s_{4}=-2
\int_{\Sigma}d^{4}\chi\sqrt{|h|}\,\textsl{g}_{\mu\sigma}[(\tau^{ij}x^{\mu}_{\,\,,i})_{|j}+\tau^{ij}
\Gamma^{\mu}_{\,\,\,\nu\lambda}x^{\nu}_{\,\,,i}x^{\lambda}_{\,\,,j}]\bar{\delta} x^{\sigma}=
-2\int_{\Sigma}d^{4}\chi\sqrt{|h|}\,\textsl{g}_{\mu\sigma}[\tau^{ij}_{\,\,\,\,|j}x^{\mu}_{\,\,,i}+\tau^{ij}
(x^{\mu}_{\,\,;ij}+\Gamma^{\mu}_{\,\,\,\nu\lambda}x^{\nu}_{\,\,,i}x^{\lambda}_{\,\,,j})]\bar{\delta} x^{\sigma}=-2
\int_{\Sigma}d^{4}\chi\sqrt{|h|}\,\textsl{g}_{\mu\sigma}(\tau^{ij}_{\,\,\,\,|j}\,x^{\mu}_{\,\,,i}-
\tau^{ij}
K^{\alpha}_{\,\,\,\,ij}n_{\alpha}^{\,\,\,\mu})\bar{\delta} x^{\sigma}$, since $x^{\mu}_{\,\,|ij}=x^{\mu}_{\,\,;ij}$,
$x^{\mu}_{\,\,;ij}+\Gamma^{\mu}_{\,\,\,\nu\lambda}x^{\nu}_{\,\,,i}x^{\lambda}_{\,\,,j}=-K^{\alpha}_{\,\,\,\,ij}
n_{\alpha}^{\,\,\,\mu}$, where $K^{\alpha}_{\,\,\,\,ij}=n^{\alpha}_{\,\,\,\,i;j}$ are the extrinsic curvatures on
the brane and $n_{\alpha}^{\,\,\,\mu}$ ($\alpha=1,...,\delta\!=\!D-4$) form a basis of normal vectors to the brane. The covariant
differentiations $|$ and $;$ correspond to $h_{ij}$ and $\textsl{g}_{\mu\nu}$ respectively, while
$\Gamma^{\mu}_{\,\,\,\nu\lambda}$ are the Christoffel symbols of
$\textsl{g}_{\mu\nu}$. Due to the arbitrariness of $\bar{\delta} x^{\mu}$ it arises
$\tau^{ij}_{\,\,\,\,|j}\,x^{\mu}_{\,\,,i}-\tau^{ij}K^{\alpha}_{\,\,\,\,ij}n_{\alpha}^{\,\,\,\mu}=0$ and since the
vectors $x^{\mu}_{\,\,,i}$, $n_{\alpha}^{\,\,\,\mu}$ are independent, two sets of equations arise
$\tau^{ij}_{\,\,\,\,|j}=0$, $\tau^{ij}K^{\alpha}_{\,\,\,\,ij}=0\Leftrightarrow\tau^{ij}(x^{\mu}_{\,\,;ij}
+\Gamma^{\mu}_{\,\,\,\nu\lambda}x^{\nu}_{\,\,,i}x^{\lambda}_{\,\,,j})=0$ (since $n_{\alpha}^{\,\,\,\mu}
n^{\beta}_{\,\,\,\mu}=\delta^{\beta}_{\alpha}$). Note that the previous equivalence of the two expressions,
one with free index $\alpha$ and the other with free index $\mu$ is due to that the vectors
$x^{\mu}_{\,\,;ij}+\Gamma^{\mu}_{\,\,\,\nu\lambda}x^{\nu}_{\,\,,i}x^{\lambda}_{\,\,,j}$ are normal to the brane.
The variation described so far is the same with the one leading
to the Nambu-Goto equation of motion. Indeed for $L=1$, it is $\tau^{ij}=\frac{1}{2}h^{ij}$ and the
first equation is empty, while the second becomes $h^{ij}K^{\alpha}_{\,\,\,\,ij}=0
\Leftrightarrow \Box_{h}x^{\mu}+\Gamma^{\mu}_{\,\,\,\nu\lambda}h^{\nu\lambda}=0$ (since $-n_{\alpha}
^{\,\,\,\mu}h^{ij}K^{\alpha}_{\,\,\,\,ij}=\Box_{h}x^{\mu}+\Gamma^{\mu}_{\,\,\,\nu\lambda}h^{\nu\lambda}$)
which is the Nambu-Goto equation of motion. Note again that the previous equivalence of the two expressions for the
Nambu-Goto equation, one with free index $\alpha$ and the other with free index $\mu$ is due to that the vector
$\Box_{h}x^{\mu}+\Gamma^{\mu}_{\,\,\,\nu\lambda}h^{\nu\lambda}$ is normal to the brane.
Similarly, the Regge-Teitelboim equation of motion \cite{Teitelboim} is a generalization where $L$ collects
the four-dimensional terms of (\ref{Stotal}), i.e. $L=\frac{r_{c}^{D-4}}{2\kappa_{D}^{2}}R-\lambda+\frac{L_{mat}}
{\sqrt{|h|}}$. It is $\tau^{ij}=-\frac{1}{2}\big(\frac{r_{c}^{D-4}}{\kappa_{D}^{2}}G^{ij}+\lambda h^{ij}
-T^{ij}\big)$, so the first equation becomes the standard conservation $T^{ij}_{\,\,\,\,\,|j}=0$ and the second
$\big(\frac{r_{c}^{D-4}}{\kappa_{D}^{2}}G^{ij}+\lambda h^{ij}-T^{ij}\big)K^{\alpha}_{\,\,\,\,ij}=0$.

In order to express the back-reaction of the brane onto the bulk and vice-versa, we now go one step further and
consider a general higher-dimensional action of the form
$s_{_{\!D}}\!=\!\int_{M}d^{D}x\sqrt{|\textsl{g}|}\,\mathcal{L}(\textsl{g}_{\mu\nu})$, where $\mathcal{L}$
is any scalar on $M$ built up from the metric $\textsl{g}_{\mu\nu}$, e.g. $\mathcal{L}=\mathcal{R}(\textsl{g}
_{\mu\nu})$. Under an arbitrary variation of the bulk metric $\delta\textsl{g}_{\mu\nu}$ the variation of
$s_{_{\!D}}$ is
$\delta s_{_{\!D}}\!=\!\int_{M}d^{D}x\sqrt{|\textsl{g}|}\,\textsc{e}^{\mu\nu}\delta\textsl{g}_{\mu\nu}$,
where
$\textsc{e}^{\mu\nu}=\frac{\delta\mathcal{L}}{\delta\textsl{g}_{\mu\nu}}+\frac{\mathcal{L}}{2}\textsl{g}^{\mu\nu}$,
and the stationarity $\delta s_{_{\!D}}=0$ under arbitrary variations $\delta\textsl{g}_{\mu\nu}$ gives the bulk
field equations $\textsc{e}^{\mu\nu}=0$. The boundary terms arising from this variation in the presence of a
defect disappear by a suitable choice for the boundary condition of $\delta\textsl{g}_{\mu\nu}$,
usually by choosing a Dirichlet boundary condition for $\delta\textsl{g}_{\mu\nu}$.
However, in the presence of the defect, inside
$\textsc{e}^{\mu\nu}$, beyond the regular terms which obey $\textsc{e}^{\mu\nu}\!=\!0$, in general there are also
non-vanishing distributional terms making the variation $\delta s_{_{\!D}}$ not identically zero. The bulk
action knows about the defect through these distributional terms. Since
$\text{distr}\,\textsc{e}^{\mu\nu}\propto \delta^{(\delta)}$, where $\delta^{(\delta)}$ is the
$\delta$-dimensional delta function with support on the defect, it is
$\text{distr}\,\textsc{e}^{\mu\nu}\delta\textsl{g}_{\mu\nu}\propto \delta^{(\delta)}\delta\textsl{g}_{\mu\nu}=
\delta^{(\delta)}\,\delta\textsl{g}_{\mu\nu}|_{brane}$, so only the variation of the bulk metric at the
brane position contributes to $\delta s_{_{\!D}}$.
More precisely, these distributional terms always appear in the parallel to the brane components and if
$\text{distr}\,\textsc{e}^{ij}=k^{ij}\delta^{(\delta)}$ the variation $\delta s_{_{\!D}}$ gets the form
$\delta s_{_{\!D}}=\int_{M}d^{D}x\sqrt{|\textsl{g}|}\,k^{ij}\delta^{(\delta)}\delta h_{ij}=
\int_{\Sigma}d^{4}\chi\sqrt{|h|}\,k^{ij}\delta h_{ij}$.
Therefore, there is an extra variation of the bulk metric at the brane position $\delta\textsl{g}_{\mu\nu}|_{brane}$
(which in the adapted frame coincides with the variation of the induced metric $\delta h_{ij}$) which is independent from the bulk
metric variation and this extra variation determines the brane equation of motion (actually, it is sensible the
brane equation of motion not to depend on the variation of the fields away from the brane).
The corresponding variation of the total action at the brane position is
$\delta (s_{_{\!D}}+s_{4})|_{brane}=\int_{\Sigma}d^{4}\chi\sqrt{|h|}\,(k^{ij}+\tau^{ij})\delta
h_{ij}$. In particular, if all the components of the variation $\delta h_{ij}$ are independent from each other,
the stationarity of the total action at the brane position $\delta (s_{_{\!D}}+s_{4})|_{brane}=0$ gives the
standard matching conditions (or standard brane equations of motion)
$k^{ij}+\tau^{ij}=0\Leftrightarrow k^{\mu\nu}\!+\!\tau^{\mu\nu}=0$, where $k^{\mu\nu}=k^{ij}x^{\mu}_{\,\,,i}
x^{\nu}_{\,\,,j}$, $\tau^{\mu\nu}=\tau^{ij}x^{\mu}_{\,\,,i}x^{\nu}_{\,\,,j}$ are parallel to the brane tensors.
Our aim is to consider a variation $\bar{\delta}x^{\mu}$ of the embedding fields and derive a meaningful
non-trivial brane equation of motion in the presence of a higher-dimensional action $s_{_{\!D}}$ on top of the
four-dimensional action $s_{4}$. As we have seen in the previous paragraph, since the bulk coordinates do not
change, $\textsl{g}_{\mu\nu}$ at the brane position transforms as
$\bar{\delta}_{x}\textsl{g}_{\mu\nu}\!=\!\textsl{g}_{\mu\nu,\lambda}\bar{\delta} x^{\lambda}$ and the induced
metric as
$\bar{\delta}_{x}h_{ij}=x^{\mu}_{\,\,,i}x^{\nu}_{\,\,,j}(\textsl{g}_{\mu\nu,\lambda}\bar{\delta}x^{\lambda}
+\textsl{g}_{\mu\lambda}\bar{\delta}x^{\lambda}_{\,\,,\nu}+\textsl{g}_{\nu\lambda}\bar{\delta}x^{\lambda}_{\,\,,\mu})$.
Now, the corresponding variation of the total action at the brane position is due to $\bar{\delta}x^{\mu}$ and
since $\bar{\delta}_{x}h_{ij}$ is a special variation of the arbitrary $\delta h_{ij}$, it will be
$\bar{\delta}_{x}(s_{_{\!D}}+s_{4})|_{brane}=\int_{\Sigma}d^{4}\chi\sqrt{|h|}\,(k^{ij}+\tau^{ij})\bar{\delta}
_{x}h_{ij}$. Following the same steps as in the previous paragraph with $\tau^{ij}$ replaced by
$k^{ij}+\tau^{ij}$, the stationarity $\bar{\delta}_{x}(s_{_{\!D}}+s_{4})|_{brane}=0$ gives, due to the
arbitrariness of $\bar{\delta} x^{\mu}$, the brane equations of motion
$(k^{ij}+\tau^{ij})_{|j}=0$, $(k^{ij}+\tau^{ij})K^{\alpha}_{\,\,\,\,ij}=0
\Leftrightarrow(k^{ij}+\tau^{ij})(x^{\mu}_{\,\,;ij}
+\Gamma^{\mu}_{\,\,\,\nu\lambda}x^{\nu}_{\,\,,i}x^{\lambda}_{\,\,,j})=0$. These can be called ``gravitating
Nambu-Goto matching conditions'' since they collect also the contribution from bulk gravity and they form a
schematic summary of our proposal. Nothing ab initio assures their consistency with the bulk field equations.

There is an equivalent way to describe this variation and get the same results. Instead of considering the
active notion above where the brane is deformed to another position defined by the displacement vector
$\bar{\delta}x^{\mu}$, we consider the passive viewpoint of a bulk coordinate change
$x^{\mu}\rightarrow x'^{\mu}=x^{\mu}+\delta x^{\mu}$, where now the brane does not change position but it is
described by different coordinates and the change of the embedding fields is $\delta x^{\mu}$. Of course,
only the value of the variation $\delta x^{\mu}$ on the brane and not the
values $\delta x^{\mu}$ away from the brane is expected to influence the corresponding variation of the
brane-bulk action at the brane position. Although a bulk action $s_{_{\!D}}$ is invariant under coordinate
transformations, the presence of the defect, i.e. of the distributional terms inside $s_{_{\!D}}$, make
$\delta_{x}s_{_{\!D}}|_{brane}\!\neq\!0$. The crucial point is how the tensor fields $\phi^{\mu}_{\nu}(x^{\rho})$
are varied. One option which is unsuccessful is the tensorial variation $\tilde{\delta}_{x}\phi^{\mu}_{\nu}=
\phi\,'^{\mu}_{\nu}(x'^{\rho})-\phi^{\mu}_{\nu}(x^{\rho})=\phi^{\lambda}_{\nu}(x^{\rho})\delta x^{\mu}_{,\lambda}
-\phi^{\mu}_{\lambda}(x^{\rho})\delta x^{\lambda}_{,\nu}$, where $\delta x^{\mu}_{\,\,\,,\nu}\equiv
(\delta x^{\mu})_{,\nu}$. The successful option is the {\textit{functional}} variation which is the change
in the functional form of $\phi^{\mu}_{\nu}$, i.e.
$\delta_{x}\phi^{\mu}_{\nu}=\phi\,'^{\mu}_{\nu}(x^{\rho})-\phi^{\mu}_{\nu}(x^{\rho})
=\phi^{\lambda}_{\nu}(x^{\rho})\delta x^{\mu}_{,\lambda}
-\phi^{\mu}_{\lambda}(x^{\rho})\delta x^{\lambda}_{,\nu}-\phi^{\mu}_{\nu,\lambda}\delta x^{\lambda}
=-\pounds_{\delta x}\phi^{\mu}_{\nu}$. So, the fields transform according to the Lie derivative
with generator the infinitesimal coordinate change and $\delta_{x}\phi^{\mu}_{\nu}$ is tensor, contrary to
$\tilde{\delta}_{x}\phi^{\mu}_{\nu}$ which is not. In particular,
the functional variation of $\textsl{g}_{\mu\nu}$ is $\delta_{x} \textsl{g}_{\mu \nu}=
\textsl{g}\,'_{\mu\nu}(x^{\rho})-\textsl{g}_{\mu\nu}(x^{\rho})
= -(\textsl{g}_{\mu \nu , \lambda} \delta x^{\lambda}+\textsl{g}_{\mu
\lambda} {\delta x^{\lambda}}_{, \nu}+\textsl{g}_{\nu \lambda} {\delta
x^{\lambda}}_{,\mu})=-\pounds_{\delta x} \textsl{g}_{\mu \nu}$. The functional change of the embedding fields
$x^{\mu}(\chi^{i})$ is $\delta_{x}x^{\mu}=x'^{\mu}|_{x^{\nu}}-x^{\mu}=x^{\mu}-x^{\mu}=0$, and for the
tangent vectors it is $\delta_{x}(x^{\mu}_{\,\,,i})=0$, or to be more formal $\delta_{x}(x^{\mu}_{\,\,,i})=
\delta_{x}t^{\mu}_{(i)}=t'^{\mu}_{(i)}(x^{\rho})-t^{\mu}_{(i)}(x^{\rho})=x^{\mu}_{\,\,,i}-x^{\mu}_{\,\,,i}=0$,
where $t^{\mu}_{(i)}(x^{\rho})=x^{\mu}_{\,\,,i}$, $t'^{\mu}_{(i)}(x'^{\rho})=x'^{\mu}_{\,\,,i}$.
Therefore, contrary to the $\bar{\delta}x^{\mu}$ variation, here the fields $x^{\mu}_{\,\,,i}$ are not
varied, but the functional variation of $h_{ij}=\textsl{g}_{\mu\nu}\,x^{\mu}_{\,\,,i}\,x^{\nu}_{\,\,,j}$
is again the same $\delta_{x}h_{ij}=(\delta_{x}\textsl{g}_{\mu\nu})x^{\mu}_{\,\,,i}x^{\nu}_{\,\,,j}
=-x^{\mu}_{\,\,,i}x^{\nu}_{\,\,,j}(\textsl{g}_{\mu \nu , \lambda} \delta x^{\lambda}+\textsl{g}_{\mu
\lambda} {\delta x^{\lambda}}_{, \nu}+\textsl{g}_{\nu \lambda} {\delta x^{\lambda}}_{,\mu})$.
The variation of the total brane-bulk action at the brane position is
$\delta_{x}(s_{_{\!D}}+s_{4})|_{brane}=\int_{\Sigma}d^{4}\chi\sqrt{|h|}\,(k^{ij}+\tau^{ij})\delta_{x}h_{ij}$
and the stationarity $\delta_{x}(s_{_{\!D}}+s_{4})|_{brane}=0$ gives, due to the
arbitrariness of $\delta x^{\mu}$, the same brane equations of motion
$(k^{ij}+\tau^{ij})_{|j}=0$, $(k^{ij}+\tau^{ij})K^{\alpha}_{\,\,\,\,ij}=0
\Leftrightarrow(k^{ij}+\tau^{ij})(x^{\mu}_{\,\,;ij}
+\Gamma^{\mu}_{\,\,\,\nu\lambda}x^{\nu}_{\,\,,i}x^{\lambda}_{\,\,,j})=0$. Alternatively, instead of directly
going to the brane coordinates, we can work with the full spacetime indices and have
$\delta_{x}s_{_{\!D}}=\int_{M}d^{D}x\sqrt{|\textsl{g}|}\,\text{distr}\,\textsc{e}^{\mu\nu}
\delta_{x}\textsl{g}_{\mu\nu}=\int_{M}d^{D}x\sqrt{|\textsl{g}|}\,k^{\mu\nu}\delta^{(\delta)}
\delta_{x}\textsl{g}_{\mu\nu}=\int_{\Sigma}d^{4}\chi\sqrt{|h|}\,k^{\mu\nu}\delta_{x}\textsl{g}_{\mu\nu}=
\int_{\Sigma}d^{4}\chi\sqrt{|h|}\,k^{ij}x^{\mu}_{\,\,,i}x^{\nu}_{\,\,,j}\delta_{x}\textsl{g}_{\mu\nu}=
\int_{\Sigma}d^{4}\chi\sqrt{|h|}\,k^{ij}\delta_{x}h_{ij}$, which is the same result as above. In the next section
we will follow this process (actually slightly modified because of the use of Lagrange multipliers)
for a codimension-2 defect in EGB bulk gravity and basically try to find $k^{ij}$.

\section{Six-dimensional setup, matching conditions and effective equations}
\label{General setup, matching conditions and effective equations}

Let us consider the general system of six-dimensional Einstein-Gauss-Bonnet gravity coupled to a localized
3-brane source. The total brane-bulk action is
\begin{eqnarray}
\!\!\!\!\!\!\!S\!=\!\frac{1}{2\kappa_{6}^{2}}\!\int_{\!M}
\!\!\!d^6x\sqrt{|\textsl{g}|}
\Big[\mathcal{R}\!-\!2\Lambda_{6}
\!+\!\alpha\big(\mathcal{R}^{2}
\!-\!4\mathcal{R}_{\mu\nu}\mathcal{R}^{\mu\nu}
\!+\!\mathcal{R}_{\mu\nu\kappa\lambda}\mathcal{R}^{\mu\nu\kappa\lambda}\big)\!\Big]
\!+\!\int_{\!\Sigma} \!d^4\chi\sqrt{|h|}
\Big(\!\frac{r_{c}^{2}}{2\kappa_{6}^{2}}R\!-\!\lambda\!\Big)
\!+\!\!\int_{\!M}\!\!\!d^6\!x\,\mathcal{L}_{mat}\!+\!\!\int_{\!\Sigma}\!\!d^{4}\!\chi\,L_{mat},\label{Stotal}
\end{eqnarray}
where $\textsl{g}_{\mu\nu}$ is the (continuous) bulk metric tensor and $h_{\mu\nu}$ is the induced metric on the
brane ($\mu,\nu,...$ are six-dimensional coordinate indices). The bulk coordinates are $x^{\mu}$ and the
brane coordinates are $\chi^{i}$ ($i,j,...$ are
four-dimensional worldvolume indices). The calligraphic quantities refer to the bulk metric, while the
regular ones to the brane metric. The brane tension is $\lambda$ and the
induced-gravity term \cite{deffayet}, if present, has a crossover length scale $r_{c}$. $\mathcal{L}_{mat}$,
$L_{mat}$ are the matter Lagrangians of the bulk and of the brane respectively.

Varying (\ref{Stotal}) with respect to the bulk metric we get the bulk equations of motion
\begin{equation}
\mathcal{G}_{\mu\nu}+2\alpha\big(\mathcal{R}
\mathcal{R}_{\mu\nu}-2\mathcal{R}_{\mu
\kappa}\mathcal{R}_{\nu}^{\,\,\,\kappa}\!-2\mathcal{R}_{\mu
\kappa\nu \lambda}\mathcal{R}^{\kappa\lambda}
+\mathcal{R}_{\mu\kappa\lambda\sigma}\mathcal{R}_{\nu}^{\,\,\,\kappa\lambda\sigma}\big)
-\frac{\alpha}{2}\big(\mathcal{R}^2\!-
4\mathcal{R}_{\kappa\lambda}\mathcal{R}^{\kappa\lambda}\!+\mathcal{R}_{\kappa\lambda\rho\sigma}\mathcal{R}
^{\kappa\lambda\rho\sigma}\big)\textsl{g}_{\mu\nu}
\!=\!\kappa_{6}^2
\mathcal{T}_{\mu\nu}-\Lambda_{6}
\textsl{g}_{\mu\nu}\,,
\label{bulk equations}
\end{equation}
where $\mathcal{G}_{\mu\nu}$ is the bulk Einstein tensor and $\mathcal{T}_{\mu\nu}$ is a regular
bulk energy-momentum tensor. We are mainly interested in a bulk with a pure cosmological constant $\Lambda_{6}$,
but as it will be seen, the existence of a non-vanishing $\mathcal{T}_{\mu\nu}$ is not very crucial. More
precisely, we define the variation $\delta \textsl{g}_{\mu\nu}$ with respect to the bulk metric to vanish on
the defect. In this variation, beyond the basic terms proportional to $\delta \textsl{g}_{\mu\nu}$
which give (\ref{bulk equations}), there appear, as usually, extra terms proportional to the second covariant
derivatives $(\delta \textsl{g}_{\mu\nu})_{;\kappa\lambda}$ (the covariant differentiation $;$ corresponds to
$\textsl{g}_{\mu\nu}$). Since we are interested in the
equations of motion of bulk gravity, we consider a hypersurface which is an infinitely thin ``tube'' around the
codimension-2 defect, and then, the extra bulk integral becomes an integral on this hypersurface with terms
proportional to $(\delta \textsl{g}_{\mu\nu})_{;\kappa}$. Adding suitable boundary terms on the hypersurface
\cite{Germani} (analogous to the Gibbons-Hawking term) to cancel the normal derivatives of
$\delta \textsl{g}_{\mu\nu}$, the surface integral of the total variation finally consists only of terms
proportional to $\delta \textsl{g}_{\mu\nu}$. Considering that the variation of
the bulk metric vanishes on this hypersurface, there is nothing left beyond the terms in equation
(\ref{bulk equations}). In the limit where the ``tube'' shrinks at the codimension-2 brane, the variation of
the bulk metric vanishes on this brane. Note that the boundary terms added on the hypersurface have nothing to
do with the equation of motion of the codimension-2 brane derived below, but they are just added to make the
variation of the bulk metric well defined. Equations (\ref{bulk equations}) in the vicinity of the brane
contain, in general, regular terms as well as divergent terms (distributional terms are different and refer
to the brane position, not to the bulk space) and will be analyzed further later.
\par
According to the standard method, the interaction of the brane with the
bulk comes from the variation $\delta \textsl{g}_{\mu\nu}$ at the brane position of the action (\ref{Stotal}),
which is equivalent to adding on the right-hand side of equation (\ref{bulk equations}) the term $\kappa_{6}^2\,
\tilde{T}_{\mu\nu}\,\delta^{(2)}$, where
$\tilde{T}_{\mu\nu}=T_{\mu\nu}\!-\!\lambda\,
h_{\mu\nu}\!-\!(r_{c}^{2}/\kappa_{6}^2)G_{\mu\nu}$. $T_{\mu\nu}$ is
the brane energy-momentum tensor, $G_{\mu\nu}$ the brane Einstein
tensor and $\delta^{(2)}$ the two-dimensional delta function with support on the defect. This approach has been
analyzed in \cite{CKP} and discussed in the Introduction.
\par
Here, we discuss an alternative approach where the interaction of the brane with bulk gravity is obtained
by varying the total action (\ref{Stotal}) with respect to $\delta x^{\mu}$, the embedding fields of the brane
position. The embedding fields are some functions
$x^{\mu}(\chi^{i})$ and their variations are $\delta x^{\mu}(x^{\nu})$.
While in the standard method the variation of the bulk metric at the brane position remains arbitrary,
here the corresponding variation is induced by $\delta x^{\mu}$ as explained in section
\ref{General arguments and introduction of the method}, it is given by
\begin{equation}
\delta \textsl{g}_{\mu \nu}=\delta_{x} \textsl{g}_{\mu \nu}=
\textsl{g}\,'_{\mu\nu}(x^{\rho})-\textsl{g}_{\mu\nu}(x^{\rho})
= -(\textsl{g}_{\mu \nu , \lambda} \delta x^{\lambda}+\textsl{g}_{\mu
\lambda} {\delta x^{\lambda}}_{, \nu}+\textsl{g}_{\nu \lambda} {\delta
x^{\lambda}}_{,\mu})=-\pounds_{\delta x} \textsl{g}_{\mu \nu}\,,
\label{delta x variation}
\end{equation}
and is obviously independent from the variation leading to (\ref{bulk equations}). The induced metric
$h_{ij}=\textsl{g}_{\mu\nu}\,x^{\mu}_{\,\,,i}\,x^{\nu}_{\,\,,j}$ enters the localized terms of the action
(\ref{Stotal}) and depends explicitly and implicitly (through $\textsl{g}_{\mu\nu}$) on the embedding fields.
Also the various bulk terms of (\ref{Stotal}) contribute implicitly to the brane variation under the variation
of the embedding fields. The result of $\delta x^{\mu}$ variation gives, as we will see, as coefficient of $\delta x^{\mu}$
a combination of vectors parallel and normal to the brane, therefore, two sets of equations will finally arise
as matching conditions instead of one. Instead of directly expressing $\delta \textsl{g}_{\mu\nu}$,
$\delta h_{ij}$ in terms of $\delta x^{\mu}$, it is more convenient to include the constraints in the action
and vary independently. So, the first constraint $h_{ij}=\textsl{g}_{\mu\nu}\,x^{\mu}_{\,\,,i}\,x^{\nu}_{\,\,,j}$
implies the independent variation of $h_{ij}$. The variation $\delta x^{\mu}$ affects the variation of the
parallel to the brane vectors $x^{\mu}_{\,\,,i}$ which in turn influences the variation of the normal vectors
$n_{\alpha}^{\,\,\,\mu}$. If $n_{\alpha}^{\,\,\,\mu}$ ($\alpha=1,2$) are arbitrary unit vectors
normal to the brane and to each other, the additional constraints $n_{\alpha\mu}x^{\mu}_{\,\,,i}=0$,
$\textsl{g}_{\mu\nu}n_{\alpha}^{\,\,\,\mu}n_{\beta}^{\,\,\,\nu}=\delta_{\alpha\beta}$ have to be added, and
$\delta n_{\alpha\mu}$ is another independent variation. Finally, the third variation
$\delta \textsl{g}_{\mu\nu}$ depends on $\delta x^{\mu}$ by (\ref{delta x variation}). Therefore,
$\delta \textsl{g}_{\mu\nu}$ are independent from
$\delta n_{\alpha\mu}$, $\delta h_{ij}$, but the various $\delta \textsl{g}_{\mu\nu}$ components
are not all independent from each other, so in the end they have to be expressed in terms of $\delta x^{\mu}$
which are independent. If $\lambda^{ij}$, $\lambda^{\alpha i}$,
$\lambda^{\alpha\beta}$ are the Lagrange multipliers corresponding to the above constraints,
the constraint action added to $S$ is
\begin{equation}
S_{c}=\int_{\Sigma}\!d^{4}\chi \,\sqrt{|h|} \,\Big [\lambda^{ij} (h_{i
j} - \textsl{g}_{\mu\nu} x^{\mu}_{\,\,,i} x^{\nu}_{\,\,,j}) + \lambda^{\alpha i}
n_{\alpha \mu} x^{\mu}_{\,\,,i} + \lambda^{\alpha\beta} (\textsl{g}_{\mu\nu}
n_{\alpha}^{\,\,\,\mu} n_{\beta}^{\,\,\,\nu} -\delta_{\alpha\beta})\Big]\,.
\label{Sconstraint}
\end{equation}
Actually, since in the action (\ref{Stotal}) the normals $n_{\alpha}^{\,\,\,\mu}$ are not explicitly
present, the inclusion of the Lagrange multipliers $\lambda^{\alpha i},\lambda^{\alpha\beta}$ is not very
significant and they will end up to vanish. However, we keep them in order to preserve a uniform treatment
with the codimension-1 case where the Gibbons-Hawking term contains the normal vector explicitly.

Variation of $S\!+\!S_{c}$ with respect to $n_{\alpha\mu},h_{ij},\textsl{g}_{\mu\nu}$ at the brane gives
\begin{eqnarray}
\delta(S\!+\!S_{c})\big|_{\!b\!r\!a\!n\!e}\!\!&=&\!\!\int_{\!\Sigma}\!d^{4}\chi \,\sqrt{|h|}\,\big(\lambda^{\alpha
i}x^{\mu}_{\,\,,i}+2\lambda^{\alpha\beta}n_{\beta}^{\,\,\,\mu}\big)\delta
n_{\alpha\mu}\!+\!\int_{\!\Sigma}\!d^{4}\chi \,\sqrt{|h|}\,\Big[\lambda^{ij}
+\frac{1}{2}(T^{ij}\!-\!\lambda h^{ij})-\frac{r_{c}^{2}}{2\kappa_{6}^{2}}G^{ij}\Big]
\delta h_{ij}\nn\\
&-&\!\!\int_{\!\Sigma}\!d^{4}\chi
\sqrt{|h|}\,\big(\lambda^{ij}x^{\mu}_{\,\,,i}x^{\nu}_{\,\,,j}\!+\!
\lambda^{\alpha\beta}n_{\alpha}^{\,\,\,\mu}n_{\beta}^{\,\,\,\nu}\big) \delta
\textsl{g}_{\mu\nu}\nn\\
&-&\!\!\frac{1}{2\kappa _{6}^{2}} \!\int_{\!M}\!\! d^{6}x \sqrt{|\textsl{g}|}
\,\Big \lbrace \mathcal{G}^{\mu\nu}\!+\!\alpha\mathcal{J}^{\mu\nu}
\!-\!\kappa_{6}^{2}\mathcal{T}^{\mu\nu}\!+\!\Lambda_{6}\textsl{g}^{\mu\nu}
\!\Big \rbrace  \delta \textsl{g}_{\mu \nu}\Big|_{\!b\!r\!a\!n\!e}\nn\\
&+&\!\!\frac{1}{\kappa_{6}^{2}}\!\int_{\!M}\!\!d^{6}x
\sqrt{|\textsl{g}|}\, \Big\lbrace  \textsl{g}^{\mu [ \kappa}\textsl{g}^{\lambda ] \nu}
+ 2\alpha\Big(\!\mathcal{R}^{\mu\nu\kappa\lambda}+2 \mathcal{R}^{\nu [ \kappa}\textsl{g}^{\lambda ] \mu}-
2 \mathcal{R}^{\mu [ \kappa}\textsl{g}^{\lambda ] \nu}
+\mathcal{R} \textsl{g}^{\mu [ \kappa}\textsl{g}^{\lambda ] \nu} \Big)\!\Big\rbrace
(\delta \textsl{g}_{\nu\kappa})_{; \lambda\mu}\Big|_{\!b\!r\!a\!n\!e},
\label{totalvariation}
\end{eqnarray}
where
\begin{equation}
\mathcal{J}^{\mu\nu}=2\mathcal{R}\mathcal{R}^{\mu\nu}
\!-\!4\mathcal{R}^{\mu\kappa}\mathcal{R}^{\nu}_{\,\,\,\kappa}
\!-\!4\mathcal{R}^{\mu\kappa\nu\lambda} \mathcal{R}_{\kappa\lambda}
\!+\!2\mathcal{R}^{\mu\kappa\lambda\sigma} \mathcal{R}^{\nu}_{\,\,\,\kappa\lambda\sigma}
-\frac{1}{2} \big(\mathcal{R}^{2}
\!-\!4\mathcal{R}_{\kappa\lambda}\mathcal{R}^{\kappa\lambda}
\!+\!\mathcal{R}_{\kappa\lambda\rho\sigma} \mathcal{R}^{\kappa\lambda\rho\sigma}
\big)\textsl{g}^{\mu\nu}.
\label{H}
\end{equation}
When $r_{c}\!\neq\! 0$, one should add in (\ref{Stotal}) the integral of the extrinsic curvature $k$ of
$\partial\Sigma$ (if $\partial\Sigma$ is not empty) to cancel some terms from the variation $\delta R$; this,
in general, does not affect the dynamics of $\Sigma$ \cite{guven}. Note that in the $\delta\textsl{g}_{\mu\nu}$
variation of $\textsl{g}_{\mu\nu}n_{\alpha}^{\,\,\,\mu} n_{\beta}^{\,\,\,\nu}$, it is $n_{\alpha\mu}$ which
is kept fixed and not $n_{\alpha}^{\,\,\,\mu}$.

The six-dimensional terms in (\ref{totalvariation}) have to be integrated out around the brane and finally give
a four-dimensional integral which contributes to the matching conditions, as will be explained in the next
section. Note that the quantity in curly brackets
appearing in the third line of (\ref{totalvariation}), though formally identical to that of equation
(\ref{bulk equations}), however it contains additional information, so this curly bracket does not vanish.
Equations (\ref{bulk equations}) refer to the bulk and also to the limit as these equations approach the brane,
while the corresponding curly bracket in (\ref{totalvariation}) refers exactly to the brane, and therefore,
it contains extra distributional terms which are not present in (\ref{bulk equations}).

Let us consider for simplicity that there is axial symmetry in the bulk, so that the bulk metric ansatz
can be written in the brane Gaussian-normal coordinates as
\begin{equation}
ds_{6}^{2}=dr^{2}+L^{2}(\chi,r)d\theta^{2}+
g_{ij}(\chi,r) d\chi^{i} d\chi^{j}\,.
\label{ansatz}
\end{equation}
The braneworld metric $h_{ij}(\chi)=g_{i j}(\chi,0)$ is assumed to be
regular everywhere with the possible exception of isolated singular
points. The angle $\theta$ has the standard periodicity $2\pi$.
Since $\theta$ is an angle, close to the brane $r\approx 0$, it has to be $L\propto r$ and the measure
of the six-dimensional integration is $\sqrt{|\textsl{g}|}\propto r$. Therefore, only terms proportional to
$\frac{\delta(r)}{r}$ inside the two curly brackets of (\ref{totalvariation}) contribute to the four-dimensional
equations of motion. Let $\mathcal{K}_{ij}(\chi,r)=\frac{1}{2}g'_{ij}(\chi,r)$ (a prime denotes
$\partial / \partial r$) be the extrinsic curvature tensor defined everywhere in the bulk.
Since the various tensors $\mathcal{G}^{\mu\nu}$, $\mathcal{R}^{\mu\nu}$,
..., shown in Appendix \ref{geometric components}, contain $L''$, $\mathcal{K}'_{ij}$, the quantities $L'$
or/and $\mathcal{K}_{ij}$ have to be discontinuous at $r=0$.

Therefore, there are two sources of discontinuity:
\newline
\indent{(a)} cone discontinuity, where the transverse space to the defect is assumed to have the standard conical
singularity structure with $L(\chi,r)\!=\!\beta(\chi)r+\mathcal{O}(r^{2})$ for $r\!\approx\! 0$. The conical
deficit is $2\pi(1-\beta)>0$, so it is typically defined $L'(\chi,0)\!=\!1$. The discontinuity of $L'$ arises
due to the values $L'(\chi,0^{+})\!=\!\beta(\chi)$ and $L'(\chi,0)\!=\!1$.
\newline
\indent{(b)} extrinsic curvature discontinuity, where there is a jump in the extrinsic tangential sector.
If the extrinsic curvature vanishes on the brane,
the corresponding jump is $\mathcal{K}_{ij}(\chi,0)\!=\!0\neq \mathcal{K}_{ij}(\chi,0^{+})$.
\newline
Combining these two sources of discontinuity we can have four cases which will give four kinds of matching conditions:
\newline
\indent{(i)} pure cone or topological matching conditions, discussed in \cite{Charmousis}, which have a geometric
origin based on the distributional version of the Chern-Gauss-Bonnet theorem \cite{Zegers}. Here, there is
a conical singularity, but the extrinsic curvature has no jump
$\mathcal{K}_{ij}(\chi,0)\!=\!\mathcal{K}_{ij}(\chi,0^{+})$.
\newline
\indent{(ii)} pure extrinsic curvature discontinuity, where there is no cone, so the deficit angle is
$\beta=1$. Although a conical singularity is usually attached to a codimension-2 defect, it is still consistent
to appear a discontinuity only in the extrinsic curvature, as in codimension-1 defects.
\newline
\indent{(iii)} cone plus extrinsic curvature discontinuity, introduced in \cite{Ruth 2004}, which assumes not only a conical
deficit based on the normal geometry but also a jump in the extrinsic tangential sector.
\newline
\indent{(iv)} ``smooth'' matching conditions, where there is neither conical singularity nor extrinsic curvature
discontinuity. Although this full smoothness may look peculiar, it will be seen
later that they are consistent back-reacted matching conditions with non-trivial solutions.
\newline
Introducing the index $\eta=0,1$ so that
$\mathcal{K}_{ij}(\chi,0)\!=\!\eta K_{ij}(\chi)\,\,\,,\,\,\,K_{ij}(\chi)\equiv\mathcal{K}_{ij}(\chi,0^{+})$,
the previous matching conditions are described as (i) $\beta\neq 1,\,\eta=1$, (ii) $\beta=1,\,\eta=0$,
(iii) $\beta\neq 1,\,\eta=0$, (iv) $\beta=1,\,\eta=1$.
\newline
More generally, we could consider that the two extrinsic curvatures $\mathcal{K}_{ij}(\chi,0),
\mathcal{K}_{ij}(\chi,0^{+})$ are proportional to each other, i.e.
$\mathcal{K}_{ij}(\chi,0)\!=\!\eta \mathcal{K}_{ij}(\chi,0^{+})$ with $\eta$ a continuous parameter, but
in the present paper we are going to restrict ourselves to the two characteristic cases $\eta=0,1$.
Actually, the case $\eta=1$ of continuous extrinsic curvature is especially interesting because it sets
in a natural way the minimal demand for consistency, either of a pure cone singularity or of a smooth
transverse section. Any value of the parameter $\eta$, other than $\eta=1$, carries into the problem the
arbitrariness of the  matrix $\mathcal{K}_{ij}(\chi,0)$, so the matching conditions (\ref{matching1}),
(\ref{matching2}) do not give the equation of motion of the defect until $\eta$ is
specified. Only for $\eta=1$ are the equations of motion uniquely defined without any other information.
Imposing a value of $\eta$, e.g. $\eta=0$, is similar to the codimension-1 case, where the Israel matching
conditions for a general non-$Z_{2}$ bulk do not give the equation of motion of the defect, but just the
discontinuity of the extrinsic curvature, and
only after the additional information of $Z_{2}$ symmetry is imposed, do the matching conditions
define the equation of motion. Analogously to the case $\eta=1$, for a codimension-1 brane the case
of continuous extrinsic curvature (which is consistent in the alternative matching conditions) gives without
any extra assumption the equation of motion. To make things worse, if we legitimately imagine that
$\mathcal{K}_{ij}(\chi,0),\,\mathcal{K}_{ij}(\chi,0^{+})$ are not proportional to each other, but
they are unrelated matrices, then the matching conditions (\ref{matching1}), (\ref{matching2}) will
carry $\mathcal{K}_{ij}(\chi,0)$ and the situation will be much more messy.

For $r>0$ the expansions of $L(\chi,r), \mathcal{K}_{ij}(\chi,r)$ in powers of $r$ are
\begin{eqnarray}
&&L(\chi,r)=\beta(\chi)\,r+\frac{1}{2}\beta_{2}(\chi)\,r^{2}+\mathcal{O}(r^{3})\label{L expansion}\\
&& \mathcal{K}_{ij}(\chi,r)=K_{ij}(\chi)+C_{ij}(\chi)\,r+\mathcal{O}(r^{2}).
\label{K expansion}
\end{eqnarray}
\noindent
Together with the values $L'(\chi,0),\mathcal{K}_{ij}(\chi,0)$, the functions
$L'(\chi,r),\mathcal{K}_{ij}(\chi,r)$ are also defined for $r=0$ and are in general discontinuous. Therefore, the
functions $L''(\chi,r),\mathcal{K}'_{ij}(\chi,r)$ obtain distributional parts (beyond the regular ones) given by
\begin{eqnarray}
&&\textrm{distr}\,L''(\chi,r)=-(1-\beta(\chi))\,\delta(r)\label{distr L}\\
&& \textrm{distr}\,\mathcal{K}'_{ij}(\chi,r)=(1-\eta)K_{ij}(\chi)\,\delta(r).\label{distr K}
\end{eqnarray}
\noindent
For convenience we include the expansion of $g_{ij}(\chi,r)$ which is defined for $r\geq 0$
\begin{equation}
g_{ij}(\chi,r)=h_{ij}(\chi)+2K_{ij}(\chi)\,r+C_{ij}(\chi)\,r^{2}+\mathcal{O}(r^{3}).
\label{g expansion}
\end{equation}

\subsection{Matching conditions} \label{Matching conditions}

As usually done when dealing with distributional sources, the matching conditions are derived by
integrating around the singular space. Here, the six-dimensional terms in (\ref{totalvariation})
are integrated over the $(r,\theta)$ transverse disk of radius $\epsilon$ in the limit
$\epsilon\rightarrow 0$. Because of the axial symmetry the angular dependence is factorized.
In Appendix \ref{geometric components} the components of the tensors $\mathcal{R}_{\mu\nu\kappa\lambda}$,
$\mathcal{R}_{\mu\nu}$, $\mathcal{G}_{\mu\nu}$ in the Gaussian-normal coordinate system (\ref{ansatz}) are
given. Some of these components contain $L'',\mathcal{K}'_{ij}$, which have regular as well as distributional
pieces (\ref{distr L}), (\ref{distr K}). Plaguing the expressions of $\mathcal{R}_{\mu\nu\kappa\lambda}$,
$\mathcal{R}_{\mu\nu}$, $\mathcal{G}_{\mu\nu}$ in the curly brackets of the six-dimensional integrals of
(\ref{totalvariation}) and expanding according to (\ref{L expansion}), (\ref{K expansion}),
(\ref{g expansion}) we find four types of terms. First, distributional terms of the form
$\frac{\delta(r)}{r}$, arising from the factors $\frac{L''}{L}$
and $\frac{L'}{L} \mathcal{K}'_{ij}$. These terms multiplied by $r$, coming from the measure of integration,
are the surviving terms which contribute to the matching conditions. Second, distributional terms of the
form $\delta(r)$, which when multiplied by $r$ vanish. Third, regular terms, i.e. terms with finite
values for $r=0$, either continuous or discontinuous, which when multiplied with the measure of integration give
a continuous function vanishing on the brane, therefore their integration obviously
vanishes in the limit $\epsilon\rightarrow 0$. Finally, singular terms, i.e. finite $\chi-$dependent terms
multiplied by $1/r$, which when multiplied by $r$ and integrated obviously vanish.
It can be seen that the last line of (\ref{totalvariation}),
which contains terms multiplied by $(\delta \textsl{g}_{\nu\kappa})_{;\lambda\mu}$, does not contribute to the
matching conditions. From the previous terms which are multiplied by $\delta \textsl{g}_{\mu\nu}$,
only the parallel to the brane components, i.e. terms multiplied
by $\delta g_{ij}$, contain distributional terms $\frac{\delta(r)}{r}$ and therefore contribute to the
matching conditions. The variational fields $\delta\textsl{g}_{\mu\nu}$ are considered, as usually,
smooth functions.

Of course, the correct variational fields have to be dimensionless. Otherwise, since $\theta$ is angle,
$L$ has dimension of length and $\delta\textsl{g}_{\theta\theta}$ has dimension of length square. Since
$\mathcal{G}^{\theta}_{\theta}\!+\alpha\mathcal{J}^{\theta}_{\theta}$ has only regular terms,
$\mathcal{G}^{\theta\theta}\!+\alpha\mathcal{J}^{\theta\theta}$ will have singular terms $1/r^{2}$. But then,
multiplying by $r$ and integrating would give divergence, which is not the case. The correct dimensionless variational
field is $\textsl{g}^{\theta\theta}\delta\textsl{g}_{\theta\theta}$ and the corresponding multiplicative term
is $\textsl{g}_{\theta\theta}(\mathcal{G}^{\theta\theta}\!+\alpha\mathcal{J}^{\theta\theta})=
\mathcal{G}^{\theta}_{\theta}\!+\alpha\mathcal{J}^{\theta}_{\theta}$
which possesses no singular terms at all. Somehow similarly, if the coordinates $\chi^{i}$ have length dimensions,
then $g_{ij}$ and $\delta g_{ij}$ are dimensionless.

Note also that although the action (\ref{Stotal}) contains distributional terms, however $S$ is finite.

The distributional term of $\mathcal{G}^{ij}$ is easily found to be
\begin{equation}
\textrm{distr}\,\mathcal{G}^{ij}=-\frac{1-\beta}{\beta}h^{ij}\frac{\delta(r)}{r}.
\label{distr Einstein}
\end{equation}
The distributional terms of $\mathcal{J}^{ij}$ are more difficult, but they can be combined as
$-\frac{4}{L}(L'\,\mathcal{W}^{ij})'-4\frac{L''}{L}G^{ij}$, where
\begin{equation}
\mathcal{W}^{ij}=\mathcal{K}^{ik}\mathcal{K}_{k}^{j}-\mathcal{K}\mathcal{K}^{ij}
-\frac{1}{2}(\mathcal{K}_{k\ell}\mathcal{K}^{k\ell}-\mathcal{K}^{2})g^{ij}
\label{calW}
\end{equation}
and $\mathcal{K}=\mathcal{K}^{i}_{i}$. Then, it is found that
\begin{equation}
\textrm{distr}\,\mathcal{J}^{ij}=4\Big(\frac{\eta\!-\!\beta}{\beta}W^{ij}
+\frac{1\!-\!\beta}{\beta}G^{ij}\Big)\frac{\delta(r)}{r},
\label{distr H}
\end{equation}
where
\begin{equation}
W^{ij}=K^{ik}K_{k}^{j}-KK^{ij}-\frac{1}{2}(K_{k\ell}K^{k\ell}-K^{2})h^{ij}=\mathcal{W}^{ij}(r=0^{+}).
\label{calW}
\end{equation}
From (\ref{distr Einstein}), (\ref{distr H}), it is seen that the matching conditions (i), (iii) have
distributional terms coming from both $\mathcal{G}^{ij}$, $\mathcal{J}^{ij}$. The distributional source of the
matching condition (ii) is only $\mathcal{J}^{ij}$. Finally, the matching condition (iv) is not related
to distributional terms.

The result for the total variation (\ref{totalvariation}) at the brane position is
\begin{eqnarray}
\delta(S\!+\!S_{c})\big|_{\!b\!r\!a\!n\!e}\!\!&=&\!\!\int_{\!\Sigma}\!d^{4}\chi \,\sqrt{|h|}\,(\lambda^{\alpha
i}x^{\mu}_{\,\,,i}+2\lambda^{\alpha\beta}n_{\beta}^{\,\,\,\mu})\,\delta
n_{\alpha\mu}
\nn\\
&+&\!\!\int_{\!\Sigma}\!d^{4}\chi \,\sqrt{|h|}\,\Big[\lambda^{ij}
+\frac{1}{2}T^{ij}\!+\!\frac{2\pi(1\!-\!\beta)\!-\!\kappa_{6}^{2}\lambda}{2\kappa_{6}^{2}}h^{ij}
\!-\!\frac{4\pi\alpha(1\!-\!\beta)}{\kappa_{6}^{2}}\Big(1\!+\!\frac{r_{c}^{2}}{8\pi\alpha(1\!-\!\beta)}\Big)G^{ij}
\!-\!\frac{4\pi\alpha(\eta\!-\!\beta)}{\kappa_{6}^{2}}W^{ij}\Big]\delta h_{ij}\nn\\
&-&\!\!\int_{\!\Sigma}\!d^{4}\chi\,
\sqrt{|h|}\,(\lambda^{\alpha\beta}n_{\alpha}^{\,\,\,\mu}n_{\beta}^{\,\,\,\nu}
+\lambda^{ij}x^{\mu}_{\,\,,i}x^{\nu}_{\,\,,j})\,\delta\textsl{g}_{\mu\nu}\,.
\label{branevariation}
\end{eqnarray}
At this point, had we considered $\delta
n_{\alpha\mu}, \delta h_{ij}, \delta\textsl{g}_{\mu\nu}$ independent, three equations algebraic in
the Lagrange multipliers would arise, and therefore all Lagrange multipliers would have been zero, leading to
the standard matching condition \cite{Ruth 2004}, \cite{CKP} (which consists of vanishing the quantity
inside the bracket of equation (\ref{matching1})). Instead, as explained above, $\delta
n_{\alpha\mu}, \delta h_{ij}$ are independent variations, but $\delta\textsl{g}_{\mu\nu}$ depends on
$\delta x^{\mu}$ which are also independent. So, $\delta(S+S_{c})\big|_{b\!r\!a\!n\!e}=0$ gives
\begin{eqnarray}
&&\lambda^{\alpha i}x^{\mu}_{\,\,,i}+2\lambda^{\alpha\beta}n_{\beta}^{\,\,\,\mu}=0
\label{n variation}\\
&&\lambda^{ij}=\frac{4\pi\alpha(\eta\!-\!\beta)}{\kappa_{6}^{2}}W^{ij}
\!+\!\frac{4\pi\alpha(1\!-\!\beta)}{\kappa_{6}^{2}}\Big(1\!+\!\frac{r_{c}^{2}}{8\pi\alpha(1\!-\!\beta)}\Big)G^{ij}
\!+\!\frac{\kappa_{6}^{2}\lambda\!-\!2\pi(1\!-\!\beta)}{2\kappa_{6}^{2}}h^{ij}
-\frac{1}{2}T^{ij}
\label{h variation}\\
&&\int_{\!\Sigma}\!d^{4}\chi\,
\sqrt{|h|}\,(\lambda^{\alpha\beta}n_{\alpha}^{\,\,\,\mu}n_{\beta}^{\,\,\,\nu}
+\lambda^{ij}x^{\mu}_{\,\,,i}x^{\nu}_{\,\,,j})\,\delta\textsl{g}_{\mu\nu}=0\,,
\label{g variation}
\end{eqnarray}
where $\delta\textsl{g}_{\mu\nu}$ obeys (\ref{delta x variation}). Since the vectors $x^{\mu}_{\,\,,i}$\,,
$n_{\alpha}^{\,\,\,\mu}$ are independent, equation (\ref{n variation}) implies
$\lambda^{\alpha i}=\lambda^{\alpha\beta}=0$. Then, equation (\ref{g variation}) with $\lambda^{ij}$
given by (\ref{h variation}) takes the form
\begin{equation}
\int_{\!\Sigma}\!d^{4}\chi\,\sqrt{|h|}\,\lambda^{ij}
(\textsl{g}_{\mu\nu,\lambda}x^{\mu}_{\,\,,i}x^{\nu}_{\,\,,j}\delta
x^{\lambda}+2\textsl{g}_{\mu\nu}x^{\mu}_{\,\,,i}x^{\lambda}_{\,\,,j}\delta
x^{\nu}_{\,\,,\lambda})\!=\!0\,. \label{trial}
\end{equation}
After an integration of (\ref{trial}) by parts and imposing $\delta x^{\mu}|_{\partial
\Sigma}=0$ it becomes
\begin{equation}
\int_{\!\Sigma}\!d^{4}\chi\,\sqrt{|h|}\,\textsl{g}_{\mu\sigma}\big[\lambda^{ij}_{\,\,\,|j}\,x^{\mu}_{\,\,,i}
+\lambda^{ij}(x^{\mu}_{\,\,;ij}\!+\!\Gamma^{\mu}_{\,\,\,\nu\lambda}x^{\nu}_{\,\,,i}x^{\lambda}_{\,\,,j})\big]
\delta x^{\sigma}=0\,.
\label{step}
\end{equation}
The covariant differentiation $|$ corresponds to $h_{ij}$ and $\Gamma^{\mu}_{\,\,\,\nu\lambda}$ are the
Christoffel symbols of $\textsl{g}_{\mu\nu}$. Due to the arbitrariness of $\delta
x^{\mu}$ and since the extrinsic curvatures of the brane $K^{\alpha}_{\,\,\,\,ij}\!=\!n^{\alpha}_{\,\,\,\,i;j}$
satisfy $-K^{\alpha}_{\,\,\,\,ij}n_{\alpha}^{\,\,\,\mu}=
x^{\mu}_{\,\,;ij}+\Gamma^{\mu}_{\,\,\,\nu\lambda}x^{\nu}_{\,\,,i}x^{\lambda}_{\,\,,j}$\,,
equation (\ref{step}) is equivalent to
\begin{equation}
\lambda^{ij}_{\,\,\,|j}\,x^{\mu}_{\,\,,i}
-\lambda^{ij}K^{\alpha}_{\,\,\,\,ij}n_{\alpha}^{\,\,\,\mu}=0\,.
\label{naked}
\end{equation}
Therefore, two matching conditions arise
\begin{eqnarray}
&&\lambda^{ij}K^{\alpha}_{\,\,\,\,ij}=0\\
&&\lambda^{ij}_{\,\,\,|j}=0\,.
\label{lambda matching}
\end{eqnarray}
Using (\ref{h variation}), these matching conditions finally take the form
\begin{eqnarray} &&
\Big [ W^{ij}
\!+\!\frac{1\!-\!\beta}{\eta\!-\!\beta}
\Big(\!1\!+\!\frac{r_{c}^2}{8\pi\alpha(1\!-\!\beta)}\!\Big)
G^{ij}\!+\!\frac{\kappa_{6}^{2}\lambda\!-\!2\pi(1\!-\!\beta)}{8\pi\alpha(\eta\!-\!\beta)}h^{\!ij}\!-\!
\frac{\kappa_{6}^{2}}{8\pi\alpha(\eta\!-\!\beta)}T^{ij}\Big]K_{ij}=0
\label{matchingW1}\\
&&T^{ij}_{\,\,\,\,\,|j}=\frac{2\pi}{\kappa_{6}^{2}}\beta_{,j}(h^{ij}\!-\!4\alpha
G^{ij})+\frac{8\pi\alpha}{\kappa_{6}^{2}}\big[(\eta\!-\!\beta)W^{ij}\big]_{|j}\,.
\label{matchingW2}
\end{eqnarray}

Since the extrinsic curvature $K_{\alpha ij}=\textsl{g}(\nabla_{i}n_{\alpha},\partial_{j})=n_{\alpha i;j}$
($\nabla$ refers, as also $;$, to \textsl{g}) in the coordinates (\ref{ansatz}) has $K_{rij}=g'_{ij}/2$,
$K_{\theta ij}=0$, {\it{the matching conditions of codimension-2 Einstein-Gauss-Bonnet gravity}} can be rewritten,
recovering the manifest normal frame indices
\begin{eqnarray} &&
\!\!\!\!\!\!\!\!\!\!\!\Big\{\!\!K^{\!\beta\ell}_{\,\,\,\,\,\ell}
K_{\!\beta}^{\,\,ij}\!-\!\!K^{\!\beta\ell i}K_{\!\beta \,\,\ell}^{\,\,j}
\!-\!\frac{1}{2}\big(\!K^{\!\beta
k}_{\,\,\,\,\,k}K_{\!\beta\,\,\ell}^{\,\,\ell}\!-\!\!K^{\!\beta
k\ell}\!K_{\!\beta k\ell}\big)h^{ij}
\!-\!\frac{1\!\!-\!\!\beta}{\eta\!-\!\!\beta}
\Big(\!1\!+\!\frac{r_{c}^2}{8\pi\alpha(1\!-\!\!\beta)}\!\Big)
G^{ij}\!-\!\frac{\kappa_{6}^{2}\lambda\!-\!2\pi(1\!\!-\!\!\beta)}{8\pi\alpha(\eta\!-\!\!\beta)}h^{\!ij}\!+\!
\frac{\kappa_{6}^{2}}{8\pi\alpha(\eta\!-\!\!\beta)}T^{ij}\!\Big\}\!K^{\!\alpha}_{\,\,ij}\!\!=\!0\nn\\
\label{matching1}\\
&&\!\!\!\!\!\!\!\!T^{ij}_{\,\,\,\,\,|j}=\frac{2\pi}{\kappa_{6}^{2}}\beta_{,j}(h^{ij}\!-\!4\alpha
G^{ij})-\frac{8\pi\alpha}{\kappa_{6}^{2}}\Big\{\!(\eta\!-\!\beta)\Big[\!K^{\alpha\ell}_{\,\,\,\,\,\,\ell}
K_{\alpha}^{\,\,\,ij}\!-\!K^{\alpha\ell i}K_{\alpha
\,\,\,\ell}^{\,\,\,j}-\frac{1}{2}(K^{\alpha
k}_{\,\,\,\,\,\,k}K_{\alpha\,\,\ell}^{\,\,\,\ell} \!-\!
K^{\alpha k\ell}K_{\alpha k\ell})h^{ij}\!\Big]\Big\}_{|j}\,.
\label{matching2}
\end{eqnarray}
Indices $\alpha, \beta,...$ are lowered/raised with the matrix
$\textsl{g}_{\alpha\beta}=\textsl{g}(n_{\alpha},n_{\beta})$ and its inverse $\textsl{g}^{\alpha\beta}$.
With respect to local rotations $n_{\alpha}\!\rightarrow\! O^{\,\,\,\beta}_{\alpha}(x^{\mu})\,n_{\beta}$,
$K_{\alpha ij}$ transforms as a vector $K_{\alpha ij}\!\rightarrow\! O^{\,\,\,\beta}_{\alpha}K_{\beta ij}$,
thus equation (\ref{matching2}) is invariant under changes of the normal frame, while (\ref{matching1})
transforms as a vector.

Equation (\ref{matching1}) has been brought in a more compact form by dividing by $1\!-\!\beta$, $\eta\!-\!\beta$.
Equations (\ref{matching1}), (\ref{matching2}) are obviously defined for the matching conditions (i), (iii);
they are also defined for the matching condition (ii) and they are still pretty complicated;
finally, for the matching condition (iv) they take the simpler form
$[\kappa_{6}^{2}(T^{ij}\!-\!\lambda h^{ij})-r_{c}^{2}G^{ij}]K^{\alpha}_{\,\,\,\,ij}\!=\!0$,
$T^{ij}_{\,\,\,\,\,|j}\!=\!0$. For a 3-brane, making a general counting, the number of the matching conditions
(\ref{matching1}), (\ref{matching2}) is $\delta+4=D$ which is 6 here, while the number of the standard matching
conditions is 10. In any case the role of the matching conditions is only fulfilled by the inclusion of the
bulk field equations.

Equation (\ref{matching1}) is the
algebraic in the extrinsic curvature matching condition. It is a cubic equation in the extrinsic curvature,
contrary to the matching condition derived according to the standard method \cite{Ruth 2004}, \cite{CKP} which
is quadratic in the extrinsic curvature. In the absence of the Gauss-Bonnet term, (\ref{matching1}) reduces to
the matching condition of codimension-2 Einstein gravity \cite{KofTom}, which is linear in extrinsic curvature.
Equation (\ref{matching1}) is the generalization of the Nambu-Goto equation of motion
when the self-gravitating brane interacts with bulk gravity. In the limiting case of no back-reaction,
a probe brane with tension $\lambda$ moving in a fixed background arises. Indeed, in the probe limit, all the
geometric quantities $h_{ij}$, $K_{\alpha ij}$, $G_{ij}$, $\beta$ get their background values (most probably the
background value of $\beta=1$) when the bulk gravity couplings go to zero (i.e. $1/\kappa_{6}^{2}\rightarrow 0$,
$\alpha/\kappa_{6}^{2}\rightarrow 0$) and the extra brane sources vanish (i.e.
$T_{ij}\rightarrow 0$, $r_{c}^{2}/\kappa_{6}^{2}\rightarrow 0$). Then,
equation (\ref{matching1}) becomes $h^{ij}K^{\alpha}_{\,\,\,\,ij}=0$ which is the Nambu-Goto equation of motion.
Inversely, whenever any extra term beyond $\lambda h^{ij}K^{\alpha}_{\,\,\,\,ij}$ (or all terms) appears
in (\ref{matching1}), (\ref{matching2}) and these equations are consistent with all the other equations,
then these matching conditions are meaningful back-reacted matching conditions.
In this spirit, the matching conditions (iv) with neither conical singularity
nor extrinsic curvature discontinuity form an unusual but interesting example. In this case, only the localized
matter and four-dimensional gravity terms are present in the action, and although the higher-dimensional bulk
terms do not have a direct imprint in the matching conditions, there is still back-reaction since the bulk
equations have also to be satisfied at the brane position. These ``smooth'' matching conditions correspond
to the Regge-Teitelboim equations of motion \cite{Teitelboim}, \cite{battye}, \cite{Davidson} with the crucial
difference, however, that there, there are no higher-dimensional gravity terms in the action and the bulk is
prefixed (Minkowski). Therefore, possible difficulties discussed in \cite{Deser} are irrelevant here, since
they emanate from the embeddibility restrictions in the given non-dynamical bulk space, while the matching
conditions here dynamically propagate in a non-trivial bulk space.
``Smooth'' matching conditions are also meaningful in codimension-1 standard treatment
\cite{kofinasR4}, without of course the $K^{\alpha}_{\,\,\,\,ij}$ contraction (where there is no balance of
distributional terms between the two sides of the distributional equation, but the right-hand side vanishes on
its own), although there, they lose their significance since there is no Nambu-Goto probe limit so that these
matching conditions to signal a minimal departure from that limit.

Equation (\ref{matching2}) is the second matching condition and expresses a non-conservation equation of
the brane energy-momentum tensor, where the energy exchange between the brane and the bulk is due to the
variability along the brane of both the deficit angle and the extrinsic geometry. In the next section,
equation (\ref{matching2}) will be written in a more convenient form, from where it will be seen that the
possible non-conservation of energy is only due to the variability of $\beta$. According to the
conventional treatment, a different non-conservation equation is also derived \cite{CKP}, not as a second
matching condition, but as a combination of the algebraic matching condition with some bulk equations evaluated
on the brane. In the absence of the Gauss-Bonnet term, (\ref{matching2}) reduces to the non-conservation
equation of codimension-2 Einstein gravity, where the energy exchange is due to a variable
deficit angle. In the probe limit, equation (\ref{matching2}) is identically satisfied.

\subsection{Effective equations} \label{Effective equations}

Having finished with the brane equations of motion arising from the distributional parts of the various
quantities, we pass to the bulk equations of motion. These bulk equations are also defined limitingly on the brane,
and therefore, additional equations have to be satisfied at the brane position beyond the matching
conditions. More precisely, when the non-distributional quantities of Appendix \ref{geometric components}
are substituted in the bulk equations (\ref{bulk equations}) (more precisely in the equations with one index up
and one down) and expand according to (\ref{L expansion}),
(\ref{K expansion}), (\ref{g expansion}), there appear two kinds of terms : (a) regular terms $\mathcal{O}(1)$,
i.e. terms with finite values for $r=0$, and (b) singular terms $\mathcal{O}(1/r)$, i.e. finite $\chi$-dependent
terms multiplied by $1/r$. Since the singular terms cannot be canceled by any regular bulk energy-momentum
tensor $\mathcal{T}_{\mu\nu}$, their $\chi$-dependent coefficients have to vanish providing some new equations
on the brane. Obviously, the regular parts which remain will vanish independently, defining additional equations
on the brane. Note that $\mathcal{T}_{\mu\nu}$ cannot blow up close to the distributional singularity,
otherwise the singularity would not be distributional.

Regarding the $ri$ components of the bulk equations, their $\mathcal{O}(1/r)$ leading terms come from terms
multiplied by $L'/L$, $L'_{|i}/L$ and yield the equation
\begin{equation}
\frac{\beta_{,j}}{\beta}\Big(W^{ij}+G^{ij}-\frac{1}{4\alpha}h^{ij}\Big)
+W^{ij}_{\,\,\,\,\,|j}=0\,.
\label{sing ri}
\end{equation}
Similarly, the $\mathcal{O}(1/r)$ terms of the $rr$ bulk equation are obtained from the terms $L'/L$
and we get the equation
\begin{equation}
\Big(W^{ij}+G^{ij}-\frac{1}{4\alpha}h^{ij}\Big)K_{ij}=0\,.
\label{sing rr}
\end{equation}
Finally, in the $\mathcal{O}(1/r)$ terms of the $ij$ components of the bulk equations, not only $L'/L$,
$L'_{|i}/L$ terms, but also terms of the form $L''/L$, contribute. The corresponding brane equations contain
$g_{ij,2}(\chi)=g''_{ij}(\chi,0)$, $\beta_{2}(\chi)=L''(\chi,0)$ where, of course, the second derivatives
refer to the regular pieces of the quantities. These equations are pretty complicated and we will give
their explicit form in the case of cosmology.

Using (\ref{sing rr}), the matching condition (\ref{matchingW1}) gets the following simpler form which
is linear and homogeneous in the extrinsic curvature and it does not contain the deficit angle
\begin{equation}
\big(\sigma_{1}G^{ij}+\sigma_{2}h^{ij}-T^{ij}\big)K_{ij}=0,
\label{simpler matching1}
\end{equation}
where
\begin{equation}
\sigma_{1}=\frac{r_{c}^{2}}{\kappa_{6}^{2}}\!+\!\frac{8\pi\alpha(1\!-\!\eta)}
{\kappa_{6}^{2}}\,\,\,\,\,\,\,,\,\,\,\,\,\,\,
\sigma_{2}=\lambda\!-\!\frac{2\pi(1\!-\!\eta)}{\kappa_{6}^{2}}\,.
\label{constants}
\end{equation}
\noindent
Note that for the matching conditions (iv), equation (\ref{sing rr}) cannot be combined with equation
(\ref{matchingW1}) to eliminate $W^{ij}$ since there is no $W^{ij}$ in (\ref{matchingW1}) in this case, however,
equation (\ref{simpler matching1}) is still valid since it coincides with (\ref{matchingW1}) for
$\eta\!=\!\beta\!=\!1$. Therefore, the proof of the consistency and the investigation of the effective
equations will also cover the case (iv).
\newline
It will also be useful to define
\begin{equation}
\sigma=\sigma_{2}+\frac{\sigma_{1}}{4\alpha}=\lambda+\frac{r_{c}^{2}}{4\alpha\kappa_{6}^{2}}\,,
\label{sigmadef}
\end{equation}
which is positive for positive brane tension $\lambda$. Note that
$\sigma_{1}=0\Leftrightarrow \eta=1,r_{c}=0\Leftrightarrow \sigma_{2}=\sigma=\lambda$.
\newline
\noindent
Using (\ref{sing ri}), the conservation equation (\ref{matchingW2}) or (\ref{matching2}) also get a simpler form
\begin{equation}
T^{ij}_{\,\,\,\,\,|j}=-\eta\frac{8\pi\alpha}{\kappa_{6}^{2}}\frac{\beta_{,j}}{\beta}
\Big(W^{ij}+G^{ij}-\frac{1}{4\alpha}h^{ij}\Big)\,.
\label{simpler matching2}
\end{equation}
For $\eta=0$ it is seen that the energy on the brane is strictly conserved, in analogy to the case $\eta=0$
of the standard approach \cite{kofinas}. However, for $\eta=1$ and a varying $\beta$, i.e. case (i), the
brane radiates in the bulk and there is inevitable energy exchange between the brane and the bulk. Similar
non-conservation of energy also occurs in the standard treatment for $\eta=1$, however, the exchange is
different \cite{CKP}.

What remains are the regular equations on the brane. The regular parts of the $ij$ equations contain
$g_{ij,3}(\chi)=g'''_{ij}(\chi,0),\beta_{3}(\chi)=L'''(\chi,0)$ and are insignificant for all the other equations.
\newline
In Appendix \ref{regular equations}, the regular parts of the $rr,ri$
bulk equations on the brane are derived. Obviously, there are manifestly regular terms inside the tensor
components $\mathcal{E}^{\mu}_{\nu}=\mathcal{G}^{\mu}_{\nu}+\alpha\mathcal{J}^{\mu}_{\nu}$ which are obtained by
setting formally $\frac{L'}{L}=\frac{L'_{|i}}{L}=\frac{L''}{L}=0$, $L=\beta$ in the unperturbed expressions of
$\mathcal{E}$'s. However, there are additional hidden regular terms coming from the expansion of the terms
containing $\frac{L'}{L}$, $\frac{L'_{|i}}{L}$, $\frac{L''}{L}$. Therefore, the $\mathcal{O}(1)$ part of the
$rr$ bulk equation, after use of (\ref{sing rr}), becomes
\begin{equation}
\mathcal{E}^{r}_{r}\big{|}_{\frac{L'}{L}=0,L=\beta}+K'-4\alpha(3W^{i}_{j}+G^{i}_{j})K_{i}^{j\,\prime}
-4\alpha G_{j}^{i\,\prime} K^{j}_{i}=\kappa_{6}^{2}\mathcal{T}^{r}_{r}-\Lambda_{6}\,,
\label{generic rr regular}
\end{equation}
where $K_{i}^{j\,\prime}$ denotes $\mathcal{K}_{i}^{j\,\prime}(\chi,0)$. The $\mathcal{O}(1)$ part of
the $ri$ bulk equation because of its complication will be given only for cosmology.
Finally, the $\theta\theta$ bulk equation contains only regular $\mathcal{O}(1)$ terms, the corresponding
brane equation contains $g_{ij,2}$ and its form is
\begin{equation}
\mathcal{E}^{\theta}_{\theta}=\kappa_{6}^{2}\mathcal{T}^{\theta}_{\theta}-\Lambda_{6}\,.
\label{thouthou}
\end{equation}
\noindent
Its explicit form will be given in the case of cosmology.

To summarize, our system of equations consists of the simplified matching conditions (\ref{simpler matching1}),
(\ref{simpler matching2}), the $\mathcal{O}(1/r)$ equations (\ref{sing ri}), (\ref{sing rr}), the
$\mathcal{O}(1/r)$ $ij$ equations, the $\mathcal{O}(1)$ equations (\ref{generic rr regular}), (\ref{thouthou})
and the $\mathcal{O}(1)$ $ri$ equation.

\section{Cosmological equations and consistency} \label{Cosmological equations and consistency}

In this section, we will study the cosmological equations for a codimension-2 brane and check their
consistency. For this purpose we consider the following bulk cosmological metric
\begin{equation}
ds_{6}^{2}=dr^{2}+L^{2}(t,r)d\theta^{2}-n^{2}(t,r)dt^{2}
+a^{2}(t,r)\gamma_{\hat{i}\hat{j}}(\chi^{\hat{\ell}})d\chi^{\hat{i}}d\chi^{\hat{j}}\,,
\label{cosmological metric}
\end{equation}
where $\gamma_{\hat{i}\hat{j}}$ is a maximally symmetric 3-dimensional metric characterized by its spatial
curvature $k=-1,0,1$. The energy-momentum tensor on the brane (beyond that of the brane tension $\lambda$)
is assumed to be the one of a perfect fluid with energy density $\rho$ and pressure $p$.

It is very convenient to define the quantities
\begin{eqnarray}
A=\frac{a'}{a}\,&,&\,N=\frac{n'}{n}\label{notation1}\\
X=H^{2}+\frac{k}{a^{2}}\,&,&\,Y=\frac{\dot{H}}{n}+H^{2}\,\,\,\,\,,\,\,\,\,\,H=\frac{\dot{a}}{na}
\label{notation2}\\
\mathcal{X}=X-A^{2}+\frac{1}{12\alpha}\,&,&\,\mathcal{Y}=Y-AN+\frac{1}{12\alpha}\,,
\label{notation3}
\end{eqnarray}
where a dot denotes differentiation with respect to $t$. The cosmic scale factor, lapse function and
Hubble parameter arise as the restrictions on the brane of the functions $a(t,r),n(t,r)$ and $H(t,r)$
respectively. Other quantities also have their corresponding values when restricted on the brane,
and since all the following equations will refer to the brane position, we will use the same symbols for the
restricted quantities without confusion.

For the metric (\ref{cosmological metric}), equation (\ref{simpler matching1}) becomes
\begin{equation}
N=fA\,,
\label{N}
\end{equation}
where
\begin{equation}
f=3\frac{p-\sigma_{2}+\sigma_{1}(X+2Y)}{\rho+\sigma_{2}-3\sigma_{1} X}\,,
\label{f}
\end{equation}
while equation (\ref{simpler matching2}) becomes
\begin{equation}
\dot{\rho}+3nH(\rho+p)=-\eta\frac{24\pi\alpha}{\kappa_{6}^{2}}\frac{\dot{\beta}}{\beta}\mathcal{X}\,.
\label{simple rho dot}
\end{equation}

We now focus on the $\mathcal{O}(1/r)$ parts of the bulk equations. Equations (\ref{sing rr}),
(\ref{sing ri}), which correspond to the $rr$, $rt$ bulk equations, get the form
\begin{eqnarray}
&&\Big(1+\frac{N}{A}\Big)\mathcal{X}+2\mathcal{Y}=0
\label{sing rr cosmo}\\
&&2A^{2}\Big[\frac{\dot{A}}{nA}+H\Big(1-\frac{N}{A}\!\Big)\Big]=\frac{\dot{\beta}}{n\beta}\mathcal{X}\,.
\label{sing ri cosmo}
\end{eqnarray}
The $\mathcal{O}(1/r)$ part of the $tt$ bulk equation, which was not computed in
Section \ref{Effective equations}, is now found to be
\begin{equation}
2\frac{a_{2}}{a}-\frac{\beta_{2}}{\beta A}\mathcal{X}=\mathcal{X}+\frac{1}{6\alpha}\,,
\label{sing tt cosmo}
\end{equation}
where $a_{2}(t)=a''(t,0),n_{2}(t)=n''(t,0),\beta_{2}(t)=L''(t,0)$. Finally, the $\mathcal{O}(1/r)$ part
of the $\hat{i}\hat{j}$ bulk equations is
\begin{equation}
\frac{n_{2}}{n}+\Big(1\!+\!\frac{N}{A}\Big)\frac{a_{2}}{a}-(\mathcal{X}+2\mathcal{Y})\frac{\beta_{2}}
{2\beta A}=\mathcal{Y}+\frac{N}{2A}\mathcal{X}+\frac{1}{12\alpha}\Big(2\!+\!\frac{N}{A}\Big)
-\frac{2\dot{\beta}}{n\beta}\Big[\frac{\dot{A}}{nA}+H\Big(1\!-\!\frac{N}{A}\Big)\Big]\,,
\label{sing ij cosmo diff}
\end{equation}
which can be written in a simpler form after using equations
(\ref{N}), (\ref{sing rr cosmo}), (\ref{sing ri cosmo}), (\ref{sing tt cosmo})
\begin{equation}
\frac{n_{2}}{n}+f\frac{a_{2}}{a}-\mathcal{Y}\frac{\beta_{2}}{\beta A}=\frac{1+f}{12\alpha}
-\mathcal{X}-\Big(\frac{\dot{\beta}}{n\beta}\Big)^{\!2}\frac{\mathcal{X}}{A^{2}}\,.
\label{sing ij cosmo}
\end{equation}

Up to now, the cosmological equations that have to be satisfied on the brane are (\ref{N}),
(\ref{simple rho dot}), (\ref{sing rr cosmo}), (\ref{sing ri cosmo}), (\ref{sing tt cosmo}),
(\ref{sing ij cosmo}). What remains are the regular equations on the brane which contain up to second
transverse derivatives. The $\theta\theta$ regular equation (\ref{thouthou}) takes the form
\begin{equation}
(\mathcal{X}+2\mathcal{Y})\frac{a_{2}}{a}+\mathcal{X}\frac{n_{2}}{n}=\mathcal{X}\mathcal{Y}
+\frac{\mathcal{X}+\mathcal{Y}}{6\alpha}+2A^{2}\Big[\frac{\dot{A}}{nA}+H\Big(1\!-\!\frac{N}{A}\Big)\Big]^{2}
+\frac{1}{12\alpha}\Big(\kappa_{6}^{2}\mathcal{T}^{\theta}_{\theta}\!-\!\Lambda_{6}\!-\!
\frac{5}{12\alpha}\Big)\,,
\label{theta regular diff}
\end{equation}
which, after use of equations (\ref{N}), (\ref{sing rr cosmo}), (\ref{sing ri cosmo}), takes the simpler form
\begin{equation}
\frac{n_{2}}{n}-f\frac{a_{2}}{a}=\frac{1-f}{12\alpha}+\mathcal{Y}+\Big(\frac{\dot{\beta}}{n\beta}\Big)^{\!2}
\frac{\mathcal{X}}{2A^{2}}+\frac{1}{12\alpha\mathcal{X}}\Big(\kappa_{6}^{2}\mathcal{T}^{\theta}_{\theta}
\!-\!\Lambda_{6}\!-\!\frac{5}{12\alpha}\Big)\,.
\label{theta regular}
\end{equation}
The $rr$ regular equation (\ref{generic rr regular}) with the help of Appendix \ref{regular equations},
equation (\ref{smyrni}), becomes
\begin{eqnarray}
&&\!\!\!\!\!\!\!\!\!\!\!\!\!\!\!\!\!\!\!\!\!\!\!\!\!\!\!\!\!
\frac{1}{n\beta}\Big(\frac{\dot{\beta}}{n}\Big)^{^{\!\centerdot}}\mathcal{X}+\frac{H\dot{\beta}}{n\beta}
(\mathcal{X}\!+\!2\mathcal{Y})+\mathcal{X}\mathcal{Y}+\frac{1}{6\alpha}(\mathcal{X}\!+\!\mathcal{Y})
+2A^{2}\Big[\Big(1\!+\!2\frac{N}{A}\Big)A'+N'\Big]-\big[(\mathcal{X}\!+\!2\mathcal{Y})A'+\mathcal{X}N'\big]\nn\\
&&\,\,\,\,\,\,\,\,\,\,\,\,\,\,\,\,\,\,\,\,\,\,\,\,\,\,\,\,\,\,\,\,\,\,\,\,\,\,\,\,\,\,\,\,\,\,\,
\,\,\,\,\,\,\,\,\,\,\,\,\,\,\,\,\,\,\,\,\,\,\,\,\,\,\,\,\,\,\,\,\,\,\,\,\,\,\,\,
-\big[(A\!+\!N)X'+2AY'\big]=\frac{1}{12\alpha}
\Big(\Lambda_{6}+\frac{5}{12\alpha}-\kappa_{6}^{2}\mathcal{T}^{r}_{r}\Big)\,.
\label{rr regular}
\end{eqnarray}
Finally, the $rt$ regular equation takes the form of (\ref{gulang})
\begin{equation}
\frac{\dot{\beta}_{2}}{n\beta}+\frac{H}{A}\frac{\dot{\beta}^{2}}{n^{2}\beta^{2}}+
\Big[\frac{\mathcal{X}'}{\mathcal{X}}
+\frac{1}{2A}\Big(\mathcal{X}-2A'+\frac{1}{6\alpha}\Big)-N-\frac{\beta_{2}}{\beta}\Big]
\frac{\dot{\beta}}{n\beta}-\frac{2A}{\mathcal{X}}\Big[\frac{\dot{A}'}{n}+(H'+HN)(A-N)+H(A'-N')\Big]
=\frac{n\kappa_{6}^{2}}{12\alpha\mathcal{X}}\mathcal{T}^{t}_{r}\,.
\label{rt regular}
\end{equation}
The $\mathcal{O}(1)$ parts of the $tt$, $\hat{i}\hat{j}$ bulk equations contain third derivatives with
respect to $r$ and form an algebraic system of two equations for the three quantities
$a_{3}(t)=a'''(t,0),\,n_{3}(t)=n'''(t,0),\,\beta_{3}(t)=L'''(t,0)$.
Therefore, these two equations are decoupled from all the other equations and do not deserve further study.

{\textit{In summary, all the cosmological equations on the brane which have to be satisfied simultaneously are}}
(\ref{N}), (\ref{simple rho dot}), (\ref{sing rr cosmo}), (\ref{sing ri cosmo}), (\ref{sing tt cosmo}),
(\ref{sing ij cosmo}), (\ref{theta regular}), (\ref{rr regular}), (\ref{rt regular}). In particular, equations
(\ref{sing tt cosmo}), (\ref{sing ij cosmo}), (\ref{theta regular}) form an algebraic system for the unknown
functions $a_{2},n_{2},\beta_{2}$, which after solved and substituted in (\ref{rr regular}), (\ref{rt regular}),
they make these equations to be satisfied identically when
$\mathcal{T}^{r}_{r}=\mathcal{T}^{\theta}_{\theta}$, $\mathcal{T}^{t}_{r}=0$ on the brane. This calculation
is shown in Appendix \ref{Algebraic system} and proves the consistency of the whole system for all kinds of
matching conditions. Therefore,
{\textit{the essential equations on the brane are}} (\ref{N}), (\ref{simple rho dot}), (\ref{sing rr cosmo}),
(\ref{sing ri cosmo}), which is a differential system of four equations for five unknowns $a,\rho,\beta,A,N$.

\section{Codimension-2 cosmological evolution and solutions} \label{cosmo evolution}

We summarize by writing again the coupled system of the essential equations which govern the cosmological
evolution. Since the quantity $N$ can be eliminated from equation (\ref{N}), the system consists of the
three essential equations (\ref{simple rho dot}), (\ref{sing rr cosmo}), (\ref{sing ri cosmo})
which are written as
\begin{eqnarray}
&&\dot{\rho}+3nH(\rho+p)=-\eta\frac{24\pi\alpha}{\kappa_{6}^{2}}\frac{\dot{\beta}}{\beta}\mathcal{X}
\label{1}\\
&&(1+f)\mathcal{X}+2\mathcal{Y}=0
\label{2}\\
&&\frac{\dot{A}}{nA}+H(1-f)=\frac{\dot{\beta}}{n\beta}\frac{\mathcal{X}}{2A^{2}}\,,
\label{3}
\end{eqnarray}
where
\begin{eqnarray}
&&\mathcal{X}=X-A^{2}+\frac{1}{12\alpha}\,\,\,\,\,\,\,,\,\,\,\,\,\,\,\mathcal{Y}=Y-fA^{2}+\frac{1}{12\alpha}
\label{4}\\
&&X=H^{2}+\frac{k}{a^{2}}\,\,\,\,\,\,\,,\,\,\,\,\,\,\,Y=\frac{\dot{X}}{2nH}+X
\label{5}\\
&&f=3\,\frac{p-\sigma_{2}+\sigma_{1}(X+2Y)}{\rho+\sigma_{2}-3\sigma_{1}X}\,\,.
\label{6}
\end{eqnarray}
Although this is a system of three equations for four unknowns $a,\rho,\beta,A$, however, it will be seen in
a while that it is integrable\,! Note also that in the system (\ref{1})-(\ref{3}), $\beta$ is present only
through the derivative factor $\dot{\beta}/\beta$ and there are no separate $\beta$ terms. Therefore,
whenever one focuses on the case $\beta$=constant, there is no difference between $\beta$=constant and
$\beta\!=\!1$. The effective system of equations (\ref{1})-(\ref{3}) is blind on the constant value of the cone
(of course, the bulk solution should know about this value). Therefore, for $\beta$=constant, the matching
condition (i) coincides with (iv) and the matching condition (iii) coincides with (ii) from the viewpoint of
the effective equations. This does not mean that the cone of the matching conditions (i), (iii)
disappears. The cone is still there, but its constant throughout the brane deficit angle does not affect the
effective equations, so, effectively, it is like ``opening'' the constant deficit angle to a smooth plane.

From equations (\ref{2}), (\ref{3}) we take
\begin{equation}
\frac{\dot{\mathcal{X}}}{n\mathcal{X}}+\frac{\dot{\beta}}{n\beta}=-(3+f)H\,\,\,\,,\,\,\,\,
3+f=3\frac{\rho+p+\sigma_{1}\frac{\dot{X}}{nH}}{\rho+\sigma_{2}-3\sigma_{1}X}\,,
\label{7}
\end{equation}
and with the help of (\ref{1}), it takes the form
\begin{equation}
\frac{\dot{\mathcal{X}}}{\mathcal{X}}+\frac{\dot{\beta}}{\beta}=
\frac{(\rho\!+\!\sigma_{2}\!-\!3\sigma_{1}X)^{^{\centerdot}}}{\rho\!+\!\sigma_{2}\!-\!3\sigma_{1}X}
+\eta\frac{24\pi\alpha}{\kappa_{6}^{2}}\frac{\dot{\beta}}{\beta}\,
\frac{\mathcal{X}}{\rho\!+\!\sigma_{2}\!-\!3\sigma_{1}X}\,,
\label{8}
\end{equation}
or equivalently
\begin{equation}
\Big(\frac{\mathcal{X}}{\rho\!+\!\sigma_{2}\!-\!3\sigma_{1}X}\Big)^{\!-1}
\,\Big(1-\eta\frac{24\pi\alpha}{\kappa_{6}^{2}}\frac{\mathcal{X}}{\rho\!+\!\sigma_{2}\!-\!3\sigma_{1}X}\Big)^{\!-1}
\,\Big(\frac{\mathcal{X}}{\rho\!+\!\sigma_{2}\!-\!3\sigma_{1}X}\Big)^{^{\!\centerdot}}+
\frac{\dot{\beta}}{\beta}=0\,.
\label{9}
\end{equation}
Equation (\ref{9}) can be integrated giving the general solution
\begin{equation}
\mathcal{X}\Big(\eta\frac{24\pi\alpha}{\kappa_{6}^{2}}+c\beta\Big)=\rho\!+\!\sigma_{2}\!-\!3\sigma_{1}X
\quad,\quad c :\,\text{integration\,constant}.
\label{10}
\end{equation}
Equation (\ref{2}) is written in terms of $X,A$ as
\begin{equation}
(1+3f)A^{2}=\frac{\dot{X}}{nH}+(3+f)\Big(X\!+\!\frac{1}{12\alpha}\Big)\,,
\label{11}
\end{equation}
and $\mathcal{X}$ is found to be
\begin{equation}
(1+3f)\mathcal{X}=2(f-1)\Big(X\!+\!\frac{1}{12\alpha}\Big)-\frac{\dot{X}}{nH}\,.
\label{12}
\end{equation}
Combining equations (\ref{10}), (\ref{12}) and replacing the quantity $3+f$ from (\ref{7}), we get
the Raychaudhuri equation of the theory containing, however, also $\beta$
\begin{equation}
\blacklozenge\quad\frac{\dot{X}}{nH}\Big(1+6\sigma_{1}\frac{X+\frac{1}{12\alpha}}
{3\sigma_{1}X\!-\!\rho\!-\!\sigma_{2}}\!+\!\frac{9\sigma_{1}}{\eta\frac{24\pi\alpha}{\kappa_{6}^{2}}\!+\!c\beta}
\Big)
+2\Big(X\!+\!\frac{1}{12\alpha}\Big)\Big(4+3\frac{\rho+p}{3\sigma_{1}X\!-\!\rho\!-\!\sigma_{2}}
\Big)
+\frac{9(\rho\!+\!p)+8(3\sigma_{1}X\!-\!\rho\!-\!\sigma_{2})}
{\eta\frac{24\pi\alpha}{\kappa_{6}^{2}}\!+\!c\beta}=0\,,
\label{ray}
\end{equation}
where $X=H^{2}+\frac{k}{a^{2}}$, $c$ is integration constant, $\eta=0,1$ and
$\sigma_{1}=\frac{r_{c}^{2}}{\kappa_{6}^{2}}+\frac{8\pi\alpha(1-\eta)}
{\kappa_{6}^{2}}$, $\sigma_{2}=\lambda-\frac{2\pi(1-\eta)}{\kappa_{6}^{2}}$,
$\sigma=\sigma_{2}+\frac{\sigma_{1}}{4\alpha}=\lambda+\frac{r_{c}^{2}}{4\alpha\kappa_{6}^{2}}$.
\newline
\noindent
Equation (\ref{1}) with the use of equation (\ref{10}) takes the form
\begin{equation}
\blacklozenge\quad\dot{\rho}+3nH(\rho+p)=\eta\frac{24\pi\alpha}{\kappa_{6}^{2}}\frac{\dot{\beta}}{\beta
\big(\eta\frac{24\pi\alpha}{\kappa_{6}^{2}}\!+\!c\beta\big)}(3\sigma_{1}X\!-\!\rho\!-\!\sigma_{2})\,.
\label{concha}
\end{equation}
{\textit{Equations (\ref{ray}), (\ref{concha}) form the general final two-dimensional system for $a,\rho$, however, still
with the indeterminacy of the function $\beta(t)$}}. Note that one integration constant $c$ enters the
equations, and actually, in the form $c\beta$. Moreover, it is seen from equations (\ref{ray}), (\ref{concha})
that under the rescaling of the deficit angle $\beta(t)\rightarrow c\beta(t)$, the constant $c$ disappears.
However, since this rescaling induces a new angle variable with values in a range different than the standard,
we leave $c$ intact in the following.  This system of equations is the full information available to
us at the brane position and constitute a non-closed system. In other words, one needs extra information coming
from the bulk geometry in order to fix one of the functions, e.g. $\beta$, and then to solve fully the system.
The solution in the bulk is no longer unique as in the case of codimension-1 brane cosmology \cite{binetruy},
\cite{Davidson} and one has a family of bulk solutions parametrized by the angular deficit function $\beta$.
Not all these bulk solutions will be acceptable since certain of them will inevitably carry singularities away
from the brane. Note that the matching conditions (ii), (iv), having $\beta=1$, will form a closed system
of equations on the brane without any undetermined function.

After $X,\rho$ have been found, one can use equations (\ref{10}), (\ref{N}) to find $A,N$.
Only in highly exceptional cases will the extrinsic curvature vanish identically.
As it can be seen in Appendix \ref{geometric components}, the six-dimensional Ricci scalar (as well as other
curvature invariants), beyond the distributional terms, it contains also singular $1/r$ terms multiplied by
the extrinsic curvature. This means that in general the bulk geometry has a genuine curvature singularity at
$r=0$ apart from the distributional one. In fact, it is expected from purely geometrical considerations that
higher codimension defects will develop curvature singularities in their zero width limit \cite{gre}.
Moreover, for the standard Schwarzschild solution of a 0-brane (point mass), the metric is singular
on the point and the curvature diverges on the defect.

In the special case of $\beta$=constant, there should be no $\beta$ in equations (\ref{ray}), (\ref{concha}) since
in this case, there is no $\beta$ in the initial system (\ref{1})-(\ref{3}). A seemingly remaining constant
$\beta$ in (\ref{ray}), (\ref{concha}) is the result of the integration process. However, the integration process
knows to put $\beta$ in the form $c\beta$, thus a constant $\beta$ is absorbed in the integration constant $c$.
Therefore, whenever we speak about constant $\beta$ in the following, we will replace $c\beta$ by $c$ in
(\ref{ray}), (\ref{concha}).

Equation (\ref{ray}), although much complicated, can be brought in a more convenient form defining for
$\sigma_{1}\neq 0$ the variable $\xi$ as
\begin{equation}
\xi=\frac{\rho+\sigma}{3\sigma_{1}X-\rho-\sigma_{2}}
\quad\Leftrightarrow\quad X=\frac{\rho}{3\sigma_{1}}+\frac{\sigma_{2}}{3\sigma_{1}}
+\frac{\rho+\sigma}{3\sigma_{1}}\frac{1}{\xi}\,.
\label{xiX}
\end{equation}
It is seen from the last expression that in principle $H^{2}$ has the standard FRW term
linear in $\rho$, a cosmological constant term and a dark energy contribution attributed to $\xi$, but more
can be said after $\xi$ is found. So, equation (\ref{ray}), using (\ref{concha}), becomes for $p=w\rho$
\begin{equation}
\frac{\dot{\xi}}{nH(\xi\!+\!1)}-\eta\frac{24\pi\alpha}{\kappa_{6}^{2}}\frac{\dot{\beta}}{nH\beta
\big(\eta\frac{24\pi\alpha}{\kappa_{6}^{2}}\!+\!c\beta\big)}=
\Big[\frac{8}{3}-3(1\!+\!w)\Big(1\!+\!\frac{\sigma}{\rho}\Big)^{\!-1}\Big]\,
\frac{3\xi\!+\!3\!+\!9\sigma_{1}\big(\eta\frac{24\pi\alpha}{\kappa_{6}^{2}}\!+\!c\beta\big)^{\!-1}}
{2\xi\!+\!3\!+\!9\sigma_{1}\big(\eta\frac{24\pi\alpha}{\kappa_{6}^{2}}\!+\!c\beta\big)^{\!-1}}
\,\frac{\xi}{\xi\!+\!1}\,.
\label{simpler}
\end{equation}
In order to get some understanding of the behavior of the system, equation (\ref{simpler}) will be integrated in
the next section for a characteristic case of $\beta$.

In the exceptional case where the denominator $3\sigma_{1}X\!-\!\rho-\!\sigma_{2}$ vanishes, some of the previous
equations are not defined. Equation (\ref{N}) gives $\sigma_{1}(X+2Y)+p-\sigma_{2}=0$, which leads to the
conservation equation $\dot{\rho}+3nH(\rho+p)=0$ and from equation (\ref{simple rho dot}) we get
$\beta$\,=\,constant. From equation (\ref{sing rr cosmo}) we can find $\frac{N}{A}$ as a function of $A,\rho$
and from (\ref{sing ri cosmo}) a differential equation for $\frac{dA}{da}$ gives $A$. The four-dimensional
cosmology in this case is the standard FRW cosmology with cosmological constant.

If the deficit angle is time-dependent instead of being exactly constant, a time-varying effective four-dimensional
gravitational constant will be induced. To make an estimate, from the coefficient of the linear term in
(\ref{xiX}) we have $8\pi G_{\!N}=\kappa_{4}^{2}=\frac{\kappa_{6}^{2}}{r_{c}^{2}+8\pi\alpha(1-\beta)}$,
which can also be taken from the relative coefficients of the terms $G^{ij}$, $T^{ij}$ in the matching
condition (\ref{matching1}). The variation of $G_{\!N}$ is constrained during the early cosmology by the
primordial abundances at the nucleosynthesis epoch approximately by
$\frac{|\dot{G}_{\!N}|}{G_{\!N}H}\lesssim 0.2$ \cite{bambi}. This constrains the variation of $\beta$ as
$\Big|\frac{\dot{\beta}}{(1-\beta+r_{c}^{2}/8\pi\alpha)H}\Big|\lesssim 0.2$
which is not a rather strong constraint. Further constraints come from the fact that the theory with varying
$\beta$ is similar to a scalar-tensor theory, and therefore, there will be strong constraints from solar
system observations. Not knowing the full family of solutions in the bulk, we choose to consider in the next
two sections \ref{constant deficit angle}, \ref{topological matching conditions} the case
where $\beta$ is constant. This case will give, at least seemingly, acceptable four
dimensional cosmologies, and has the merit that the system of equations is closed and does not depend on
undetermined functions. In Section \ref{constant deficit angle} we derive the general cosmology for
$\sigma_{1}\neq 0$, while in section \ref{topological matching conditions} the general cosmology for
the complementary case $\sigma_{1}=0$.

\subsection{Cosmology with $\sigma_{1}\neq 0$} \label{constant deficit angle}

As explained above, one characteristic and meaningful way to close the system and capture some features of its
behavior is to assume a constant deficit angle $\beta(t)\!=\!\text{constant}$. Then, equation (\ref{concha})
gets the standard conservation form
\begin{equation}
\dot{\rho}+3nH(\rho+p)=0\,,
\label{standard conservation}
\end{equation}
and equation (\ref{simpler}), using (\ref{standard conservation}) to convert the derivative with respect
to time to derivative with respect to $\rho$, becomes integrable
\begin{equation}
\frac{2\xi+3+9\sigma_{1}\big(\eta\frac{24\pi\alpha}{\kappa_{6}^{2}}\!+\!c\big)^{\!-1}}
{3\xi+3+9\sigma_{1}\big(\eta\frac{24\pi\alpha}{\kappa_{6}^{2}}\!+\!c\big)^{\!-1}}
\,\frac{1}{\xi}\,d\xi=d\rho\,\Big[\frac{1}{\rho+\sigma}-\frac{8}{9(1\!+\!w)}\frac{1}{\rho}\Big]\,,
\label{integrable}
\end{equation}
with general solution
\begin{equation}
\quad\quad\quad\quad\quad\quad\,\,\,\frac{1}{\tilde{c}}\,\rho^{\frac{8}{3(1+w)}}\,
(\rho+\sigma)^{-3}\,\xi^{3}-3\xi=2\gamma
\,\,\,\,\,\,\,\,,\,\,\,\,\,\,\,\,\tilde{c} :\,\text{integration\,constant}\,\,,\,\,
\gamma=\frac{3}{2}+\frac{9\sigma_{1}}{2}\Big(\eta\frac{24\pi\alpha}{\kappa_{6}^{2}}+c\Big)^{\!-1}\,.
\label{solution1}
\end{equation}
Equation (\ref{solution1}) is a cubic for $\xi$ and can be solved
analytically giving the function $\xi(\rho)$, and therefore the Hubble evolution $H^{2}(\rho)$. Since the
cubic has various branches, there are also branches for the cosmic evolution.

{\textit{\underline{Branch I}}} : $\tilde{c}(\rho+\sigma)>0$ and $\tilde{c}\,(\rho+\sigma)^{3}
\leq\gamma^{2}\rho^{\frac{8}{3(1+w)}}$.
\newline
There is one real solution for $\xi$
\begin{equation}
\xi=2\,\text{sgn}(\gamma)\,\rho^{-\frac{4}{3(1+w)}}\sqrt{\tilde{c}(\rho+\sigma)^{3}}\,
\cosh\!\!\Big[\frac{1}{3}\text{arccosh}\Big(
\frac{|\gamma|\,\rho^{\frac{4}{3(1+w)}}}{\sqrt{\tilde{c}(\rho+\sigma)^{3}}}\Big)\Big]\,.
\label{xisol3}
\end{equation}
For positive brane tension $\lambda>0$, it is $\sigma>0$, therefore the first inequality above becomes
$\tilde{c}>0$ and the Hubble equation is
\begin{equation}
H^{2}+\frac{k}{a^{2}}=\frac{\rho}{3\sigma_{1}}+\frac{\sigma_{2}}{3\sigma_{1}}
+\text{sgn}(\gamma)\,\frac{c_{\ast}\rho^{\frac{4}{3(1+w)}}}
{6\sigma_{1}\sqrt{\rho+\sigma}}\,\cosh^{-1}\!\!\Big[\frac{1}{3}\text{arccosh}\Big(
\frac{|\gamma| c_{\ast}\rho^{\frac{4}{3(1+w)}}}{\sqrt{(\rho+\sigma)^{3}}}\Big)\Big]\,,
\label{buika3}
\end{equation}
where $c_{\ast}=\tilde{c}^{-1/2}>0$ and $\cosh^{-1}{\!x}=\frac{1}{\cosh{x}}$. So, equation (\ref{buika3})
contains two integration constants $c_{\ast}\!>\!0,\,\gamma$. For negative brane tension some
slight difference will occur in (\ref{buika3}) according to (\ref{xisol3}).
\newline
As it is seen from the second inequality above, this solution does not accept the regime $\rho\rightarrow 0$,
but already the linear $\rho$ term of FRW is present. Note that the coefficient of this term is
determined by the Gauss-Bonnet coupling $\alpha/\kappa_{6}^{2}$ and the possible induced gravity
coupling $r_{c}^{2}/\kappa_{6}^{2}$, and not by the higher-dimensional gravitational constant $1/\kappa_{6}^{2}$
of the Einstein term. The Einstein term contributes only to the effective cosmological constant and to the extra
correction term of the Hubble equation. This remark is valid also for the other branches below.
\newline
The second inequality above also shows that the energy density $\rho$ cannot become infinite (for $w\!>\!-1/9$),
which means that an infinite-density singularity $a=0$ is not encountered for the solution (\ref{buika3}).
This is true independent of the spatial curvature $k$, or the equation of state. More precisely, for $w=1/3$, which is the
realistic equation of state at the early universe, the condition for the solution (\ref{buika3}) to exist is
$0<\tilde{c}<\frac{4\gamma^{2}}{27\sigma}$ (or for $c_{\ast}$ that $c_{\ast}>\frac{\sqrt{27\sigma}}{2|\gamma|}$).
Note that this effect of avoidance of the infinity is not the same with that of codimension-1 cosmology of the
standard approach \cite{KMP},
where the infinite density singularity is removed by the combined effect of the Gauss-Bonnet and the induced
gravity term, while none of the two terms separately can do this. Here, the coupling $r_{c}$ can be set to
zero (along with $\eta=0$) and still the finite behavior occurs. The maximum energy density $\rho_{M}$ is
found (for $w=1/3$) by the equation $z^{3}\!+\!(3\!-\!\ell^{2}/3)z\!=\!\ell\!-\!1\!-\!2\ell^{3}/27$, where
$z=\frac{\rho_{M}}{\sigma}+\frac{\ell}{3}$, $\ell=3-\frac{\gamma^{2}}{\tilde{c}\sigma}$ (the previous condition
for $\tilde{c}$ means $\ell<-15/4$). In order for $\rho_{M}$ to be the maximum energy density
and not the minimum, the appropriate solution of the cubic equation for $z$ must satisfy $z>2+\ell/3$ for
any $\ell$. To summarize, for a radiation brane the solution (\ref{buika3}) exists only if $\tilde{c}$
(or $c_{\ast}$) satisfies the previous condition, and then necessarily it has finite energy density ever.
\newline
Not only the energy density is finite at early times, but there is a range of the integration constant $c$
(or $\gamma$) for which there is accelerated expansion near the minimum scale factor. This, in principle,
serves as a geometric form of inflation alternative to the scalar field inflation. More precisely, this happens
for $-1/18<\gamma<0$ (or in terms of $c$ for $-3\sigma_{1}<\eta\frac{24\pi\alpha}{\kappa_{6}^{2}}
+c<-81\sigma_{1}/28$). Indeed, using the proper time on the brane, the acceleration parameter $\ddot{a}/a$
is given by the formula $\ddot{a}/a=X+\dot{X}/(2H)$. Using (\ref{xiX}) to express $\dot{X}$ in terms of
$\dot{\xi}$ and (\ref{simpler}) to substitute $\dot{\xi}$ in terms of $\xi(\rho)$, we find
\begin{equation}
3\sigma_{1}\frac{\ddot{a}}{a}=\Big[\frac{1\!+\!9w}{12}\frac{3\xi+2\gamma}{\xi(\xi+\gamma)}
-\frac{1\!+\!3w}{2}\frac{\xi+1}{\xi}\Big]\rho+\sigma_{2}-\frac{3\xi+\gamma}{\xi(\xi+\gamma)}
\frac{\sigma}{3}\,.
\label{kota}
\end{equation}
Equation (\ref{kota}) is valid for any $w$ (for example it can be used for $w=0$) and is true also for
the other branches of solutions to be discussed below, not only the current one. For the solution (\ref{buika3}) and
$w=1/3$, the value of $\xi$ at the initial scale factor is found from (\ref{xisol3}) to be $\xi_{M}=2\gamma$
and the corresponding value of the acceleration is
$3\frac{\sigma_{1}}{\sigma}\frac{\ddot{a}}{a}|_{_{M}}=(\frac{\sigma_{2}}{\sigma}-\frac{7}{18\gamma})
+(1+\frac{1}{18\gamma})(\frac{\ell}{3}-z)$, therefore for the range of $\gamma$ given above,
$\frac{\ddot{a}}{a}|_{_{M}}$ is positive and finite. Note that this result also holds independently of
the spatial curvature $k$ of the universe.
\newline
Finally, the four-dimensional Ricci scalar on the brane is given by $R/6=\dot{H}+2H^{2}+k/a^{2}=X+\ddot{a}/a$,
therefore, form the previous result for $\frac{\ddot{a}}{a}|_{_{M}}$, we obtain that $R$ is finite and there is
no initial singularity.
\newline
To conclude, the solution (\ref{buika3}) with $w=1/3$, for a range of the integration constants $c_{\ast},\gamma$
avoids a cosmological singularity (both in density and curvature) and undergoes accelerated expansion
near the minimum scale factor. There remains however one crucial caveat: there is no exit from acceleration,
as it can be seen that at the minimum energy density there is also acceleration.

{\textit{\underline{Branch II}}} : $\tilde{c}(\rho+\sigma)>0$ and $\tilde{c}\,(\rho+\sigma)^{3}
\geq\gamma^{2}\rho^{\frac{8}{3(1+w)}}$.
\newline
There are three real solutions for $\xi$
\begin{equation}
\xi=2\rho^{-\frac{4}{3(1+w)}}\sqrt{\tilde{c}(\rho+\sigma)^{3}}\,\cos\!\Big[\frac{1}{3}\arccos\!\Big(
\frac{\gamma\,\rho^{\frac{4}{3(1+w)}}}{\sqrt{\tilde{c}(\rho+\sigma)^{3}}}\Big)\!+\!\frac{2\pi m}{3}\Big]
\,\,\,,\,\,\,m=0,1,2\,.
\label{xisol2}
\end{equation}
For positive brane tension $\lambda>0$, it is $\sigma>0$, therefore the first inequality above becomes
$\tilde{c}>0$ and the Hubble equation is
\begin{equation}
H^{2}+\frac{k}{a^{2}}=\frac{\rho}{3\sigma_{1}}+\frac{\sigma_{2}}{3\sigma_{1}}
+\frac{c_{\ast}\rho^{\frac{4}{3(1+w)}}}{6\sigma_{1}\sqrt{\rho+\sigma}}
\,\cos^{-1}\!\!\Big[\frac{1}{3}\arccos\!\Big(
\frac{\gamma c_{\ast}\rho^{\frac{4}{3(1+w)}}}{\sqrt{(\rho+\sigma)^{3}}}\Big)\!+\!\frac{2\pi m}{3}\Big]
\,\,\,,\,\,\,m=0,1,2\,,
\label{buika2}
\end{equation}
where $c_{\ast}=\tilde{c}^{-1/2}>0$ and $\cos^{-1}{\!x}=\frac{1}{\cos{x}}$. So, equation (\ref{buika2})
contains two integration constants $c_{\ast}\!>\!0,\,\gamma$ or alternatively, we could consider the two
integration constants $c_{\ast}>0, \,c^{\ast}=\gamma c_{\ast}$. For negative brane tension some
slight difference will occur in (\ref{buika2}) according to (\ref{xisol2}). The solution (\ref{buika2})
accepts the regime $\rho\rightarrow 0$, where for $w=0$ it is
\begin{equation}
H^{2}+\frac{k}{a^{2}}\approx \frac{\rho}{3\sigma_{1}}+\frac{\sigma_{2}}{3\sigma_{1}}
+\frac{c_{\ast}}{6\sigma_{1}\!\sqrt{\sigma}}\cos^{-1}\!\!\Big[\frac{\pi(1\!+\!4m)}{6}\Big]
\,\rho^{\frac{4}{3}}\,.
\label{linearized2}
\end{equation}
Depending on the value of $\tilde{c}$ the solution can evolve from an initial big bang to $\rho
\rightarrow 0$, or from a finite value of the energy density to $\rho\rightarrow 0$.

{\textit{\underline{Branch III}}} : $\tilde{c}(\rho+\sigma)<0$.
\newline
There is one real solution for $\xi$
\begin{equation}
\xi=-2\rho^{-\frac{4}{3(1+w)}}\sqrt{-\tilde{c}(\rho+\sigma)^{3}}\,
\sinh\!\!\Big[\frac{1}{3}\text{arcsinh}\Big(
\frac{\gamma\,\rho^{\frac{4}{3(1+w)}}}{\sqrt{-\tilde{c}(\rho+\sigma)^{3}}}\Big)\Big]\,.
\label{xisol1}
\end{equation}
For positive brane tension $\lambda>0$, it is $\sigma>0$, therefore the inequality above becomes $\tilde{c}<0$
and the Hubble equation is
\begin{equation}
H^{2}+\frac{k}{a^{2}}=\frac{\rho}{3\sigma_{1}}+\frac{\sigma_{2}}{3\sigma_{1}}
-\frac{c_{\ast}\rho^{\frac{4}{3(1+w)}}}
{6\sigma_{1}\sqrt{\rho+\sigma}}\,\sinh^{-1}\!\!\Big[\frac{1}{3}\text{arcsinh}\Big(
\frac{\gamma c_{\ast}\rho^{\frac{4}{3(1+w)}}}{\sqrt{(\rho+\sigma)^{3}}}\Big)\Big]\,,
\label{buika1}
\end{equation}
where $c_{\ast}=|\tilde{c}|^{-1/2}>0$ and $\sinh^{-1}{\!x}=\frac{1}{\sinh{x}}$. So, equation (\ref{buika1})
contains two integration constants $c_{\ast}\!>\!0,\,\gamma$ or alternatively, we could consider the two
integration constants $c_{\ast}>0, \,c^{\ast}=\gamma c_{\ast}$. For negative brane tension some
slight difference will occur in (\ref{buika1}) according to (\ref{xisol1}). The solution (\ref{buika1}) accepts
the regime $\rho\rightarrow 0$, where for $w=0$ it is
\begin{equation}
H^{2}+\frac{k}{a^{2}}\approx \frac{\rho}{3\sigma_{1}}\Big(1-\frac{3}{2\gamma}\Big)+\frac{1}{3\sigma_{1}}
\Big(\sigma_{2}-\frac{3\sigma}{2\gamma}\Big)-\frac{2\gamma c_{\ast}^{2}}{27\sigma_{1}\sigma^{2}}
\rho^{\frac{8}{3}}\,.
\label{linearized1}
\end{equation}
Although in principle $\gamma$ has any sign, the linearized regime (\ref{linearized1}), in comparison to
the standard FRW equation, gives the constraints $8\pi G_{N}=\kappa_{4}^{2}=\frac{1}{\sigma_{1}}
\big(1-\frac{3}{2\gamma}\big)$ and $\Lambda_{\text{eff}}=\frac{1}{\sigma_{1}}\big(\sigma_{2}-\frac{3\sigma}
{2\gamma}\big)$. Combining these relations we get $\frac{3}{8\alpha\gamma}=\kappa_{4}^{2}
[\lambda-2\pi(1\!-\!\eta)M_{6}^{4}]-\Lambda_{\text{eff}}$, where $\kappa_{6}^{2}=M_{6}^{-4}$. Since
$\kappa_{4}^{2}\sim 10^{-31}\text{TeV}^{-2}$, $\Lambda_{\text{eff}}\sim 10^{-90}\text{TeV}^{2}$,
$M_{6}\gtrsim \text{TeV}$, for any reasonable $\lambda$ the term $\Lambda_{\text{eff}}$ is
insignificant. Finally, $\gamma$ can indeed be either positive or negative depending on the value of
$\lambda$, and the correction term of (\ref{buika1}) can also have either sign. Moreover, setting the
effective cosmological constant $\Lambda_{\text{eff}}$ to zero, in case we want
the vacuum to be the Minkowski space, we get $\kappa_{4}^{2}=(4\alpha\lambda+r_{c}^{2}M_{6}^{4})^{-1}$.

In general, the solutions found above need a more systematic analysis to study their cosmological behaviour.
Usually one adjusts the parameters of a solution such that in the absence of matter to recover the Minkowski
background. However, this is not necessary since any curved spacetime is locally Minkowski and all local
physics constraints are satisfied. For example, the Randall-Sundrum fine-tuning in five dimensions assures a
Minkowski background by exactly vanishing the effective cosmological constant \cite{rs}, \cite{binetruy}. In this
case, the today acceleration would possibly be attributed to some dark energy component. If, on the contrary, the
effective cosmological constant of a model does not vanish, the background is de-Sitter and it contributes to the
today acceleration. For the branches I, II above, the effective cosmological constant is
$\frac{\sigma_{2}}{\sigma_{1}}$ and is no need to vanish (for $\eta=0$) or cannot vanish (for $\eta=1$).
Its value, together with the correction term, could possibly define the dark energy today.

Let's finish with an intriguing comment. For a 3-brane in $D$ dimensions, out of the scales $\kappa_{D}^{2}
=M_{D}^{2-D}$, $\lambda$, $r_{c}$ it is constructed the following scale with dimensions of effective
cosmological constant, $\Lambda_{\text{eff}}=\frac{\kappa_{D}^{2}\lambda}{r_{c}^{D-4}}=\frac{\lambda}{M_{D}^{D-2}}
\frac{1}{r_{c}^{D-4}}$. The induced gravity crossover scale $r_{c}$ distinguishes the four-dimensional from
the higher-dimensional regime. For a reasonable value of the brane tension $\lambda\sim M_{D}^{4}$ (e.g. $\lambda\sim
\text{TeV}^{4}$ a typical cut-off) which might also have theoretical explanation as it is connected to the
fundamental mass scale, it arises $\Lambda_{\text{eff}}\sim\frac{1}{(M_{D}r_{c})^{D-6}}\frac{1}{r_{c}^{2}}$.
For $D=6$ it is $\Lambda_{\text{eff}}\sim\frac{1}{r_{c}^{2}}$. If the induced gravity crossover scale is
of the cosmic horizon size $r_{c}\sim H^{-1}$, then $\Lambda_{\text{eff}}\sim H^{2}$, i.e. for today
$\Lambda_{\text{eff}}$ has the observed magnitude! Since the induced gravity term is generically induced
by quantum corrections coming from the bulk gravity and its coupling with matter living on the brane,
it wouldn't be absurd to suppose that the accumulative contribution of all the matter inside our horizon
gives a connection of $r_{c}$ to $H^{-1}$. But if $\Lambda_{\text{eff}}\sim H^{2}$, why only today
$\Lambda_{\text{eff}}$ emerges and not in the past? To make another conjecture, only today
a reasonable number of galaxies with high energetic interiors exist from where the quantum loops
generate the induced gravity term. On the contrary, any other dimensionality $D$ fails to support
the observed value of $\Lambda_{\text{eff}}$. For $D=5$ it is $\Lambda_{\text{eff}}\sim \frac{M_{D}}{r_{c}}$
and in order to be $\Lambda_{\text{eff}}\sim H_{0}^{2}$ it must be
$r_{c}\sim\frac{M_{D}}{H_{0}}H_{0}^{-1}$, which for $M_{D}\sim \text{TeV}$ gives $r_{c}\sim 10^{45}H_{0}^{-1}$
(too high to be realistic). For $D=7$ it is $\Lambda_{\text{eff}}\sim \frac{1}{M_{D}r_{c}^{3}}$
and in order to be $\Lambda_{\text{eff}}\sim H_{0}^{2}$ it must be
$r_{c}\sim(\frac{H_{0}}{M_{D}})^{\frac{1}{3}}H_{0}^{-1}$, which for $M_{D}\sim \text{TeV}$ gives
$r_{c}\sim 10^{-15}H_{0}^{-1}\sim A.U.$ (this may look marginally legitimate but it is probably
already excluded since high precision measurements on solar system astronomy occur at distances $30A.U.$
which is larger than $A.U.$; more importantly, a higher value of $M_{D}$, which is more reasonable, reduces $r_{c}$).
For $D=8$, it is $\Lambda_{\text{eff}}\sim \frac{1}{M_{D}^{2}r_{c}^{4}}$
and in order to be $\Lambda_{\text{eff}}\sim H_{0}^{2}$ it must be
$r_{c}\sim\sqrt{\frac{H_{0}}{M_{D}}}H_{0}^{-1}$, which for $M_{D}\sim \text{TeV}$ gives
$r_{c}\sim 10^{-27}H_{0}^{-1}\sim 10^{-12}A.U.$ (obviously incompatible). Incompatibility worsens with yet
lower $r_{c}$ for higher $D$. In all the solutions (\ref{buika3}), (\ref{buika2}), (\ref{buika1}), or even
in (\ref{xiX}), there appears the effective cosmological constant $\Lambda_{\text{eff}}=\frac{\sigma_{2}}
{\sigma_{1}}=\frac{\kappa_{6}^{2}\lambda-2\pi(1-\eta)}{r_{c}^{2}+8\pi\alpha(1-\eta)}$. Already the scale
$\frac{\kappa_{6}^{2}\lambda}{r_{c}^{2}}$ starts becoming visible, but for $\eta=1$ it is exactly
$\Lambda_{\text{eff}}=\frac{\sigma_{2}}{\sigma_{1}}=\frac{\kappa_{6}^{2}\lambda}{r_{c}^{2}}$.
Therefore, we have a consistent
formalism which naturally embodies the effective cosmological constant scale $\frac{\kappa_{6}^{2}\lambda}
{r_{c}^{2}}$ which can suppress the large value of vacuum energy to the small observed cosmological constant
(for the standard treatment a relevant scale $\Lambda_{\text{eff}}=\frac{\kappa_{6}^{2}\lambda-2\pi(1-\beta)}
{r_{c}^{2}+8\pi\alpha(1-\beta)}$ also arises, as it is seen from (\ref{england}) by setting $\rho\!=\!p\!=\!0$,
$Y\!=\!X$). However, even if all the above arguments have some physical relevance, there is a point that
creates a serious phenomenological difficulty. This is the constraint from the four-dimensional Newton's constant
$G_{\!N}$. From the solutions (\ref{buika3}), (\ref{linearized2}) it has to be $G_{\!N}\sim \sigma_{1}^{-1}=\frac{\kappa_{6}^{2}}{r_{c}^{2}}$,
and then, $r_{c}$ is constrained by $M_{6}$ and all the previous numerology collapses. However, we give
an idea how this difficulty might be evaded. In the spirit of equations (\ref{linearized1}), (\ref{hip hop}),
a possible dependence of $G_{\!N}$ on some extra integration constant $G_{\!N}=G_{\!N}(\sigma_{1},\textsf{c})$
(even a whole time-dependence of $G_{\!N}$ due to $\beta(t)$ would create for today an analogous result) could
liberate $r_{c}$ from $M_{6}$ and charge the value of $G_{\!N}$ to this integration constant $\textsf{c}$.
For example, for the solution (\ref{linearized1}) it is
$8\pi G_{\!N}=\kappa_{4}^{2}=\frac{\kappa_{6}^{2}}{r_{c}^{2}}(1-\frac{3}{2\gamma})$, and therefore
$\frac{1}{\kappa_{4}^{2}}\neq \frac{r_{c}^{2}}{\kappa_{6}^{2}}$, as opposed to what one usually imagines from the
action (\ref{Stotal})! To say it in a different way, in the present theory, Newton's constant is not determined
solely from the parameters ($M_{6},r_{c}$) of the action, but in general depends also from the
integration constants of the considered solution. Although in the concrete example of equation (\ref{linearized1})
the dependence of $G_{\!N}$ on $\gamma$ does not save the argument, further investigation of the theory and
its solutions could shed more light on this issue.

\subsection{Cosmology with $\sigma_{1}= 0$} \label{topological matching conditions}

The case $\sigma_{1}\!=\!0$ corresponds to $\eta=1$, without the induced gravity term $r_{c}=0$. Therefore, it
corresponds to the matching conditions (i), (iv) with $r_{c}=0$. Then, the conservation equation (\ref{concha})
for general $\beta(t)$ can be rewritten as
\begin{equation}
\Big(\frac{\rho\!+\!\lambda}{\frac{24\pi\alpha}{\kappa_{6}^{2}\,\beta}\!+\!c}\Big)
^{\centerdot}+3nH\,\frac{\rho+p}{\frac{24\pi\alpha}{\kappa_{6}^{2}\,\beta}\!+\!c}=0\,
\label{rewrite conservation}
\end{equation}
and displays an interplay between the energy density and the deficit angle during the evolution. This equation
looks like the standard energy conservation equation for redefined energy densities
of matter and brane tension, i.e. $\rho\rightarrow \rho f(\beta)$ and $\rho_{\lambda}\rightarrow
\rho_{\lambda} f(\beta)$, where $\rho_{\lambda}=\lambda$ and
$f(\beta)=\big(\frac{24\pi\alpha}{\kappa_{6}^{2}\,\beta}\!+\!c\big)^{-1}$. Then, for $p=w\rho$ it is
$p\rightarrow pf(\beta)$ and since $p_{\lambda}=-\lambda$ it is $p_{\lambda}\rightarrow p_{\lambda}
f(\beta)$. It is the same to say that the total energy density $\tilde{\rho}=\rho+\lambda$ is redefined
$\tilde{\rho}\rightarrow \tilde{\rho}f(\beta)$, and therefore, for the total pressure $\tilde{p}=p-\lambda$
it is $\tilde{p}\rightarrow \tilde{p}f(\beta)$. Furthermore, the Raychaudhuri equation (\ref{ray}) becomes
\begin{equation}
\frac{\dot{X}}{nH}+2\Big(\!X\!+\!\frac{1}{12\alpha}\Big)\frac{\rho-3p+4\lambda}{\rho+\lambda}
+\frac{\rho+9p-8\lambda}{\frac{24\pi\alpha}{\kappa_{6}^{2}}\!+\!c\beta}=0\,.
\label{alli mia}
\end{equation}
Equation (\ref{rewrite conservation}) is written as
\begin{equation}
\frac{dy}{d\Omega}+3(1\!+\!w)y-3(1\!+\!w)\lambda\Big(\frac{24\pi\alpha}{\kappa_{6}^{2}\,\beta}+c\Big)^{-1}=0\,,
\label{monaxo}
\end{equation}
where
\begin{equation}
y=\frac{\rho\!+\!\lambda}{\frac{24\pi\alpha}{\kappa_{6}^{2}\,\beta}\!+\!c}\,\,\,\,\,,\,\,\,\,\,
\Omega=\ln{\frac{a}{a_{0}}}\,,
\label{turk}
\end{equation}
and $a_{0}$ is, for example, the today scale factor. Therefore, for $\beta=\beta(a)$, (\ref{monaxo}) is a linear
differential equation for $y$ and can be integrated giving $\rho(a)$. Then, (\ref{alli mia}) becomes a linear
differential equation for $\tilde{X}$
\begin{equation}
\frac{d\tilde{X}}{d\Omega}+2\frac{(1\!-\!3w)\rho\!+\!4\lambda}{\rho+\lambda}\tilde{X}
+\big[(1\!+\!9w)\rho\!-\!8\lambda\big]\Big(\frac{24\pi\alpha}{\kappa_{6}^{2}}\!+\!c\beta\Big)^{-1}=0
\,\,\,\,\,,\,\,\,\,\,\tilde{X}=X\!+\!\frac{1}{12\alpha}\,,
\label{rwta}
\end{equation}
giving $H(a)$ or $H(\rho)$. For example, for a slowly varying $\beta$ around today, one could assume that
$\beta(\Omega)\approx\beta_{0}+\nu\,\Omega^{2}$.

Here, we will restrict ourselves to the case of constant deficit angle during the
universe evolution, $\beta(t)\!=\!\text{constant}$. The absence, however, of the induced gravity term makes the
derived cosmology rather simple. As we have explained, since $\beta$ is constant, the quantity $c\beta$ in
equations (\ref{rewrite conservation}), (\ref{alli mia}) should be replaced by $c$. Equation
(\ref{rewrite conservation}) becomes the standard conservation law. Equation (\ref{alli mia}) with the use of
the conservation equation reduces to the linear differential equation
\begin{equation}
\frac{dX}{d\rho}-\frac{2}{3(1\!+\!w)}\,\frac{(1\!-\!3w)\rho+4\lambda}{\rho
(\rho+\lambda)}\Big(\!X\!+\!\frac{1}{12\alpha}\Big)=\frac{1}{\frac{24\pi\alpha}{\kappa_{6}^{2}}\!+\!c}
\frac{(1\!+\!9w)\rho-8\lambda}{3(1\!+\!w)\rho}\,,
\label{linear}
\end{equation}
with general solution
\begin{equation}
H^{2}+\frac{k}{a^{2}}=\frac{\kappa_{6}^{2}\,\rho}{24\pi\alpha(1\!-\!\bar{c})}
+\Big(\frac{\kappa_{6}^{2}\,\lambda}{24\pi\alpha(1\!-\!\bar{c})}-\frac{1}{12\alpha}\Big)
+\frac{\tilde{c}}{(\rho+\lambda)^{2}}\,\rho^{\frac{8}{3(1+w)}}\,,
\label{hip hop}
\end{equation}
where $\bar{c}=-\frac{\kappa_{6}^{2}}{24\pi\alpha}c$ is a redefinition of the integration constant $c$, and
$\tilde{c}$ is another integration constant.
So, equation (\ref{hip hop}) contains two integration constants $\bar{c},\tilde{c}$. Note that
the four-dimensional Newton constant is determined by an integration constant and not solely from the parameters
of the theory.

Setting for this cosmology $\kappa_{4}^{2}=\frac{\kappa_{6}^{2}}{8\pi\alpha(1-\bar{c})}$ the effective
four-dimensional gravitational constant, we have
\begin{equation}
H^{2}+\frac{k}{a^{2}}=\frac{\kappa_{4}^{2}}{3}\rho+\Big(\frac{\kappa_{4}^{2}\lambda}{3}-\frac{1}{12\alpha}\Big)
+\frac{\tilde{c}}{(\rho+\lambda)^{2}}\,\rho^{\frac{8}{3(1+w)}}\,.
\label{couver}
\end{equation}
By fine-tuning the brane tension to the value $\lambda=\frac{1}{4\alpha\kappa_{4}^{2}}$, the effective
cosmological constant vanishes and (\ref{couver}) becomes
\begin{equation}
H^{2}+\frac{k}{a^{2}}=\frac{\kappa_{4}^{2}}{3}\rho
+\frac{\tilde{c}}{\big(\rho+\frac{1}{4\alpha\kappa_{4}^{2}}\big)^{2}}\,\rho^{\frac{8}{3(1+w)}}\,,
\label{couvera}
\end{equation}
which includes a non-trivial correction beyond the standard linear to energy density term.

\section{Special cases, Einstein limit and Comparison with standard approach} \label{specom}

\subsection{Special cases} \label{special cases}

It is instructive to see the limit of a few spacial cases. These cases need special treatment since several
expressions are divided by $A,H$ etc.

(a) {\it{Minkowski brane:}} An exactly Minkowski four-dimensional spacetime is no need to be the solution of a viable gravitational
theory since any curved spacetime solution is locally Minkowski, and therefore, satisfies all known local
physics. However, the present theory possesses an exact brane Minkowski vacuum, where $H=0$, $k=0$,
$\rho=0$. In this case, the two boundary conditions $\eta=0,1$ reduce to a single case with vanishing extrinsic
curvature $A=N=0$, the deficit angle $\beta$ comes out to be an integration constant, while for the second order
coefficients it holds $\beta_{2}=0$, $3A'+N'=\kappa_{6}^{2}\mathcal{T}^{r}_{r}-\Lambda_{6}$ with
$\mathcal{T}^{r}_{r}=\mathcal{T}^{\theta}_{\theta}$, $\mathcal{T}^{t}_{r}=0$. Here, the regular
parts of the $ij$ bulk equations on the brane do not contain $a''',n'''$, as it happens when $A,N \neq 0$,
but they contain $A',N',L'''$, therefore, there is one additional equation for $A',N'$, namely
$A'-N'=\frac{\kappa_{6}^{2}}{6}(3\mathcal{T}_{t}^{t}-\mathcal{T}_{\hat{i}}^{\hat{i}})$, and finally $A',N'$
are uniquely determined $A'=\frac{\kappa_{6}^{2}}{24}(6\mathcal{T}_{r}^{r}+3\mathcal{T}^{t}_{t}
-\mathcal{T}^{\hat{i}}_{\hat{i}})\!-\!\frac{\Lambda_{6}}{4}$,
$N'=\frac{\kappa_{6}^{2}}{8}(2\mathcal{T}_{r}^{r}-3\mathcal{T}^{t}_{t}
+\mathcal{T}^{\hat{i}}_{\hat{i}})-\frac{\Lambda_{6}}{4}$. Note that in this special case the brane tension
$\lambda$ does not enter, so irrespectively of its value, a Minkowski brane is achieved without fine-tuning.

If the bulk has only $\Lambda_{6}<0$, one can find
the exact bulk solution of EGB theory which is the extension of the above Minkowski brane \cite{CKP}, \cite{chapap}.
For $\alpha<\frac{10}{24|\Lambda_{6}|}$, we start with the static black hole solution of EGB gravity
\cite{bouldeser} with horizon of toroidal topology
\begin{equation}
ds_{6}^{2}=-\Delta^{2}(R)d\tau^{2}+\frac{dR^{2}}{\Delta^{2}(R)}+R^{2}\delta_{ij}d\zeta^{i}d\zeta^{j}
\,\,\,,\,\,\,
\Delta^{2}(R)=\frac{R^{2}}{12\alpha}\left[1-\sqrt{1+24\alpha\left(\frac{\Lambda_{6}}{10}
+\frac{\mu}{R^{5}}\right)}\,\right]\label{deser}
\end{equation}
($\mu$ is integration constant) which possesses one singularity at $R=0$ shielded by an horizon at
$R_{h}=(\frac{10\mu}{-\Lambda_{6}})^{\frac{1}{5}}$. Making the double Wick rotation $\theta=-i\tau$,
$t=-i\zeta^{0}$ and defining $\chi^{\hat{i}}=\zeta^{\hat{i}}$ and $r$ instead of $R$ by $\frac{dR}{dr}=\Delta(R)$,
the solution (\ref{deser}) gets the Gaussian-normal form (\ref{ansatz})
\begin{equation}
ds_{6}^{2}=dr^{2}+L^{2}(r)d\theta^{2}+R^{2}(r)(-dt^{2}+d\vec{\chi}_{3}^{2})\,.
\label{exact bulk cosmocon}
\end{equation}
Integrating equation $\frac{dR}{dr}=\Delta(R)$ around $R=R_{h}$, we find $r\simeq 2\sqrt{\frac{-2}{5\Lambda_{6}}}
\sqrt{1-\frac{R_{h}^{5}}{R^{5}}}$, therefore the brane should be located at the horizon. Moreover, since the
induced metric on the horizon should be the Minkowski metric, it has to be $R_{h}=1$, i.e. $\mu=-\Lambda_{6}/10$,
and it is found that $L(r)=\Delta(R(r))=-\frac{\Lambda_{6}}{4}r+\mathcal{O}(r^{3})$,
$R(r)=1-\frac{\Lambda_{6}}{8}r^{2}+\mathcal{O}(r^{4})$. Comparing the metric (\ref{exact bulk cosmocon})
with (\ref{cosmological metric}), it is found that $A=N=0$, $\beta_{2}=0$,
$A'=N'=-\frac{\Lambda_{6}}{4}$, in accordance with the values found above for a Minkowski brane.
The constant deficit angle $\beta$ which from the brane viewpoint of the effective equations
remained undetermined, now the embedding of the brane in the exact bulk solution specifies its value to be
$\beta=\frac{|\Lambda_{6}|}{4}$ if the angle $\theta$ has the standard normalization $0\leq\theta<2\pi$.

For vanishing bulk matter content and $\Lambda_{6}=0$ the bulk solution of Einstein or EGB gravity
which is consistent with a Minkowski brane
is the locally flat geometry with a constant deficit angle $ds_{6}^{2}=dr^{2}+\beta^{2}r^{2}d\theta^{2}
-dt^{2}+d\vec{\chi}_{3}^{2}$. This bulk solution has $A=N=\beta_{2}=A'=N'=0$, as required by the previous brane
equations. The only difference of this configuration with the standard treatment is that
here the brane tension $\lambda$ remains an arbitrary parameter, while in the standard approach the equation
$\mathcal{G}_{ij}+\alpha \mathcal{J}_{ij}=-\kappa_{6}^{2}\lambda h_{ij}\frac{\delta(r)}{2\pi\beta r}$ is more
restrictive and implies the well-known fine-tuning $\lambda\!=\!\frac{2\pi}{\kappa_{6}^{2}}(1-\beta)$
\cite{Vilenkin}, \cite{gott}
(this fine-tuning is the same as in Einstein gravity, since for a Minkowski brane the Gauss-Bonnet term does
not contribute). Such a sort of relation we would like to arise as a direct calculation of an
appropriately defined energy of the gravitational field, and indeed, one such definition that has successfully
passed various other tests is given by the teleparallel representation of Einstein gravity \cite{maluf}.
It gives for the gravitational field of the previous locally flat bulk solution the energy per unit spatial
volume of the defect, inside a cylinder of arbitrary radius around the defect, equal to
$\varepsilon_{g}\!=\!\frac{2\pi}{\kappa_{6}^{2}}(1-\beta)$, which is same with the energy per unit length of a
cosmic string. Since the radius of the cylinder is insignificant, it may be concluded that the whole energy is
concentrated along the axis $r=0$. Therefore, in the standard approach the energy is localized on the brane and
is due to the brane tension which adjusts $\beta$, while in the current approach the energy is again practically
localized on the brane in the form of gravitational energy which now adjusts $\beta$ and has the same value
as before. So, now the gravitational energy stored at the defect, instead of the brane tension itself, adjusts
the deficit angle and the two descriptions are physically similar.

(b) {\it{Vanishing extrinsic curvature:}} Since one of the two matching conditions, equation (\ref{matching1}),
is contracted with the extrinsic curvature, in the limit of vanishing extrinsic curvature this equation is
satisfied identically. Therefore, the situation is expected rather exceptional. One such example was the Minkowski
brane above. For cosmology with $A=N=0$ the equations on the brane give the exact conservation equation,
$\beta$\,=\,constant, $\beta_{2}=0$, $\mathcal{T}^{r}_{r}=\mathcal{T}^{\theta}_{\theta}$,
$\mathcal{T}^{t}_{r}=0$. Additionally, the regular parts of the $ij$ bulk equations on the brane do not contain
$a''',n'''$, as it happens when $A,N \neq 0$, but they contain $A',N',L'''$. These two equations, together with the
$\theta\theta$ equation (\ref{theta regular diff}), can be solved for $A',N',\beta_{3}$. Therefore, $H$ remains
arbitrary and any four-dimensional cosmology can be solution, under the condition that it is embedded
geodesically in the bulk. Since there is no curvature singularity at $r=0$ in this case, maybe the stability
issue could shed more light on the physical relevance of such geodesic embeddings.

(c) { \it{Only tension on the brane:}} This means that $\rho=p=0$, and therefore, this case cannot be obtained
from the solutions of section \ref{cosmo evolution}. Physically, this situation could
approximate periods of the history of the universe where inflation or dark energy dominate over matter.
From equation (\ref{1}) it is concluded that $\beta$\,=\,constant. Then, equations (\ref{2}), (\ref{3})
do not contain $\beta$, thus, (\ref{ray}), (\ref{simpler}) should not contain $\beta$, and therefore,
$c\beta$ is replaced by $c$ in (\ref{ray}), (\ref{simpler}).

For $\sigma_{1}\neq 0$ the equation which governs the evolution is (\ref{simpler}) and gets the form
\begin{equation}
\frac{\dot{\xi}}{nH}=\frac{4\xi}{3}\frac{3\xi+2\gamma}{\xi+\gamma}\,,
\label{bouno}
\end{equation}
where $\gamma=\frac{3}{2}+\frac{9\sigma_{1}}{2}\Big(\eta\frac{24\pi\alpha}{\kappa_{6}^{2}}\!+\!c\Big)^{\!-1}$,
with general solution
\begin{equation}
\frac{1}{\tilde{c}\,a^{8}}\,\xi^{3}-3\xi=2\gamma
\label{ert}
\end{equation}
($\tilde{c}$ integration constant). It is also defined $c_{\ast}=|\tilde{c}|^{-1/2}>0$.
Equation (\ref{ert}) is a cubic for $\xi$ and can be solved
analytically giving the function $\xi(a)$, and then the Hubble evolution $H^{2}(a)$. Since the cubic has
various branches, there are also branches for the cosmic evolution, all containing the two integration constants
$c_{\ast}\!>\!0,\,\gamma$.
\noindent
\newline
For $\tilde{c}>0$ and $a^{8}\leq \gamma^{2}/\tilde{c}$, there is one real solution
\begin{equation}
H^{2}+\frac{k}{a^{2}}=\frac{\sigma_{2}}{3\sigma_{1}}+\text{sgn}(\gamma)\frac{\sigma c_{\ast}}{6\sigma_{1}a^{4}}
\,\cosh^{-1}\!\!\Big[\frac{1}{3}\text{arccosh}\Big(\frac{|\gamma| c_{\ast}}{a^{4}}\Big)\Big]\,.
\label{buika3tension}
\end{equation}
\noindent
\newline
For $\tilde{c}>0$ and $a^{8}\geq \gamma^{2}/\tilde{c}$, there are three real solutions
\begin{equation}
H^{2}+\frac{k}{a^{2}}=\frac{\sigma_{2}}{3\sigma_{1}}+\frac{\sigma c_{\ast}}{6\sigma_{1}a^{4}}
\,\cos^{-1}\!\!\Big[\frac{1}{3}\arccos\!\Big(\frac{\gamma c_{\ast}}{a^{4}}\Big)\!+\!\frac{2\pi m}{3}\Big]
\,\,\,,\,\,\,m=0,1,2\,,
\label{buika2tension}
\end{equation}
and accept the regime $a\rightarrow \infty$
\begin{equation}
H^{2}+\frac{k}{a^{2}}\approx \frac{\sigma_{2}}{3\sigma_{1}}
+\frac{\sigma c_{\ast}}{6\sigma_{1}}\cos^{-1}\!\!\Big[\frac{\pi(1\!+\!4m)}{6}\Big]
\,\frac{1}{a^{4}}\,.
\label{linearized2tension}
\end{equation}
\noindent
\newline
For $\tilde{c}<0$ there is one real solution
\begin{equation}
H^{2}+\frac{k}{a^{2}}=\frac{\sigma_{2}}{3\sigma_{1}}-\frac{\sigma c_{\ast}}{6\sigma_{1}a^{4}}
\,\sinh^{-1}\!\!\Big[\frac{1}{3}\text{arcsinh}\Big(\frac{\gamma c_{\ast}}{a^{4}}\Big)\Big]\,,
\label{buika1tension}
\end{equation}
and accepts the regime $a\rightarrow \infty$
\begin{equation}
H^{2}+\frac{k}{a^{2}}\approx \frac{1}{3\sigma_{1}}\Big(\sigma_{2}-\frac{3\sigma}{2\gamma}\Big)
-\frac{2\sigma\gamma c_{\ast}^{2}}{27\sigma_{1}}\,\frac{1}{a^{8}}\,.
\label{linearized1tension}
\end{equation}
\noindent
\newline
Note that the previous solutions for $c_{\ast}=k=0$ can give a de-Sitter brane.

For $\sigma_{1}=0$, the relevant equation is (\ref{ray}) which becomes
\begin{equation}
\frac{\dot{X}}{nH}+8\Big(\!X\!+\!\frac{1}{12\alpha}\Big)-\frac{8\lambda}
{\frac{24\pi\alpha}{\kappa_{6}^{2}}\!+\!c}=0\,,
\label{poto}
\end{equation}
with general solution
\begin{equation}
H^{2}+\frac{k}{a^{2}}=\frac{\tilde{c}}{a^{8}}
+\frac{\kappa_{6}^{2}\,\lambda}{24\pi\alpha(1\!-\!\bar{c})}-\frac{1}{12\alpha}\,,
\label{frouto}
\end{equation}
where $\bar{c}=-\frac{\kappa_{6}^{2}}{24\pi\alpha}c$ is a redefinition of the integration constant $c$, and
$\tilde{c}$ is another integration constant.
So, equation (\ref{frouto}) contains two integration constants $\bar{c},\tilde{c}$. For
$\tilde{c}=k=0$ the solution (\ref{frouto}) can also give a de-Sitter brane, which is alternatively
obtained by setting directly $H$\,=\,constant in (\ref{poto}).

\subsection{The Einstein limit} \label{Einstein}

In the limit of Einstein bulk gravity the Gauss-Bonnet term is absent $\alpha=0$. We give the effective equations
for a general axially symmetric configuration.

The matching condition (\ref{matching1}) becomes
\begin{equation}
\Big{\{}\!\kappa_{6}^{2}T^{ij}-\big[\kappa_{6}^{2}\lambda-2\pi(1-\beta)\big]h^{ij}
-r_{c}^{2}G^{ij}\!\Big{\}}K_{ij}=0\,,
\label{einstein matching1}
\end{equation}
while the matching condition (\ref{matching2}) gets the form
\begin{equation}
T^{ij}_{\,\,\,\,\,|j}=\frac{2\pi}{\kappa_{6}^{2}}h^{ij}\beta_{,j}\,.
\label{einstein matching2}
\end{equation}
Among the two terms $\frac{L''}{L}$, $\frac{L'}{L}\mathcal{K}_{ij}'$ which contribute to the distributional terms
$\frac{\delta(r)}{r}$, only the first is present in the Einstein tensor. Note that the index $\eta$ disappears
in (\ref{einstein matching1}), (\ref{einstein matching2}). In general, the term $\frac{L'}{L}\mathcal{K}_{ij}'$
attributes to the matching conditions extra terms due to the possible discontinuity of $\mathcal{K}_{ij}$. Since
this term is absent here, there is no point to consider $\mathcal{K}_{ij}$ discontinuous, therefore the Einstein
limit corresponds to $\eta=1$.

The $\mathcal{O}(1/r)$ part of the $ri$ bulk equation, namely equation (\ref{sing ri}), becomes
\begin{equation}
\beta_{,i}=0\,,
\label{einstein sing ri}
\end{equation}
while the $\mathcal{O}(1/r)$ part of the $rr$ bulk equation, i.e. equation (\ref{sing rr}), is
\footnote{In \cite{BCL}, a self-gravitating string in Einstein gravity was locally described by a thin tube of matter
represented by a smoothed conical metric, and under some constraint on the model of the string in the limit
where the thickness becomes negligible the central line of the string was shown to follow the Nambu-Goto
dynamics.}
\begin{equation}
K=0\,.
\label{einstein sing rr}
\end{equation}
The $\mathcal{O}(1/r)$ part of the $ij$ equations, using the expression of $\mathcal{G}_{ij}$
of Appendix \ref{geometric components}, takes the form
\begin{equation}
K_{ij}-Kh_{ij}-\frac{\beta_{2}}{\beta}h_{ij}=0\,.
\label{einstein sing ij}
\end{equation}

The first matching condition (\ref{einstein matching1}), using (\ref{einstein sing rr}), takes the form
\begin{equation}
\Big(\frac{r_{c}^{2}}{\kappa_{6}^{2}}G^{ij}-T^{ij}\Big)K_{ij}=0
\label{einstein simple matching1}
\end{equation}
and coincides with (\ref{simpler matching1}) if indeed it is set $\eta=1$ in the constants (\ref{constants}).
\newline
\noindent
The second matching condition (\ref{einstein matching2}), using (\ref{einstein sing ri}), becomes
\begin{equation}
T^{ij}_{\,\,\,\,\,|j}=0\,.
\label{einstein simple matching2}
\end{equation}
From equations (\ref{einstein sing rr}), (\ref{einstein sing ij}) we get
\begin{equation}
K_{ij}=\frac{\beta_{2}}{\beta}h_{ij}\,.
\label{einstein simple sing ij}
\end{equation}
Contracting (\ref{einstein simple sing ij}) with $h^{ij}$ and using again (\ref{einstein sing rr})
we take
\begin{equation}
\beta_{2}=0\,\,\,\,\,,\,\,\,\,\,K_{ij}=0\,.
\label{kati}
\end{equation}
Therefore, the algebraic matching condition (\ref{einstein simple matching1}) or (\ref{einstein matching1})
is trivially satisfied and what remains is $\beta$=\,constant, the conservation equation
(\ref{einstein simple matching2}) and equations (\ref{kati}). The vanishing of the total extrinsic curvature
means that in the Einstein limit the brane is a special case of Nambu-Goto, it is geodesic, i.e.
$x^{\mu}_{\,\,\,;ij}\!+\!\Gamma^{\mu}_{\,\,\,\nu\lambda}x^{\nu}_{\,\,\,,i}x^{\lambda}_{\,\,\,,j}=0$. Of course, this
happens ``on-shell'', it is the result of all the equations, not only the matching conditions. On the contrary,
in the EGB theory we examined, in general the brane even ``on-shell'' is not geodesic. It would be interesting
to see if in Einstein theory this geodesic result is relaxed when the ansatz of axial symmetry is abandoned.
If this is the case, the codimension-2 Einstein gravity will be not only consistent, but also non-trivial.
Note that the probe limit of a theory is a different thing and is checked ``off-shell'', from the matching
conditions only, since in the probe limit there is no bulk dynamics and the bulk equations are empty.

Continuing with the remaining equations, the regular part of the $rr$ bulk equation is
\begin{equation}
2K'+K^{2}-K_{ij}K^{ij}-R+2\frac{\Box{\beta}}{\beta}+K\frac{\beta_{2}}{\beta}=
2\kappa_{6}^{2}\mathcal{T}^{r}_{r}-2\Lambda_{6}\,,
\label{einstein  regular rr}
\end{equation}
where we denote $K_{ij}'=\mathcal{K}_{ij}'(r=0)$ and equation (\ref{einstein regular rr}), due to
(\ref{einstein sing ri}), (\ref{kati}), becomes
\begin{equation}
R-2K'=2\Lambda_{6}-2\kappa_{6}^{2}\mathcal{T}^{r}_{r}\,.
\label{einstein simple  regular rr}
\end{equation}
Similarly, the regular part of the $ri$ bulk equation is
\begin{equation}
K^{j}_{i|j}-K_{|i}+\frac{\beta_{,j}}{\beta}\Big(\frac{\beta_{2}}{2\beta}\delta^{j}_{i}+K^{j}_{i}\Big)
-\frac{\beta_{2,i}}{\beta}=\kappa_{6}^{2}\mathcal{T}_{i}^{r}\,,
\label{einstein  regular ri}
\end{equation}
which due to (\ref{einstein sing ri}), (\ref{kati}) becomes identically satisfied  whenever
$\mathcal{T}^{r}_{i}=0$ on the brane.
\newline
\noindent
Concerning the $\theta\theta$ bulk equation one obtains
\begin{equation}
2K'+K_{ij}K^{ij}+K^{2}-R=2\kappa_{6}^{2}\mathcal{T}^{\theta}_{\theta}-2\Lambda_{6}\,,
\label{einstein thouthou}
\end{equation}
which due to (\ref{kati}) becomes
\begin{equation}
R-2K'=2\Lambda_{6}-2\kappa_{6}^{2}\mathcal{T}^{\theta}_{\theta}\,.
\label{einstein simple  thouthou}
\end{equation}
Equation (\ref{einstein simple  regular rr}) is compatible with equation (\ref{einstein simple  thouthou})
whenever $\mathcal{T}^{r}_{r}=\mathcal{T}^{\theta}_{\theta}$ on the brane. Therefore, all the equations
of Einstein gravity on the brane up to now are $\beta$\,=\,constant, $\beta_{2}=0$, $T^{ij}_{\,\,\,\,\,|j}=0$,
$K_{ij}=0$, $R-2K'=2\Lambda_{6}-2\kappa_{6}^{2}\mathcal{T}^{r}_{r}$ and what remains is the regular part of the
$ij$ bulk equations.

The regular part of the $ij$ bulk equations is
\begin{equation}
G_{ij}+2K_{ik}K^{k}_{j}+KK_{ij}-2K_{ij}'+\frac{3\beta_{2}}{2\beta}K_{ij}-\frac{\beta_{|i|j}}{\beta}
+\Big(2K'+\frac{1}{2}K_{k\ell}K^{k\ell}+\frac{1}{2}K^{2}+\frac{\Box{\beta}}{\beta}
+K\frac{\beta_{2}}{2\beta}-\frac{\beta_{2}^{2}}
{2\beta^{2}}+\frac{\beta_{3}}{\beta}\Big)h_{ij}=\kappa_{6}^{2}\mathcal{T}_{ij}-\Lambda_{6}h_{ij}\,,
\label{laptop}
\end{equation}
where $\beta_{3}(\chi)=L'''(\chi,0)$. Equation (\ref{laptop}), due to (\ref{einstein sing ri}), (\ref{kati}),
becomes
\begin{equation}
G_{ij}-2K_{ij}'+\Big(2K'+\frac{\beta_{3}}{\beta}\Big)h_{ij}=\kappa_{6}^{2}\mathcal{T}_{ij}-\Lambda_{6}h_{ij}\,.
\label{simple laptop}
\end{equation}
Contracting (\ref{simple laptop}) with $h^{ij}$ we find $\beta_{3}$
\begin{equation}
\frac{4\beta_{3}}{\beta}=R-6K'-4\Lambda_{6}+\kappa_{6}^{2}\mathcal{T}_{ij}h^{ij}\,,
\label{beta3}
\end{equation}
and substituting back in (\ref{simple laptop}) we get
\begin{equation}
G_{ij}-2K_{ij}'+\frac{1}{4}(R+2K')h_{ij}=\kappa_{6}^{2}\mathcal{T}_{k\ell}\Big(\delta^{k}_{i}\delta^{\ell}_{j}
-\frac{1}{4}h^{k\ell}h_{ij}\Big).
\label{car}
\end{equation}
Of course, not all $ij$ equations of (\ref{car}) are independent, but independent are all but one.
The non-trivial final equations to be satisfied are (\ref{einstein simple regular rr}) and (\ref{car}).
For cosmology, equation (\ref{einstein simple regular rr}) gets the form
\begin{equation}
3A'+N'-3(X+Y)=\kappa_{6}^{2}\mathcal{T}^{r}_{r}-\Lambda_{6}
\label{gata}
\end{equation}
and coincides with equation (\ref{theta regular diff}) for $\alpha=0$.  Equation (\ref{car}) for cosmology
contains one independent equation which is
\begin{equation}
A'-N'+Y-X=\frac{\kappa_{6}^{2}}{6}\big(3\mathcal{T}^{t}_{t}-\mathcal{T}^{\hat{i}}_{\hat{i}}\big)\,.
\label{dog}
\end{equation}
Therefore, the two equations (\ref{gata}), (\ref{dog}) can be solved algebraically for $A',N'$ and
$X,Y$ remain undetermined, thus, the scale factor remains undetermined. The energy density obeys the
standard conservation.

The indeterminacy from the brane viewpoint of one unknown function for cosmology is the result of the
codimension-2 geometry. Certainly here, in Einstein gravity, the fact that the scale factor itself remains
undetermined is more inconvenient compared to the indeterminacy of the general EGB cosmology which can be
rendered to the deficit angle. But still the important thing is the consistency of Einstein gravity in the
present formulation, in clear contrast to the inconsistency of Einstein gravity according to the standard
treatment \cite{Israel 1977}, \cite{Geroch}, \cite{Garfinkle}, \cite{Frolov}, \cite{cline}.
Our main purpose is to raise the interest to the investigation of more realistic, alternative matching conditions,
not to give an answer on the selection of the appropriate bulk boundary conditions, or the appropriate
codimensionality, or the appropriate gravitational theory in order to pick up the unique final cosmology.

\subsection{Comparison with the cosmology of the standard approach} \label{comparison}

It was explained thoroughly in the Introduction that according to the standard approach the equations
of motion of a defect are derived by taking the variation of the brane-bulk action with respect to the bulk
metric at the brane position. This is the extension of what is done with the Israel matching conditions
and for the codimension-2 Einstein-Gauss-Bonnet theory this was performed in \cite{Ruth 2004}, \cite{CKP},
\cite{Charmousis}. The aim here is to compare the cosmological equations of the standard
treatment discussed in \cite{CKP} with the cosmological equations (\ref{1})-(\ref{3}) of the present analysis.

In \cite{CKP}, the cases (i), (iii) were examined.
Together with the matching conditions originating from the distributional terms
$\delta(r)/r$, additional matching conditions were considered arising from the distributional terms
$\delta(r)$ (also considered in \cite{Soda}). For the case (i), these extra matching conditions are identically
satisfied. However, for the case (iii) these conditions provide non-trivial
constraints. The result of the analysis in \cite{CKP} for cosmology was that while the case (i) is
consistent, the case (iii) is not consistent with a codimension-two brane, but an additional codimension-one
brane is needed. However, this result is not correct because the distributional terms $\delta(r)$ should not be
considered for deriving extra matching conditions. This becomes obvious from the variational point of view where
the volume element $rdrd\theta$ multiplies
$\delta(r)$ and vanishes it. Otherwise, looking at an equation containing two sorts of distributions it cannot be
said if the correct thing is to leave the equation as it is getting two sets of matching conditions,
or to multiply the equation with $r$ getting one set of matching conditions, or why not to multiply the
equation with $r^{2}$ getting no matching conditions at all. To say it in a similar way, it is not obvious
what is the correct regularization and the answer is provided by the variational principle which
naturally supplies the correct regularization. Therefore, the correct result is that both cases (i), (iii)
are consistent with a codimension-2 brane in EGB theory according to the standard approach. We can add here that
case (ii) is also consistent since mathematically it does not differ significantly from cases (i), (iii).
In the following, the essential cosmological equations of the standard approach are summarized in order to compare
with the current treatment. These equations for the matching conditions (i), (ii), (iii) are
\begin{eqnarray}
&&\dot{\rho}+3nH(\rho\!+\!p)=\eta\frac{\dot{\beta}}{\beta(\eta\!-\!\beta)}
\Big(3\frac{r_{c}^{2}}{\kappa_{6}^{2}} X\!-\!\rho\!-\!\lambda\Big) \label{energyexchangestandard}\\
&&A^{2}=\frac{1}{\eta\!-\!\beta}\Big(1\!-\!\beta\!+\!\frac{r_{c}^{2}}{8\pi
\alpha}\Big)X-\frac{\kappa_{6}^{2}(\rho\!+\!\lambda)}{24\pi
\alpha (\eta\!-\!\beta)}+\frac{1\!-\!\beta}{12\alpha(\eta\!-\!\beta)}\label{roa}\\
&&fA^{2}=\frac{1}{\eta\!-\!\beta}
\Big(1\!-\!\beta\!+\!\frac{r_{c}^{2}}{8\pi\alpha}\Big)Y
+\frac{\kappa_{6}^{2}(\rho\!+\!3p\!-\!2\lambda)}{48\pi\alpha(\eta\!-\!\beta)}
+\frac{1\!-\!\beta}{12\alpha(\eta\!-\!\beta)} \label{rua}\\
&&2A^{2}\Big[\frac{\dot{A}}{nA}\!+\!H(1-f)\Big]
=\frac{\dot{\beta}}{n\beta}\Big(\!X\!-\!A^{2}\!+\!\frac{1}{12\alpha}\!\Big)\label{moka}\,,
\end{eqnarray}
where, also here, $N$ has been replaced by $N=fA$. The definitions for $X,Y,f$ are the same with those
in equations (\ref{5}), (\ref{6}).

For the matching condition (iv), equations (\ref{roa}), (\ref{rua})
are replaced by the cosmological version of the four-dimensional Einstein equations
$r_{c}^{2}G_{ij}\!-\!\kappa_{6}^{2}(T_{ij}\!-\!\lambda h_{ij})=0$, therefore the strict conservation
equation replaces (\ref{energyexchangestandard}). Additionally, equations (\ref{sing rr cosmo}),
(\ref{sing ri cosmo}) are valid with zero right-hand side. Solving (\ref{sing rr cosmo}) for $N/A$
(now we do not have the equation $N=fA$) and substituting in (\ref{sing ri cosmo}), there arises
a differential equation for $A$. Although we do not plan here to study exhaustively the second
order equations to check the consistency, however, it seems most probable that matching conditions
(iv) will still be consistent in the standard treatment.

Equations (\ref{energyexchangestandard})-(\ref{moka}) form a system of four equations for four unknowns
$a,\rho, \beta, A$, however, one equation, for example the matching condition
(\ref{rua}) is redundant, so there is again one indeterminacy. Indeed, differentiating the matching condition
(\ref{roa}) with respect to time and using (\ref{energyexchangestandard}), (\ref{moka}), equation (\ref{rua})
is obtained. Although redundant, this equation is left inside the set of equations because it will be useful
in the following. The conservation equation (\ref{energyexchangestandard}) is the corresponding or (\ref{1}),
but the two right hand sides differ significantly. However, for $\eta=0$ or $\beta$\,=\,constant they both
reduce to the standard conservation equation. Equation (\ref{moka}) is identical to equation (\ref{3}). The
crucial difference between the two theories is the matching condition (\ref{roa}) compared to equation (\ref{2}),
which is a sort of equivalent to the new, algebraic in extrinsic curvature, matching condition (\ref{matching1}).
While equation (\ref{roa}) contains only $H$, equation (\ref{2}) contains also $\dot{H}$. In some sense it can
be said that in the standard approach the matching condition is already integrated, while on the contrary, the
present theory is more complicated and accepts more general solutions with more integration constants.

For example, to be more explicit, for $\beta$\,=\,constant and $\sigma_{1}=0$, equation (\ref{moka}) can be
integrated for $A$. Then, equation (\ref{roa}) gives directly the Hubble parameter in terms of one integration
constant (we do not take into account the additional integration constant for $\rho$ from
(\ref{energyexchangestandard})) and this solution was obtained in \cite{CKP}
\begin{equation}
H^{2}+\frac{k}{a^{2}}=\frac{\kappa_{6}^{2}\,\rho}{24\pi\alpha(1\!-\!\beta)}
+\Big(\frac{\kappa_{6}^{2}\,\lambda}{24\pi\alpha(1\!-\!\beta)}-\frac{1}{12\alpha}\Big)
+\frac{\tilde{c}}{(\rho+\lambda)^{2}}\,\rho^{\frac{8}{3(1+w)}}\,.
\label{old hip hop}
\end{equation}
On the other hand, this same solution for $A$ when substituted in (\ref{2}), a Raychaudhuri equation for
$\dot{H}$ is obtained. Its integration gives $H$ in terms of two integration constants and this is the
solution (\ref{hip hop}) obtained above. Maybe in this simplified case the two solutions (\ref{old hip hop}),
(\ref{hip hop}) look similar, but the main difference has already been pointed out.

In the general case of non-constant $\beta$ and any $\sigma_{1}$, for the standard treatment, the Raychaudhuri
equation can be easily derived by combining the two algebraic in $A$ matching conditions (\ref{roa}), (\ref{rua})
\begin{equation}
\Big(1\!-\!\beta\!+\!\frac{r_{c}^{2}}{8\pi\alpha}\Big)(Y\!-\!fX)+\frac{\kappa_{6}^{2}}{48\pi\alpha}
\big[(1\!+\!2f)\rho+3p\big]+\frac{1\!-\!f}{12\alpha}\Big(1\!-\!\beta\!-\!\frac{\kappa_{6}^{2}\lambda}{2\pi}\Big)
=0\,.\label{england}
\end{equation}
The analogue of equation (\ref{england}) in the alternative approach is equation (\ref{ray}).
The difference is not only that (\ref{ray}) already contains one extra integration constant, but the whole
structure of the two equations is quite different. To say it in a different way, in the alternative approach,
there is only one equation (\ref{2}) algebraic in $A$ (also containing $H,\dot{H}$), instead of the two equations
(\ref{roa}), (\ref{rua}) algebraic in $A$ in the standard approach. Differentiating this equation with respect
to time and using (\ref{3}) to get rid off $\dot{A}$, we could obtain an equation for $\ddot{H}$. Again, the
difference with (\ref{england}) is obvious. The successful treatment performed in section \ref{cosmo evolution}
managed to derive equation (\ref{ray}) which contains only $\dot{H}$, but with the cost of one integration
constant $c$. Equation (\ref{england}) together with the conservation equation (\ref{energyexchangestandard})
constitute for the standard approach the two-dimensional system for $a,\rho$ with the indeterminacy of $\beta(t)$.
It would be interesting, for example, to be integrated for $\beta$\,=\,constant and $\sigma_{1}\neq 0$.

\section{Conclusions} \label{Conclusions}

In this paper, we take up again the question of the dynamics of a self-gravitating brane. We restrict
ourselves to the case where the bulk metric is regular on the brane, as e.g. happens in the braneworld
scenario. While an equation of the general form
bulk gravity tensor equals some smooth matter content or some matter content of a ``thick'' brane is
certainly correct, we claim that it cannot be correct in the shrink limit of distributional branes.
A different treatment of the delta function characterizing the defect is needed for extracting its equation
of motion. If this is so, the Israel matching conditions, as well as their generalizations where the Einstein
bulk gravity tensor is replaced by Lovelock extensions and the branes have differing codimensions, cannot be
adequate.

Our reasoning is based on two points: First, the incapability of the conventional matching conditions to
accept the Nambu-Goto probe limit. Even the geodesic limit of the Israel matching conditions is not an
acceptable probe limit since being the geodesic equation a kinematical fact it should be preserved independent
of the gravitational theory or the codimension of the defect, which however is not the case for these matching conditions.
Second, in the $D$-dimensional spacetime we live (maybe $D=4$), classical defects of any possible codimension
could in principle be constructed (even in the lab), and therefore, they should be compatible. The standard
matching conditions fail to accept codimension-2 and 3 defects for $D=4$ (which represents effectively the
spacetime at certain length and energy scales) and most probably fail to accept high enough codimensional
defects for any $D$ since there is no corresponding high enough Lovelock density to support them.

We make a proposal that the problem is not the distributional character of the defects, neither the
gravitational theory used, but the equations of motion of the defects.
The proposed matching conditions might move towards the correct direction of finding realistic matching
conditions since they always have the Nambu-Goto probe limit, independently of the gravity theory and
independently of the dimension of spacetime or codimension of the brane. Moreover, with these
matching conditions, defects of any codimension seem to be consistent for any (second order) gravity theory.
These alternative matching conditions arise by promoting the embedding fields of the defect to the
fundamental entities. Instead of varying the brane-bulk action with respect to the bulk metric at the brane
position and derive the standard matching conditions, we vary with respect to the brane embedding fields
in a way that takes into account the gravitational back-reaction of the brane to the bulk.

In the present paper we have considered in detail the case of a 3-brane in six-dimensional Einstein-Gauss-Bonnet
gravity, derived the generic alternative matching conditions and proved the consistency for an axially
symmetric cosmological configuration. Of course, same or similar results are true for other codimension-2 defects
in other spacetime dimensions. The consistency of such branes in Einstein gravity was also discussed.
Additionally, however, a 3-brane could represent our world in the braneworld scenario, therefore, we have investigated
the cosmological equations and found solutions for the cosmic evolution. From the technical point of view,
compared to the standard equations, the main difference is that the equations here, and accordingly
their solutions, have more richness, more complicated structure and contain more integration
constants. In all the cosmologies found assuming a constant deficit angle, there is the standard FRW term linear
in energy density, a cosmological constant term and an extra correction/dark energy term. In particular,
one of these solutions for a radiation brane and for a range of the integration constants avoids a
cosmological singularity (both in density and curvature) and undergoes accelerated expansion near the
minimum scale factor.

Depending on the existence or not of a conical singularity or of a discontinuous extrinsic curvature,
there can be four possible cases as matching conditions. Actually, one of these cases is the ``smooth''
matching condition with smooth transverse section (no cone) and smooth extrinsic tangential section, which
is still consistent and possesses interesting solutions.
In general, for a codimension-2 cosmological configuration, either here or in the standard treatment, the system
of the effective equations from the brane viewpoint is non-closed and we need extra information coming from the bulk geometry in
order to fix one of the functions. On the theoretical side it would be interesting to have particular bulk
solutions setting the boundary conditions and fixing the deficit angle, whose evolution would then leave
an imprint on the cosmological evolution equations. For a pure conical brane such a dynamical deficit angle
will make the brane to radiate in the bulk.

Codimension-2 braneworlds in six-dimensional gravity or supergravity have attracted considerable interest in
relation to the cosmological constant problem (for a review see \cite{burg}). They are proposed to offer a
mechanism for understanding the smallness of the vacuum energy since in this scenario, a codimension-2 object
induces a conical singularity, and the cancelation occurring between the brane tension $\lambda$ and the bulk
gravitational degrees of freedom gives rise to a vanishing effective cosmological constant. In the
approach of the present paper there is no such a relation between the brane tension and the conical deficit in
order to obtain a Minkowski brane, but a Minkowski brane is obtained by a physically similar balance between the
gravitational energy stored at the brane and the deficit angle. However, furthermore, in the presence of the induced
gravity term, the proposed formalism embodies naturally the effective cosmological constant scale
$\kappa_{6}^{2}\lambda/r_{c}^{2}$, which for $\lambda\sim M_{6}^{4}$ (e.g. $\text{TeV}^{4}$) and
$r_{c}\sim H_{0}^{-1}$ gives the observed value $H_{0}^{2}$ of the cosmological constant. The corresponding
scale in any other bulk dimension with $\lambda\sim M_{D}^{4}$ fails to provide the observed order of magnitude
of the cosmological constant. Even if the constraint provided by the four-dimensional Newton's constant $G_{\!N}$
is not easily satisfied, there is still hope, since in the present theory $G_{\!N}$ is not determined solely
from the parameters of the action, but in general depends also on the integration constants of the considered
solution.

\[ \]
{{\bf Acknowlegements}} M.I. is supported by START project Y 435-N16 (FWF).

\appendix

\section{Geometric Components} \label{geometric components}

The non-vanishing components of the necessary geometric quantities of the metric (\ref{ansatz}) are
\begin{eqnarray}
\Gamma^{r}_{\theta \theta}\!=\!-LL^{\prime}\,\,,\,\,
\Gamma^{r}_{ij}\!=\!-\mathcal{K}_{ij}\,\,,\,\,
\Gamma^{\theta}_{r \theta}\!=\!\frac{L^{\prime}}{L}\,\,,\,\,
\Gamma^{\theta}_{\theta i}\!=\!\frac{L_{|i}}{L}\,\,,\,\,
\Gamma^{i}_{\theta\theta}\!=\!-Lg^{ij}L_{|j}\,\,,\,\,
\Gamma^{i}_{rj}\!=\!\mathcal{K}^{i}_{j}\,\,,\,\,
\Gamma^{i}_{jk}\!=\!\frac{1}{2}g^{i\ell}(g_{\ell j,k}\!+\!g_{\ell k,j}\!-\!g_{jk,\ell})
\label{non-vanishing Christoffels}
\end{eqnarray}
(where $\mathcal{K}^{i}_{j}=g^{ik}\mathcal{K}_{kj}$)
\begin{eqnarray}
&&\mathcal{R}_{rirj}=-\mathcal{K}'_{ij}+\mathcal{K}_{ik}\mathcal{K}^{k}_{j}\,\,\,,\,\,\,
\mathcal{R}_{\theta i \theta j}=-L (L_{| i | j}+L' \mathcal{K}_{ij})\,\,\,,\,\,\,
\mathcal{R}_{r \theta r \theta}=-LL''\,\,\,,\,\,\,
\mathcal{R}_{ijk\ell}=R_{ijk\ell}+\mathcal{K}_{i\ell}\mathcal{K}_{jk}-\mathcal{K}_{ik}\mathcal{K}_{j\ell}\nn\\
&&\mathcal{R}_{\theta r \theta i}=L (L_{| j}\mathcal{K}^{j}_{i}-L'_{\,| i})\,\,\,,\,\,\,
\mathcal{R}_{rijk}=\mathcal{K}_{ij | k}-\mathcal{K}_{ik | j}
\label{non-vanishing Riemanns}
\end{eqnarray}
(where $|$ denotes the covariant derivative with respect to the metric $g_{ij}$)
\begin{eqnarray}
&&\mathcal{R}_{rr} = - \frac{L''}{L}-\mathcal{K}'-\mathcal{K}_{ij}\mathcal{K}^{ij}\,\,\,,\,\,\,
\mathcal{R}_{\theta\theta}= - L(L''+\square L+L'\mathcal{K} )\,\,\,,\,\,\,
\mathcal{R}_{ri} =\frac{L_{| j}}{L} \mathcal{K}^{j}_{i}-\frac{L'_{\,| i}}{L}
+\mathcal{K}^{j}_{i | j}-\mathcal{K}_{| i}\nn\\
&&\mathcal{R}_{ij}=R_{ij}-\mathcal{K}'_{ij}+2\mathcal{K}_{ik}\mathcal{K}_{j}^{k}
-\mathcal{K}\mathcal{K}_{ij}-\frac{L_{| i | j}}{L}-\frac{L'}{L}\mathcal{K}_{ij}
\label{non-vanishing Riccis}
\end{eqnarray}
(where $\mathcal{K}=\mathcal{K}_{i}^{i}$ and $\square$ is the Laplacian operator of the metric $g_{ij}$)
\begin{equation}
\mathcal{R}=R-2 \frac{L''}{L}-2\mathcal{K}'-\mathcal{K}_{ij}\mathcal{K}^{ij}
-\mathcal{K}^{2}-2\frac{\square L}{L}-2\frac{L'}{L}\mathcal{K}
\label{Ricci scalar}
\end{equation}
\begin{eqnarray}
&&\!\!\!\!\!\!\!\!\!\!\!\!\!\!
\mathcal{G}_{ri}=\frac{L_{| j}}{L}\mathcal{K}^{j}_{i}-\frac{L'_{\,|i}}{L}+\mathcal{K}^{j}_{i | j}
-\mathcal{K}_{| i}\,\,\,\,,\,\,\,\,
\mathcal{G}_{rr}=\frac{1}{2} \mathcal{K}^{2}-\frac{1}{2} \mathcal{K}_{ij}\mathcal{K}^{ij}
+\frac{\square L}{L}+\frac{L'}{L} \mathcal{K}-\frac{1}{2} R
\nn\\
&&\!\!\!\!\!\!\!\!\!\!\!\!\!\!
\mathcal{G}_{\theta\theta}=L^{2}\Big(\mathcal{K}'+\frac{1}{2}\mathcal{K}_{ij}\mathcal{K}^{ij}
+\frac{1}{2}\mathcal{K}^{2}-\frac{1}{2}R\Big)\nn\\
&&\!\!\!\!\!\!\!\!\!\!\!\!\!\!
\mathcal{G}_{ij}=G_{ij}-\mathcal{K}'_{ij}+2\mathcal{K}_{ik}\mathcal{K}_{j}^{k}
-\mathcal{K}\mathcal{K}_{ij}-\frac{L_{| i | j}}{L}-\frac{L'}{L}\mathcal{K}_{ij}
+\Big(\frac{L''}{L}+\mathcal{K}'+\frac{1}{2}\mathcal{K}_{k\ell}\mathcal{K}^{k\ell}+\frac{1}{2}\mathcal{K}^{2}
+\frac{L'}{L}\mathcal{K}+\frac{\square L}{L}\Big)g_{ij}\,.
\label{non-vanishing Einsteins}
\end{eqnarray}

\section{Derivation of regular equations} \label{regular equations}

Using the results of Appendix \ref{geometric components}, we can find analogous but more complicated
expressions for the tensor components of
$\mathcal{E}^{\mu}_{\nu}=\mathcal{G}^{\mu}_{\nu}+\alpha\mathcal{J}^{\mu}_{\nu}$.
In some of these components there are terms proportional to $\frac{L'}{L},\frac{L'_{|i}}{L},\frac{L''}{L}$.
All the other terms of $\mathcal{E}^{\mu}_{\nu}$ are manifestly regular $\mathcal{O}(1)$
terms and can be obtained by setting formally $\frac{L'}{L}=\frac{L'_{|i}}{L}=\frac{L''}{L}=0$,
$L=\beta$ in the $\mathcal{E}^{\mu}_{\nu}$
expressions. Now, the terms containing $\frac{L'}{L},\frac{L'_{|i}}{L},\frac{L''}{L}$ are certainly the
sources of the singular $\mathcal{O}(1/r)$ terms. However, these same terms, when expanded in powers of $r$,
give additional hidden regular $\mathcal{O}(1)$ terms. In this Appendix we derive the regular $rr,ri$ bulk
equations on the brane (\ref{rr regular}), (\ref{rt regular}).

More precisely, in the $rr$ equation there are terms with $\frac{L'}{L}$, while in the $ri$ equation there are
terms with $\frac{L'}{L},\frac{L'_{|i}}{L}$. A typical expansion of these terms is of the form
\begin{eqnarray}
&&\frac{L'}{L}f(\chi,r)=f\frac{1}{r}+f'+\frac{f\beta_{2}}{2\beta}+\mathcal{O}(r)\\
&&\frac{L'_{|i}}{L}f(\chi,r)=f\frac{\beta_{,i}}{\beta}\frac{1}{r}+f'\frac{\beta_{,i}}{\beta}
+f\Big(\frac{\beta_{2,i}}{\beta}-\frac{\beta_{,i}}{\beta}\frac{\beta_{2}}{2\beta}\Big)+\mathcal{O}(r)
\label{twra}\\
&&\frac{L''}{L}f(\chi,r)=f\frac{\beta_{2}}{\beta}\frac{1}{r}+\frac{\beta_{2}}{\beta}
\Big(f'-f\frac{\beta_{2}}{2\beta}\Big)+f\frac{\beta_{3}}{\beta}+\mathcal{O}(r)\,,\label{meta}
\end{eqnarray}
where $f,f'$ are the values on the brane and $\beta_{2}(\chi)=L''(\chi,0)$, $\beta_{3}(\chi)=L'''(\chi,0)$.

In the quantity $\mathcal{E}^{r}_{r}$ the term which multiplies $\frac{L'}{L}$ is
$f=\mathcal{K}\!-\!4\alpha\big(\mathcal{W}^{i}_{j}+G^{i}_{j}\big)\mathcal{K}^{j}_{i}$, i.e. at the brane
position it is
$\mathcal{E}^{r}_{r}=\mathcal{E}^{r}_{r}\big{|}_{\frac{L'}{L}=0,L=\beta}+\frac{L'}{L}f$.
Therefore, using the identity
$\mathcal{W}_{j}^{i\,\prime}\mathcal{K}^{j}_{i}=2\mathcal{W}^{i}_{j}\mathcal{K}_{i}^{j\,\prime}$,
we get the $\mathcal{O}(1)$ part of the $rr$ equation
\begin{equation}
\mathcal{E}^{r}_{r}\big{|}_{\frac{L'}{L}=0,L=\beta}+K'-4\alpha(3W^{i}_{j}+G^{i}_{j})K_{i}^{j\,\prime}
-4\alpha G_{j}^{i\,\prime} K^{j}_{i}+\big[K-4\alpha\big(W^{i}_{j}+G^{i}_{j}\big)K^{j}_{i}\big]\frac{\beta_{2}}
{2\beta}=\kappa_{6}^{2}\mathcal{T}^{r}_{r}-\Lambda_{6}\,,
\label{nero}
\end{equation}
where $K_{i}^{j\,\prime}$ denotes $\mathcal{K}_{i}^{j\,\prime}(\chi,0)$. However, from the $\mathcal{O}(1/r)$
part of the $rr$ equation
(\ref{sing rr}) it is $f=0$ on the brane and (\ref{nero}) takes a simpler form
\begin{equation}
\mathcal{E}^{r}_{r}\big{|}_{\frac{L'}{L}=0,L=\beta}+K'-4\alpha(3W^{i}_{j}+G^{i}_{j})K_{i}^{j\,\prime}
-4\alpha G_{j}^{i\,\prime} K^{j}_{i}=\kappa_{6}^{2}\mathcal{T}^{r}_{r}-\Lambda_{6}\,.
\label{mallia rr}
\end{equation}
\newline
For cosmology it is
\begin{equation}
\mathcal{E}^{r}_{r}\big{|}_{\frac{L'}{L}=0,L=\beta}=-12\alpha\Big[\frac{1}{n\beta}
\Big(\frac{\dot{\beta}}{n}\Big)^{^{\!\centerdot}}\mathcal{X}+\frac{H\dot{\beta}}{n\beta}
\big(\mathcal{X}+2\mathcal{Y}\big)
+\mathcal{X}\mathcal{Y}+\frac{1}{6\alpha}\Big(\mathcal{X}\!+\!\mathcal{Y}\!-\!\frac{5}{24\alpha}\Big)\Big],
\label{mallia cosmo}
\end{equation}
so equation (\ref{nero}) is
\begin{eqnarray}
&&\!\!\!\!\!\!\!\!\!\!\!\!\!\!\!\!\!\!\!\!\!\!\!\!\!\!
\frac{1}{n\beta}\Big(\frac{\dot{\beta}}{n}\Big)^{^{\!\centerdot}}\mathcal{X}+\frac{H\dot{\beta}}{n\beta}
(\mathcal{X}\!+\!2\mathcal{Y})+\mathcal{X}\mathcal{Y}+\frac{1}{6\alpha}(\mathcal{X}\!+\!\mathcal{Y})
+2A^{2}\Big[\Big(1\!+\!2\frac{N}{A}\Big)A'+N'\Big]-\big[(\mathcal{X}\!+\!2\mathcal{Y})A'+\mathcal{X}N'\big]\nn\\
&&\,\,\,\,\,\,\,\,\,\,\,\,\,\,\,\,\,\,\,\,\,\,\,\,\,\,\,
-\big[(A\!+\!N)X'+2AY'\big]-A\Big[\Big(1\!+\!\frac{N}{A}\Big)\mathcal{X}+2\mathcal{Y}\Big]
\frac{\beta_{2}}{2\beta}=\frac{1}{12\alpha}
\Big(\Lambda_{6}+\frac{5}{12\alpha}-\kappa_{6}^{2}\mathcal{T}^{r}_{r}\Big)\,,
\label{lamia}
\end{eqnarray}
while its simpler form (\ref{mallia rr}) is
\begin{eqnarray}
&&\!\!\!\!\!\!\!\!\!\!\!\!\!\!\!\!\!\!\!\!\!\!\!\!\!\!\!\!\!
\frac{1}{n\beta}\Big(\frac{\dot{\beta}}{n}\Big)^{^{\!\centerdot}}\mathcal{X}+\frac{H\dot{\beta}}{n\beta}
(\mathcal{X}\!+\!2\mathcal{Y})+\mathcal{X}\mathcal{Y}+\frac{1}{6\alpha}(\mathcal{X}\!+\!\mathcal{Y})
+2A^{2}\Big[\Big(1\!+\!2\frac{N}{A}\Big)A'+N'\Big]-\big[(\mathcal{X}\!+\!2\mathcal{Y})A'+\mathcal{X}N'\big]\nn\\
&&\,\,\,\,\,\,\,\,\,\,\,\,\,\,\,\,\,\,\,\,\,\,\,\,\,\,\,\,\,\,\,\,\,\,\,\,\,\,\,\,\,\,\,\,\,\,\,
\,\,\,\,\,\,\,\,\,\,\,\,\,\,\,\,\,\,\,\,\,\,\,\,\,\,\,\,\,\,\,\,\,\,\,\,\,\,\,\,
-\big[(A\!+\!N)X'+2AY'\big]=\frac{1}{12\alpha}
\Big(\Lambda_{6}+\frac{5}{12\alpha}-\kappa_{6}^{2}\mathcal{T}^{r}_{r}\Big)\,.
\label{smyrni}
\end{eqnarray}
Using (\ref{sing ri cosmo}) it is easy to find $H'=\frac{\dot{\beta}}{2n\beta A}\mathcal{X}-AH$, and then
progressively the quantities $X',\dot{\mathcal{X}},\dot{H}'=(H')^{^{\centerdot}},Y'$. Replacing $X',Y'$ in
(\ref{smyrni}) and using (\ref{N}) we take the equivalent to the regular $rr$ equation
\begin{eqnarray}
&&\!\!\!\!\!\!\!\!\!\!
\frac{\dot{\beta}^{2}}{n^{2}\beta^{2}}\Big(2\mathcal{X}+\frac{\mathcal{X}^{2}}{2A^{2}}\Big)
+\mathcal{X}\mathcal{Y}+\frac{1}{6\alpha}(\mathcal{X}\!+\!\mathcal{Y})+2(1\!+\!f)A^{2}(X\!+\!Y)
+2A^{2}\big[(1\!+\!2f)A'+N'\big]-\big[(\mathcal{X}\!+\!2\mathcal{Y})A'+\mathcal{X}N'\big]\nn\\
&&\quad\quad\quad\quad\quad\quad\quad\quad\quad\quad\quad\quad\quad\quad\quad\quad\quad
\quad\quad\quad\quad\quad\quad\quad\quad\quad\quad\quad\quad\quad\quad
=\frac{1}{12\alpha}
\Big(\Lambda_{6}+\frac{5}{12\alpha}-\kappa_{6}^{2}\mathcal{T}^{r}_{r}\Big)\,.
\label{limani}
\end{eqnarray}

For the $rt$ regular equation, due to the complication, we give directly the cosmological expressions, while the
$r\hat{i}$ equation vanishes identically. In the quantity $\mathcal{E}^{t}_{r}$
the term which multiplies $\frac{L'}{L}$ is
$\tilde{f}=-24\alpha \frac{A}{n}\big[\frac{\dot{A}}{n}+H(A-N)\big]$, and the term which multiplies
$\frac{\dot{L}'}{L}$
is $\bar{f}=\frac{12\alpha\mathcal{X}}{n^{2}}$, i.e. at the brane position it is
$\mathcal{E}^{t}_{r}=\mathcal{E}^{t}_{r}\big{|}_{\frac{L'}{L}=\frac{\dot{L}'}{L}=0,L=\beta}
+\frac{L'}{L}\tilde{f}+\frac{\dot{L}'}{L}\bar{f}$.
\newline
It is
\begin{equation}
\mathcal{E}^{t}_{r}\big{|}_{\frac{L'}{L}=\frac{\dot{L}'}{L}=0,L=\beta}=12\alpha\Big[\frac{\dot{A}}{n^{2}}
+\frac{H}{n}(A-N)\Big]\Big(\mathcal{X}+\frac{2H\dot{\beta}}{n\beta}+\frac{1}{6\alpha}\Big)-12\alpha
\mathcal{X}\frac{N\dot{\beta}}{n^{2}\beta}\,,
\label{das}
\end{equation}
so the $\mathcal{O}(1)$ part of the $rt$ equation is
\begin{eqnarray}
&&\Big[\frac{\dot{A}}{n}+H(A-N)\Big]\Big(\mathcal{X}+\frac{2H\dot{\beta}}{n\beta}
+\frac{1}{6\alpha}+4AN-2A'-\frac{A\beta_{2}}{\beta}\Big)-2A\Big[\frac{\dot{A}'}{n}+(H'+H N)(A-N)
+H(A'-N')\Big]\nn\\
&&+(\mathcal{X}'-3N\mathcal{X})\frac{\dot{\beta}}{n\beta}+\mathcal{X}
\Big(\frac{\dot{\beta}_{2}}{n\beta}-\frac{\dot{\beta}}{n\beta}\frac{\beta_{2}}{2\beta}\Big)
=\frac{n\kappa_{6}^{2}}{12\alpha}\mathcal{T}^{t}_{r}\,.
\label{extra}
\end{eqnarray}
Using the $\mathcal{O}(1/r)$ part of the $rt$ bulk equation, i.e. equation (\ref{sing ri cosmo}),
equation (\ref{extra}) gets the simpler form
\begin{equation}
\frac{\dot{\beta}_{2}}{n\beta}+\frac{H}{A}\frac{\dot{\beta}^{2}}{n^{2}\beta^{2}}+
\Big[\frac{\mathcal{X}'}{\mathcal{X}}-\frac{\beta_{2}}{\beta}-\frac{A'}{A}
+\frac{1}{2A}\Big(\mathcal{X}+\frac{1}{6\alpha}\Big)-N\Big]
\frac{\dot{\beta}}{n\beta}-\frac{2A}{\mathcal{X}}\Big[\frac{\dot{A}'}{n}+(H'+HN)(A-N)+H(A'-N')\Big]
=\frac{n\kappa_{6}^{2}}{12\alpha\mathcal{X}}\mathcal{T}^{t}_{r}
\label{gulang}
\end{equation}
and using $H'=\frac{\dot{\beta}}{2n\beta A}\mathcal{X}-AH$ in  (\ref{gulang}), we get the equivalent to the
regular $rt$ equation
\begin{equation}
\frac{\dot{\beta}_{2}}{n\beta}+\frac{H}{A}\frac{\dot{\beta}^{2}}{n^{2}\beta^{2}}+
\Big[\frac{\mathcal{X}'}{\mathcal{X}}-\frac{\beta_{2}}{\beta}-\frac{A'}{A}
+\frac{1}{2A}\Big(\mathcal{X}+\frac{1}{6\alpha}\Big)-A\Big]
\frac{\dot{\beta}}{n\beta}-\frac{2A}{\mathcal{X}}\Big[\frac{\dot{A}'}{n}+H(A'\!-\!N')\!-\!H(A\!-\!N)^{2}\Big]
=\frac{n\kappa_{6}^{2}}{12\alpha\mathcal{X}}\mathcal{T}^{t}_{r}\,.
\label{gulang2}
\end{equation}

\section{Consistency of the system} \label{Algebraic system}

In this appendix we solve algebraically the system of equations (\ref{sing tt cosmo}), (\ref{sing ij cosmo}),
(\ref{theta regular}) for the unknown $a_{2},n_{2},\beta_{2}$. Then, we substitute these expressions in equations
(\ref{rr regular}), (\ref{rt regular}) which are the regular parts of the $rr$, $rt$ equations and
check that these are automatically satisfied. This process shows the consistency of the whole system for
all kinds of matching conditions.
More precisely, equation (\ref{rr regular}) is an algebraic equation on the set of variables
$a_{2},n_{2},\beta_{2}$, while on the other hand, equation (\ref{rt regular}) is a differential (Bianchi)
equation with respect to the second order variables. Of course, the manipulation of the various terms is
quite complicated and it needs to be organized systematically.

The solution of equations (\ref{sing tt cosmo}), (\ref{sing ij cosmo}), (\ref{theta regular}) for
$a_{2},n_{2},\beta_{2}$ is
\begin{eqnarray}
&&(1+3f)\frac{a_{2}}{a}=\frac{1+3f}{12\alpha}+f\mathcal{X}-\frac{3\mathcal{X}}{2A^{2}}
\Big(\frac{\dot{\beta}}{n\beta}\Big)^{\!2}-\frac{1}{12\alpha\mathcal{X}}
\Big(\kappa_{6}^{2}\mathcal{T}^{\theta}_{\theta}-\Lambda_{6}-\frac{5}{12\alpha}\Big)
\label{sol a}\\
&&(1+3f)\frac{n_{2}}{n}=\frac{1+3f}{12\alpha}-\frac{1\!+\!4f\!+\!f^{2}}{2}\mathcal{X}+\frac{\mathcal{X}}{2A^{2}}
\Big(\frac{\dot{\beta}}{n\beta}\Big)^{\!2}+\frac{1+2f}{12\alpha\mathcal{X}}
\Big(\kappa_{6}^{2}\mathcal{T}^{\theta}_{\theta}-\Lambda_{6}-\frac{5}{12\alpha}\Big)
\label{sol n}\\
&&(1+3f)\frac{\beta_{2}}{\beta A}=-(1+f)-\frac{3}{A^{2}}
\Big(\frac{\dot{\beta}}{n\beta}\Big)^{\!2}-\frac{1}{6\alpha\mathcal{X}^{2}}
\Big(\kappa_{6}^{2}\mathcal{T}^{\theta}_{\theta}-\Lambda_{6}-\frac{5}{12\alpha}\Big)\,.
\label{sol beta}
\end{eqnarray}
Both the regular $rr$, $rt$ equations (\ref{rr regular}), (\ref{rt regular}), or the equivalent equations
(\ref{limani}), (\ref{gulang2}), contain $A',N'$ instead of
$a_{2},n_{2}$. So, it is better to write (\ref{sol a}), (\ref{sol n}) as
\begin{eqnarray}
&&(1+3f)A'=(1+3f)\Big(\frac{1}{12\alpha}-A^{2}\Big)+f\mathcal{X}-\frac{3\mathcal{X}}{2A^{2}}
\Big(\frac{\dot{\beta}}{n\beta}\Big)^{\!2}-\frac{1}{12\alpha\mathcal{X}}
\Big(\kappa_{6}^{2}\mathcal{T}^{\theta}_{\theta}-\Lambda_{6}-\frac{5}{12\alpha}\Big)
\label{sol A}\\
&&(1+3f)N'=(1+3f)\Big(\frac{1}{12\alpha}-N^{2}\Big)-\frac{1\!+\!4f\!+\!f^{2}}{2}\mathcal{X}
+\frac{\mathcal{X}}{2A^{2}}\Big(\frac{\dot{\beta}}{n\beta}\Big)^{\!2}+\frac{1+2f}{12\alpha\mathcal{X}}
\Big(\kappa_{6}^{2}\mathcal{T}^{\theta}_{\theta}-\Lambda_{6}-\frac{5}{12\alpha}\Big)\,.
\label{sol N}
\end{eqnarray}
From (\ref{sol A}), (\ref{sol N}) we get
\begin{equation}
fA'-N'=\frac{f-1}{12\alpha}+f(f-1)A^{2}+\frac{f+1}{2}\mathcal{X}-\frac{\mathcal{X}}{2A^{2}}
\Big(\frac{\dot{\beta}}{n\beta}\Big)^{\!2}-\frac{1}{12\alpha\mathcal{X}}
\Big(\kappa_{6}^{2}\mathcal{T}^{\theta}_{\theta}-\Lambda_{6}-\frac{5}{12\alpha}\Big)\,.
\label{fA N}
\end{equation}

As far as equation (\ref{limani}) is concerned, we use (\ref{N}), (\ref{sing rr cosmo}) to express
$\mathcal{Y}$ in terms of $f,\mathcal{X}$ and $X+Y$ in terms of $f,\mathcal{X},A^{2}$. We also use (\ref{sol A}),
(\ref{fA N}) to find $A',N'$ in (\ref{limani}). Finally, we have everything in
terms of $f,\mathcal{X},A^{2}$ and (\ref{limani}) becomes identically satisfied if
$\mathcal{T}^{r}_{r}=\mathcal{T}^{\theta}_{\theta}$ on the brane. This shows the non-trivial consistency
of the regular $rr$ equation (\ref{rr regular}).

The manipulation of equation (\ref{gulang2}) is more tricky and the starting point is to take the combination
(\ref{sol beta})\,$-\frac{2}{\mathcal{X}}$(\ref{sol A}). Then, we get
\begin{equation}
\frac{\beta_{2}}{\beta A}-\frac{2}{\mathcal{X}}A'=\frac{2}{\mathcal{X}}\Big(A^{2}-\frac{1}{12\alpha}\Big)-1\,.
\label{goya}
\end{equation}
Differentiating (\ref{goya}) with respect to time and using the expression for $\dot{\mathcal{X}}$ with
respect to $\mathcal{X},\mathcal{Y},\beta$, we get some of the difficult terms appearing in (\ref{gulang2})
\begin{equation}
\frac{\dot{\beta}_{2}}{n\beta}+\Big(\frac{\mathcal{X}'}{\mathcal{X}}
-\frac{\beta_{2}}{\beta}\Big)\frac{\dot{\beta}}{n\beta}
-\frac{2A}{\mathcal{X}}\frac{\dot{A}'}{n}=\frac{\beta_{2}}{\beta A}
\frac{\dot{A}}{n}+\frac{X'}{\mathcal{X}}\frac{\dot{\beta}}{n\beta}-\frac{4AA'}{\mathcal{X}^{2}}
H(\mathcal{Y}-\mathcal{X})-\frac{2A}{\mathcal{X}^{2}}\frac{\dot{\mathcal{X}}}{n}\Big(A^{2}-
\frac{1}{12\alpha}\Big)+\frac{4A^{2}}{\mathcal{X}}\frac{\dot{A}}{n}\,\,.
\label{forte}
\end{equation}
Using equations (\ref{fA N}), (\ref{goya}), (\ref{2}), (\ref{3}) in (\ref{forte}) in order to find the coefficient
of $A'$, we obtain extra terms of
(\ref{gulang2})
\begin{eqnarray}
&&\!\!\!\!\!\!\!\!\!\!\!\frac{\dot{\beta}_{2}}{n\beta}+\Big(\frac{\mathcal{X}'}{\mathcal{X}}
-\frac{\beta_{2}}{\beta}-\frac{A'}{A}\Big)\frac{\dot{\beta}}{n\beta}
-\frac{2A}{\mathcal{X}}\Big[\frac{\dot{A}'}{n}+H(A'\!-\!N')\Big]=\Big[\frac{1}{\mathcal{X}}\Big(6A^{2}\!-\!
\frac{1}{6\alpha}\Big)\!-\!1\Big]\frac{\dot{A}}{n}+\frac{2AH}{\mathcal{X}}\Bigg[(1\!+\!3f)A'
\!+\!\frac{\mathcal{X}}{A^{2}}\frac{\dot{\beta}^{2}}{n^{2}\beta^{2}}\nn\\
&&\quad\quad\quad\quad\quad-\frac{X}{H}\frac{\dot{\beta}}{n\beta}-\frac{\dot{\mathcal{X}}}{nH\mathcal{X}}
\Big(A^{2}\!-\!\frac{1}{12\alpha}\Big)+(1-f)\Big(fA^{2}\!+\!\frac{1}{12\alpha}\Big)-\frac{f+1}{2}\mathcal{X}
+\frac{1}{12\alpha\mathcal{X}}\Big(\kappa_{6}^{2}\mathcal{T}^{\theta}_{\theta}\!-\!\Lambda_{6}\!-\!
\frac{5}{12\alpha}\Big)\Bigg]\,.
\label{milan}
\end{eqnarray}
Finally, substituting again $\dot{A}$ from (\ref{3}), $(1\!+\!3f)A'$ from (\ref{sol A}) and
$\dot{\mathcal{X}}$ in (\ref{gulang2}), this becomes an identity, given that $\mathcal{T}^{t}_{r}=0$ on the
brane. This shows the non-trivial consistency of the regular $rt$ equation (\ref{rt regular}).

%%%%%%%%%%%%%%%%%%%%%%%%%%%% BIBLIOGRAPHY %%%%%%%%%%%%%%%%%%%%%%%%%%%%%%%%%%%%%%

\end{document}